\documentclass[largeformat]{interact}

\usepackage{xcolor}
\usepackage{multirow}
\usepackage{ulem}
\usepackage{amsfonts, amsmath, amssymb, bbold}

\usepackage{epstopdf}
\usepackage[caption=false]{subfig}

\usepackage[numbers,sort&compress,merge]{natbib}
\bibpunct[, ]{[}{]}{,}{n}{,}{,}

\theoremstyle{plain}

\theoremstyle{definition}

\theoremstyle{remark}

\newcommand{\bra}[1]{\langle #1 |}
\newcommand{\ket}[1]{| #1 \rangle}

\definecolor{cream}{RGB}{222,217,201}
\definecolor{CadmiumOrange}{rgb}{0.93, 0.53, 0.18}
\newcommand{\hl}[1]{\textcolor{black}{#1}}

\begin{document}

\title{Magnetic coupling between nuclear motion and nuclear spins in molecules}

\author{
\name{Matthias Diez\textsuperscript{a,b}, Johannes K. Krondorfer\textsuperscript{a}, Albert Hirtenfelder\textsuperscript{a} and Andreas W. Hauser\textsuperscript{a}\thanks{andreas.w.hauser@gmail.com}}
\affil{\textsuperscript{a}Institute of Experimental Physics, Graz University of Technology, Petersgasse 16, A-8010 Graz;
\textsuperscript{b}Institute of Physics, University of Graz, NAWI Graz, 
Universit{\"a}tsplatz 5, A-8010 
Graz}
}

\maketitle

\begin{abstract}
Among the possible types of magnetic dipole interactions in molecular systems, couplings between nuclear motion and the nuclear spin have probably received the least attention in molecular spectroscopy. Although very small in comparison to effects related to electron spin, this type of hyperfine interaction plays an important role in the NMR spectroscopy of molecular systems. While measurement and prediction of spin-rotation tensors are a common place, vibrationally induced effects still lack a comprehensive description.

In this article we develop a generic, theoretical framework that is well embedded in modern electronic structure theory and inspired by the Breit-Pauli Hamiltonian for electronic interactions, distinguishing between nuclear spin-orbit and spin-other-orbit contributions. We show that the interaction of nuclear spins with pseudorotational excitations of highly symmetric molecules may lead to experimentally accessible hyperfine splittings in NMR spectra, triggered by infrared light.

\end{abstract}


\section{Introduction}
The interaction of molecules with magnetic fields has long served as a sensitive probe of their internal structure and dynamics. Beyond the fundamental interactions of electron and nuclear spins with external magnetic fields, magnetic effects can also arise from couplings between nuclear motion -- such as rotation and vibration -- with external fields or spin degrees of freedom. In closed-shell molecules, where the typically dominant electronic Zeeman effect vanishes, dynamically induced magnetic effects stemming from nuclear motion become accessible.

Magnetic interactions induced by nuclear motion have been studied since the 1950s for rotational and the 1970s for vibrational Zeeman effects, both theoretically and experimentally.\cite{RAMSEY-1950,PhysRev.112.1929,PhysRev.85.24,Moss1972,Flygare1974,Braun1981} These effects originate from the magnetic moments generated by moving nuclei, leading to measurable energy shifts in external magnetic fields, as observed in high-resolution rotational spectra and magnetic vibrational circular dichroism experiments.\cite{Devine1984,Wang1993,Wang1994, Keiderling-Bour} While the understanding of rotationally induced magnetization is conceptually straight-forward, a magnetic response due to vibrational excitation necessitates molecular pseudorotation as a prerequisite: only in molecules featuring suitable, degenerate vibrational modes, a coherent excitation with a defined phase lag may induce circular nuclear motion that carries angular momentum and produces a magnetic moment. The underlying mechanism has long been recognized in molecular physics~\cite{Bersuker2006,Domcke15,Domcke17}, where it plays a role in intramolecular rearrangements such as the Berry mechanism in highly symmetric molecules.\cite{Berry1960} More recently, the same principle has been re-branded within the solid-state community as an effect of `dynamical multiferroicity'~\cite{Juraschek17,Juraschek19,Juraschek20}, where degenerate phonon modes in periodic structures have been suggested as triggers for an overall magnetization of bulk materials upon vibrational excitation. Although studied in separate contexts, the underlying physics of both phenomena is the same. In a recent work on the class of phthalocyanines,~ \cite{Wihelmer2024} a topical platform for molecular magnetism, we demonstrated that both phenomena can be understood within a common framework of vibrationally induced magnetism. However, we also illustrated that simplified, semi-classical treatments based on fractional charges are not accurate enough on the molecular level. Hence, we developed a generic, theoretical framework, embedded in electronic structure theory, and suggested the usage of nuclear spins as intramolecular probes of the actual field geometry generated by nuclear motion. The latter step linked our efforts back to much earlier approaches and achievements within molecular NMR spectroscopy, but requires a careful revision and extension of known theories.

\begin{figure}[!h]
    \centering
    \includegraphics[width=0.5\linewidth]{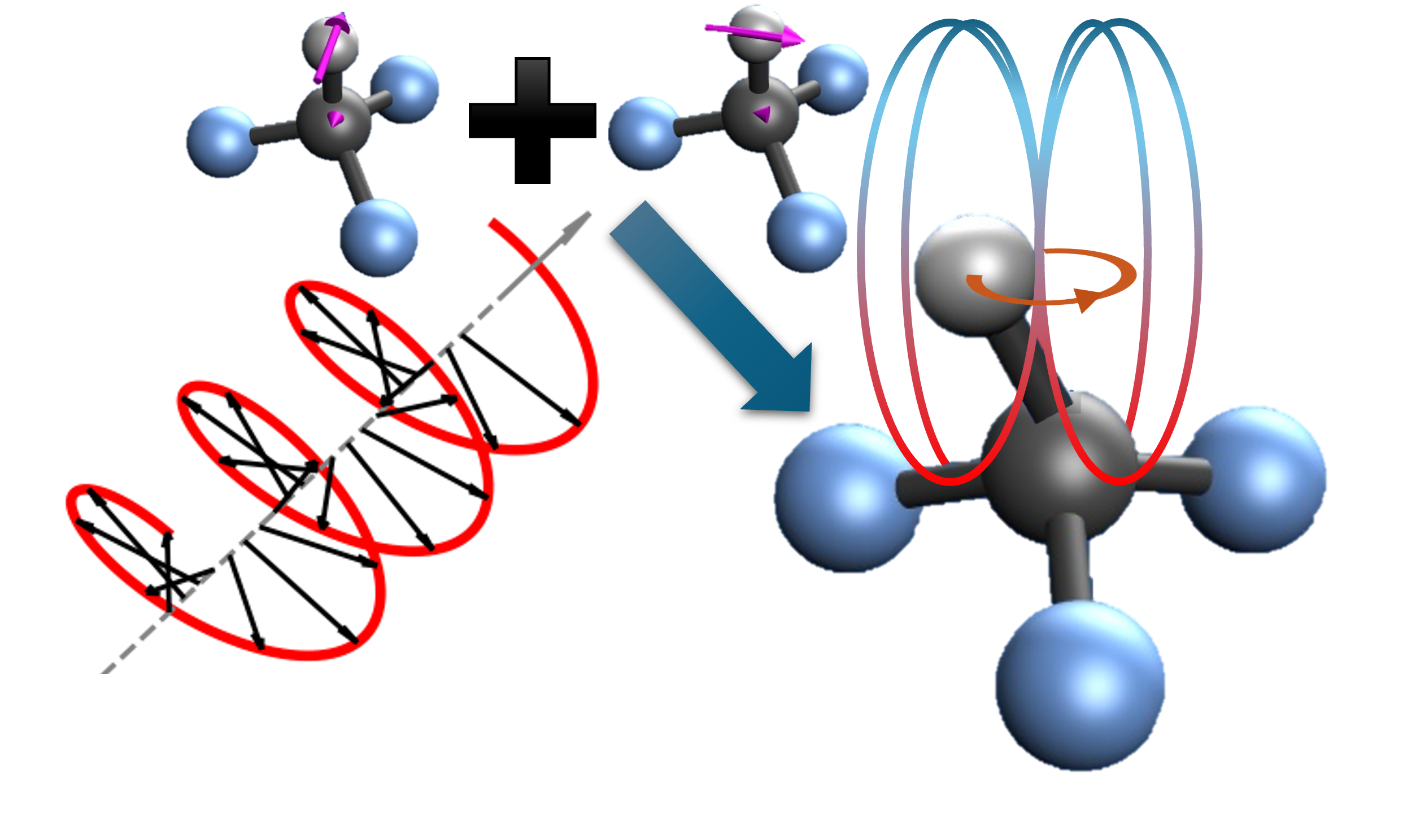}
    \caption{\hl{Scheme of spin-vibration coupling in trifluoromethane. The excitation of two degenerate vibrational modes (atomic displacement vectors are shown in purple) with a phase shift of $\pi/2$ leads to a circular motion of the hydrogen nucleus (gray), which then generates a local magnetic field that can couple to nearby nuclear spins.}}
    \label{fig:Schematic graphic}
\end{figure}

Besides magnetic interactions with external fields, nuclear motion can also couple to internal degrees of freedom, most notably to the nuclear spin.\cite{Flygare1974, Gregory-2016} In the case of molecular rotation, this gives rise to nuclear spin-rotation coupling, a well-known hyperfine interaction that plays a central role in the interpretation of high-resolution rotational and NMR spectra. A formally analogous interaction between vibrational motion and nuclear spin has been studied in CH\textsubscript{4}~\cite{Uehara-Kiyoji}, but treatments in other molecules have only been proposed by us recently.\cite{Wihelmer2024} 
\hl{As illustrated in Figure~\ref{fig:Schematic graphic}, the excitation of two infrared-active, degenerate vibrations with circularly polarized light may lead to a more or less circular motion of certain nuclei. Those nuclei then carry angular momentum, and the corresponding magnetic dipole moment couples to the individual nuclear spins. Concrete details on the actual experimental realization of such an optical excitation were only given recently for a closely related effect in the bulk, where chiral phonons are coupled to the electron spin in a paramagnetic material.\cite{ChiralPhonons2023}}

More generally over the past decades, considerable effort has gone into improving the theoretical and computational treatment of molecular magnetic properties, including rotational g-tensors and spin-rotation constants.~\cite{Wick1948,Ramsey1951, Gauss1996, Ruud2014} Nowadays, the use of (rotational) London orbitals and origin-independent electronic Hamiltonians are state-of-the-art for an accurate evaluation of magnetic response properties.\cite{Helgaker1991,Gauss1996, Helgaker1998, Benchmarking2013, Wong2023} However, while these advances have enabled highly accurate calculations of rotational magnetic effects, an analogous treatment of nuclear spin-vibration coupling has yet to be developed. In particular, there is currently no general framework that describes how a magnetic field, generated by vibrational nuclear motion, interacts with nuclear spin degrees of freedom, a question also highly relevant in the field of quantum information materials.\cite{Zhou2024nat,Mattioni2024}

In this work, we present a unified and generic theoretical treatment of nuclear spin-rotation and nuclear spin-vibration coupling, derived entirely from first principles. Our formalism extends and generalizes earlier approaches,
by explicitly identifying the contributions of nuclear spin-orbit coupling (NSOC) and nuclear spin-other-orbit coupling (NSOOC) in close analogy to the well-known Breit-Pauli Hamiltonian~\cite{Breit-1932} for electrons. This is possible since any type of relevant magnetic interaction is a form of dipole coupling, with the magnetic dipole being intrinsically related either to the angular momentum or spin of a charged particle, be it electrons or atomic nuclei. As a key result, we demonstrate that spin-vibration coupling gives rise to vibrationally induced level splittings, representing a novel type of hyperfine interaction. These changes, although small, may be accessible in NMR experiments e.g. as additional chemical shifts and could open new pathways for nuclear spin control via vibrational excitation. We illustrate the theory via benchmark calculations on trihalomethanes, benzene, triazine, and 1-3-5-substituted benzene derivatives. These molecules are selected for their high symmetry and the fact that, for all of them, a vibrational Zeeman effect has been  confirmed by experiment.~ \cite{Devine1984, DEVINE1986, Wang1994}  Thereby, we provide both a testing ground for our theoretical work and guidance for future experimental investigation.

\section{\label{sec:theory} Theoretical background}

We begin by summarizing the theoretical background and introduce a notation and nomenclature that is kept throughout the entire manuscript. The indices $a,b,c$ refer to individual nuclei, $i,j,k$ refer to electrons, and $\alpha,\beta,\gamma$ denote Cartesian coordinates. 
We consider a molecule consisting of $N_{\mathrm{N}}$ nuclei and $N_{\mathrm{e}}$ electrons. Their respective positions in a Cartesian coordinate system will be denoted by $\bm{R}_{a}$ and $\bm{r}_{i}$, and their masses by $m_{a}$ and $m_i$. The complete set of positions within a given molecule is denoted as $\bm{R}$ and $\bm{r}$ for nuclei and electrons, respectively. Furthermore, we introduce the distance vector $\bm{r}_{i,a} = \bm{r}_i - \bm{R}_a$ pointing from electron $i$ to nucleus $a$, as well as the vector $\bm{R}_{a,b} = \bm{R}_a - \bm{R}_b$. The total molecular Hamiltonian $H$ is the sum of an electronic part $H_{\mathrm{e}}$ and the kinetic energy of the nuclei $T_{\mathrm{N}}$. An arbitrary eigenstate $\Psi$ of the molecular Hamiltonian with energy $E$ may be expanded in the basis of electronic excitations $\psi^{\mathrm{e}}_n(\bm{r}|\bm{R})$ with purely nuclear coefficients $\chi_{n}(\bm{R})$,
\begin{equation}
     \Psi(\bm{r},\bm{R}) = \sum_n\psi^{\mathrm{e}}_n(\bm{r}|\bm{R}) \chi_{n}(\bm{R})\,,
\end{equation}
where $\psi^{\mathrm{e}}_n(\bm{r}|\bm{R})$ obeys the electronic Schr\"odinger equation for fixed nuclear position
\begin{equation}
    H_{\mathrm{e}}({\bm{R}}) \psi_n^{\mathrm{e}}({\bm{r}}|{\bm{R}}) = E_n^{\mathrm{e}}(\bm{R}) \psi_n^{\mathrm{e}}({\bm{r}}|{\bm{R}})\,,
\end{equation}
with the potential energy surface $E_n^{\mathrm{e}}(\bm{R})$ as its corresponding, geometry-dependent eigenvalue. This is possible as the ${\psi^{\mathrm{e}}_n}$ form a complete set of functions for fixed $\bm{R}$. The total stationary Schr\"odinger equation then reads
\begin{equation}
    E \Psi = \sum_n\left[E^{\mathrm{e}}_n\psi^{\mathrm{e}}_n \chi_{n} + \psi^{\mathrm{e}}_n  T_{\mathrm{N}} \chi_{n} \right] + \sum_n \left[(T_{\mathrm{N}} \psi^{\mathrm{e}}_n)  \chi_{n}-\sum_{a=1}^{N_{\mathrm{N}}} \frac{\hbar^2}{m_a} ({\nabla}_a \psi^{\mathrm{e}}_n)\cdot{\nabla}_a\chi_{n}\right]\,.
    \label{eq:total-Schroedinger}
\end{equation}
Usually, the second term is neglected, and the zeroth-order total molecular wave function may be written in the Born-Oppenheimer approximation as a product $\Psi_{n,\nu}^{(0)}(\bm{r},\bm{R}) = \psi_n^e(\bm{r}\vert\bm{R})\chi_{n,\nu}(\bm{R})$.~\cite{Born1927} Here $n$ refers to the $n$-th electronic eigenstate, whereas $\nu$ denotes a not yet further specified set of quantum numbers describing nuclear motion.
With these approximations, the nuclear Schr\"odinger equation reduces to
\begin{equation}
    \left[T_{\mathrm{N}} + E^{\mathrm{e}}_n(\bm{R})\right] \chi_{n,\nu}= E_{n,\nu} \chi_{n,\nu}\,,
    \label{eq:Schrödinger-nuclei}
\end{equation}
with $E_{n,\nu}$ denoting the total energy for the system in electronic state $n$ and nuclear state $\nu$.

\subsection{Watson Hamiltonian}
\hl{A practical description of molecular rotation or pseudo-rotation requires the choice of an appropriate coordinate system. Preferably, the latter should also be convenient for the coupling of nuclear and electronic degrees of freedom, providing a useful starting point for the perturbation of the electronic wavefunction due to nuclear motion.} The solution of Equation~\eqref{eq:Schrödinger-nuclei} for a harmonic expansion of the potential energy surface in the electronic ground state leads to the definition of normal coordinates. Performing a coordinate transformation of the nuclear Cartesian set into normal coordinates and Euler angles, as well as choosing the Eckart frame~\cite{Eckart1935}, the kinetic energy operator {of the nuclei} $T_{\mathrm{N}}$ takes on the so-called Watson form \cite{Watson1968},
\begin{equation}
    T_{\mathrm{N}} =
    \frac{1}{2}(\bm{J} - \bm{G} - \bm{L}_e) \Theta^{-1} (\bm{J} - \bm{G} - \bm{L}_e) 
    + \frac{1}{2}\sum_{r=1}^{3N_{\mathrm{N}}-6} P_r^2 - \frac{\hbar^2}{8} \sum_{\alpha = 1}^{3} \Theta^{-1}_{\alpha \alpha},\,
    \label{eq:Watson-Hamiltonian}
\end{equation}
with $\bm{J}$, $\bm{G}$ and $\bm{L_e}$ denoting the total, vibrational and electronic angular momenta with respect to the rotating axis of the molecular frame. Nuclear momenta in normal coordinate form are denoted as $P_r$, normal coordinates as $Q_r$. For small displacements of the nuclei from their equilibrium position, $\Theta$, the instantaneous inertia tensor of the molecule, can be considered constant and may be replaced by the inertia tensor $\Theta_{0}$ of the undistorted molecule with respect to its center of mass. Throughout this work, the center of mass is chosen as the coordinate origin. The last term in Equation~\eqref{eq:Watson-Hamiltonian} represents a small, mass-like correction term which is usually neglected. \cite{Meyer2002} The vibrational angular momentum $\bm{G}$ may be expressed as a sum over all pairs $t = (t_1,t_2)$ of vibrational modes,
\begin{equation}
    \bm{G} = \sum_{t} \bm{\zeta}_t G_t\,, \quad\text{with} \quad \bm{\zeta}_t = \sum_{a=1}^{N_\mathrm{N}}\bm{l}_{a,t_1}\times\bm{l}_{a,t_2}\,,
\end{equation}
denoting the Coriolis coupling constant, which describes the area spanned by the two normalized, mass-weighted eigenvectors $\bm{l}_{a,t_i}$ of the respective vibrations (see Ref.~\cite{wilson1980molecular} or section S1 in the SI(=Supplementary Information) for further details), and $G_t = Q_{t_1} P_{t_2} - Q_{t_2} P_{t_1}$ being the vibrational angular momentum of pair $t$ in normal coordinates.

\subsubsection{Vibrational eigenstates with vibrational angular momentum}
In harmonic approximation, the expectation value of $G_t$ with respect to Cartesian eigenfunctions vanishes. However, in the case of two degenerate vibrational states, one may form linear combinations of the respective Cartesian eigenfunctions yielding a non-zero expectation value of the vibrational angular momentum $G_t$. Within this subspace, it is possible to choose a cylindrical basis in which $G_t$ is diagonal. These cylindrical basis functions may be written as
\begin{equation}
    \ket{\chi_t} = \ket{\nu_t, l_{t}},
\end{equation}
where $\nu_t = 0, 1, 2, ...$ is the vibrational quantum number and $l_{t} = -\nu_t, -\nu_t +2, ... \, \nu_t-2, \nu_t$ is the angular momentum quantum number. The action of the vibrational Hamiltonian $H_t$ in harmonic approximation in the subspace of the degenerate modes, and the action of the corresponding angular momentum operator are
\begin{equation}
    H_t \ket{\nu_t, l_{\nu}} = \hbar \omega_t(1 + \nu_t) \ket{\nu_t, l_{t}}\,,
\end{equation}
\begin{equation}
    G_t \ket{\nu_t, l_{t}} = \hbar l_{t} \ket{\nu_t, l_{t}}\,.
\end{equation}
Details of the two-dimensional harmonic oscillator in a cylindrical basis can be found in Section~S2 of the SI, together with information on molecular systems of even higher symmetry. While higher degeneracies of the vibrational subspaces are possible, the highest degeneracy explicitly considered in this article will be twofold for practical reasons. \hl{In the case of threefold degenerate vibrational states three linearly independent normal mode vectors may be chosen, and three Coriolis coupling constants can be determined, which are not necessarily linearly independent. However, if they are, the vibrational angular momentum may point in an arbitrary direction. This additional flexibility would complicate the overall analysis, but is of little practical relevance, since the wave vector of the incoming, circularly polarized light then sets the actual plane of nuclear motion.}

\subsubsection{Corrected electronic eigenstates}
 The Eckart frame is an orthonormal body-fixed frame in which the coupling of rotational and vibrational motion is minimal.
 At the equilibrium geometry, it fully separates the six external degrees of freedom, corresponding to overall translation and rotation of a molecule, from the 3N-6 internal degrees or vibrational modes. \cite{Eckart1935} Within the Eckart frame, the coupling term $-(\bm{J}-\bm{G)} \Theta_0^{-1}\bm{L}_{\mathrm{e}}$ appears in Equation ~\eqref{eq:Watson-Hamiltonian}. If added to the electronic Hamiltonian, we obtain $H_{\mathrm{e}}'=H_{\mathrm{e}} - (\bm{J}-\bm{G)} \Theta_0^{-1}\bm{L}_{\mathrm{e}}$. Within the Born-Oppenheimer picture, we may interpret $\bm{J} - \bm{G}$ as purely nuclear operators which do not act on the electronic state. Treating this coupling as a small perturbation to the electronic Hamiltonian, we therefore obtain a correction to the electronic eigenfunction in first-order perturbation theory of the form
\begin{equation}
    \ket{\psi_n^{\mathrm{e}(1)}} = -(\bm{J} - \bm{G}) \Theta_0^{-1} \sum_{k\neq n} \ket{\psi^{\mathrm{e}(0)}_k} \dfrac{\bra{\psi^{\mathrm{e}(0)}_k} \bm{L}_e\ket{\psi^{\mathrm{e}(0)}_n}}{E_n^{\mathrm{e}}-E^{\mathrm{e}}_k}\,, 
\end{equation}
where $\bm{J} - \bm{G}$ has been pulled out of the integral over the electron coordinates, \hl{neglecting the action of the nuclear momenta on the electronic states}. For the corrected states one obtains $\ket{\psi^{\mathrm{e}}_n}  = \ket{\psi^{\mathrm{e}(0)}_n} +\ket{\psi^{\mathrm{e}(1)}_n} + \mathcal{O}(\|\Theta_0^{-1}(\bm{J}-\bm{G})\|^2)$, as it was considered in Ref.~\cite{Moss1972} in order to calculate the total magnetic moment of a pseudorotating molecule.
In first-order correction, the potential energy surface remains unaffected. Hence, the nuclear wave functions are not perturbed, and the total wavefunction is the product of the corrected electronic state and the uncorrected nuclear state. The expectation value of an observable $O$, to first order, is given as 
\begin{equation}
    \bra{\Psi} O \ket{\Psi} = \bra{\chi_{n,\nu}} \left[ \bra{\psi_n^{\mathrm{e}(0)}} O\ket{\psi_n^{\mathrm{e}(0)}} + \bra{\psi_n^{\mathrm{e}(1)}} O\ket{\psi_n^{\mathrm{e}(0)}}+\bra{\psi_n^{\mathrm{e}(0)}} O\ket{\psi_n^{\mathrm{e}(1)}} \right]\ket{\chi_{n,\nu}}\,. 
\end{equation}

If the action of the nuclear momenta on the electronic states is not neglected, the non-adiabatic terms need to be taken into account as shown in Ref.~\cite{Braun1981}.

\subsection{Nuclear spin orbit coupling}
\begin{figure}[!h]
\centering
\subfloat[Spin-orbit-coupling\label{fig:soc}]{%
\resizebox*{5cm}{!}{\includegraphics{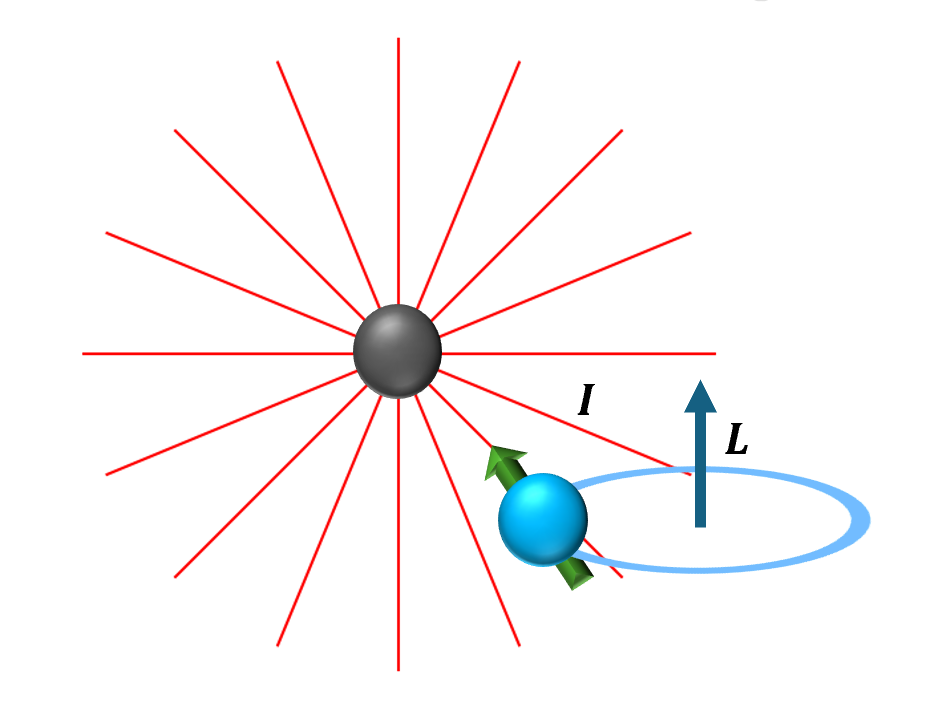}}}\hspace{5pt}
\subfloat[Spin-other-orbit coupling\label{fig:sooc}]{%
\resizebox*{5cm}{!}{\includegraphics{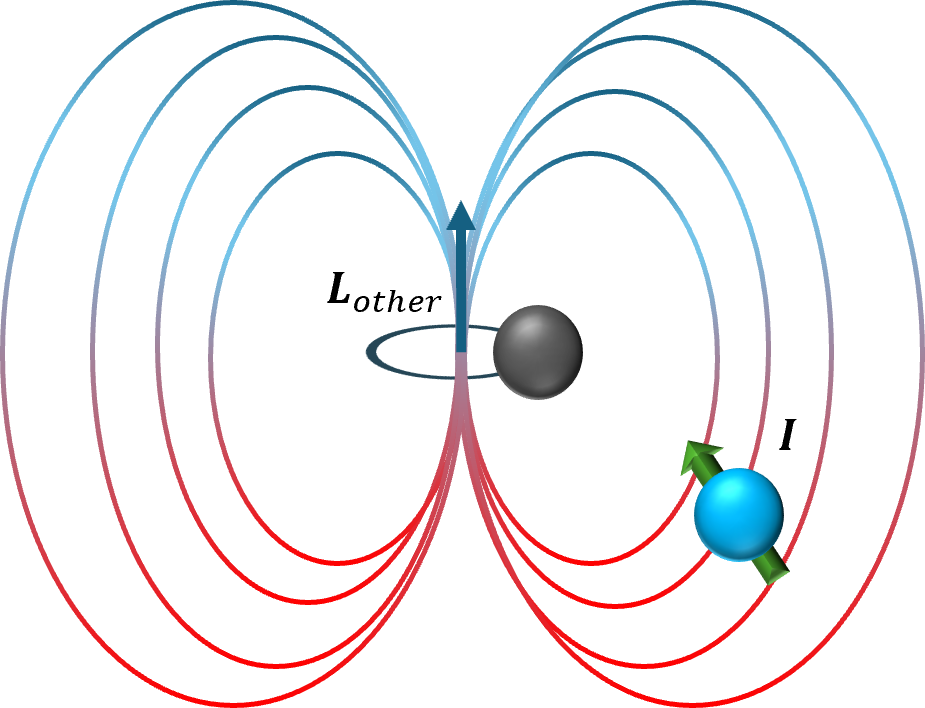}}}\par
\caption{(a) A moving particle with spin $\bm{I}$ and an associated magnetic dipole moment acquires, to first order in $1/c$, an effective electric dipole moment. Its interaction with a static electric field gives rise to spin-orbit coupling (SOC).\newline
(b) A charged particle with angular momentum $\bm{L}_\mathrm{other}$ generates a magnetic field that interacts with the spin $\bm{I}$ of another particle, resulting in spin-other-orbit coupling (SOOC).} \label{fig:Spin-orbit}
\end{figure}

Having discussed how the total wavefunction can be corrected for nuclear motion, specifically for rotational and pseudo-rotational motion, we turn to the fundamental interaction principles between charged moving particles and nuclear spins. The two basic mechanisms by which a nuclear spin $I$ may couple to nuclear motion are illustrated in Figure~\ref{fig:Spin-orbit}. Here we discuss the resulting interaction Hamiltonians \hl{which we motivate by classical electrodynamics. }

The first mechanism, illustrated in Figure~\ref{fig:soc}, is spin-orbit coupling (SOC). Here, the nuclear spin $\bm{I}_a$ of a nucleus moving with momentum $\bm{P}_a$ interacts with the electric field $\bm{E}(\bm{R}_a)$ at its position. In the laboratory frame, the motion of a magnetic dipole leads \hl{according to the theory of relativity} to a small \hl{static} electric dipole moment which interacts with this field. \hl{\cite{Fisher1971} The current distribution of a magnetic dipole with magnetic moment $\bm{\mu}_a$ at rest is $\bm{j}_{a}(\bm{r}) = -(\bm{\mu}_{a} \times \bm{\nabla}) \delta (\bm{r} - \bm{R}_{a}) $. Performing a Lorentz transformation from the momentary rest frame of the particle to the laboratory frame, one obtains the charge density of an electric dipole
\begin{equation}
    \rho (\bm{r}) = - \bm{\nabla} \cdot \left [\frac{\bm{v}_{a} \times \bm{\mu}_{a}}{c^2} \delta (\bm{r} - \bm{R}_{a}) \right] + \mathcal{O}\left(\frac{v^2}{c^2}\right)\,,
\end{equation}
where the electric dipole moment of the moving particle can be identified as $\bm{d}_{a} = \frac{1}{c^2} \bm{v}_{a} \times \bm{\mu}_{a} = \frac{1}{c^2m_a} \bm{P}_a\times\bm{\mu}_a$.}
The total electric field at the nucleus arises from the surrounding charges and is given by
\begin{equation}
    \bm{E}(\bm{R}_a) = \frac{e}{4\pi\varepsilon_0} \left[ \sum_{i=1}^{N_{\mathrm{e}}} \frac{\bm{r}_{i,a}}{r^3_{i,a}} + \sum_{b\neq a}^{N_\mathrm{N}} Z_b\frac{\bm{R}_{a,b}}{R^3_{a,b}} \right]\,.
\end{equation}
The corresponding Hamiltonian due to the interaction of $\bm{E}(\bm{R}_a)$ with dipole moment $\bm{d}_a$ for nuclear spin-orbit coupling then reads
\begin{equation}
    H_{\mathrm{NSOC}} =   -\sum_{a=1}^{N_{\mathrm{N}}} \gamma_{a} f_{a} \bm{I}_{a} \cdot \frac{(\bm{E}(\bm{R}_{{a}}) \times \bm{P}_{a})}{m_{a} c^2}\,,
    \label{eq:NSOC-Hamiltonian}
\end{equation} 
where $ f_{a} =  ( 1 - \frac{Z_{a}e \hbar}{2m_{a} \gamma_{a}})$ accounts for Thomas precession \cite{THOMAS1926}, $\gamma_{a}$ denotes the gyromagnetic factor of the respective nuclear magnetic moment, and $c$ is the velocity of light.

  The second mechanism, illustrated in Figure~\ref{fig:sooc}, is spin-other-orbit coupling (SOOC). In this case, the nuclear spin couples to the magnetic field generated by the orbital motion of other particles. This interaction is fundamentally of magnetic dipole–dipole character and includes both electronic and nuclear contributions.  \hl{Alternatively, this can be seen as the interaction of a magnetic dipole with the magnetic fields generated by a distribution of currents.
The current density of a classical point-like particle $i$ at position $\bm{r}_i$ is given by $\bm{j}(\bm{r}) = \frac{Q_i}{m_i} \bm{p}_{{i}} \delta(\bm{r}-\bm{r}_i)$. The magnetic field generated by the motion of this particle can then, according to the law of Biot-Savart, be written as
\begin{equation}
    \bm{B}(\bm{r}) = -\frac{\mu_0}{4 \pi} \frac{Q_i}{m_i}\frac{(\bm{r} - \bm{r}_i) \times \bm{p}_{{i}}} {|\bm{r} - \bm{r}_i|^3}\,.
\end{equation}
Hence, the magnetic field at the position of a nucleus $a$ due to the motions of all electrons and all other nuclei in a molecule may be written as
\begin{equation}
    \bm{B}(\bm{R}_{a}) =  \frac{\mu_0}{4 \pi} e \left[ - \frac{1}{m_\mathrm{e}}\sum_{i=1}^{N_{\mathrm{e}}} \frac{\bm{L}_{i,a}} {r_{i,a}^3}  + 
    \sum_{b \neq a}^{N_{\mathrm{N}}} \frac{Z_{b}}{m_{b}}\frac{\bm{L}_{b,a}} {R_{a,b}^3}
    \right ],
    \label{eq:mag.field-current}
\end{equation}}
with the vacuum permeability $\mu_0$, and relative angular momenta $\bm{L}_{i,a} = \bm{r}_{i,a}\times\bm{p}_i$ and $\bm{L}_{b,a} = \bm{R}_{b,a}\times\bm{P}_b$, of electrons and nuclei, respectively.
\hl{These magnetic fields interact with the magnetic moment $\bm{\mu}_{a}= \gamma_a \bm{I}_{a}$ of nucleus $a$}.
 The corresponding Hamiltonian is given by
\begin{equation}
  H_{\mathrm{NSOOC}} = \sum_{a=1}^{N_\mathrm{N}} \gamma_{a} \bm{I}_{a} \cdot \frac{\mu_0}{4 \pi} e \left[ \frac{1}{m_e}\sum_{i=1}^{N_{\mathrm{e}}} \frac{\bm{L}_{i,a}} {r_{i,a}^3} -
    \sum_{b \neq a}^{N_\mathrm{N}} \frac{Z_{b}}{m_{b}}\frac{\bm{L}_{b,a}} {R_{a,b}^3}
    \right ]\,.
\end{equation}
Both terms $H_{\mathrm{NSOOC}}$ and $ H_{\mathrm{NSOC}}$ describe relativistic effects, which, in the case of electrons, may be treated within the Breit-Pauli Hamiltonian.\cite{Breit-1932, Howard1970} \hl{We note that the transfer and generalization of this approach to particles with different masses and spins is an obvious, yet \textit{ad hoc} measure we take. Also note that, in an external magnetic field, the momentum of the electrons has to be replaced by $\bm{p}_i + e\bm{A}(\bm{r}_i)$, with $\bm{A}$ as the vector potential at the position of the electron.}


\subsubsection{Taylor expansion of the spin-interaction Hamiltonian}
For small rotational and pseudo-rotational excitations, we only assume small deviations of the nuclei from their equilibrium positions $\bm{R}_{a} = \bm{R}^0_{a} + \delta\bm{ R}_{a}$.We use the Taylor series expansion
\begin{equation}
    \frac{\bm{x} + \delta \bm{x}}{|\bm{x} + \delta \bm{x}|^3} \approx \frac{\bm{x}}{|\bm{x}|^3} + \left[\frac{ \delta \bm{x}}{|\bm{x}|^3} - 3
    \frac{ \bm{x} (\bm{x} \cdot \delta \bm{x})}{|\bm{x}|^5}\right]\,,
    \label{eq:Taylor-expansion}
\end{equation}
which remains valid for $|\delta \bm{x}| \ll |\bm{x}|$. 
Applied to the electric field $\bm{E}(\bm{R}_a)$, we obtain
\begin{equation}
    \bm{E}(\bm{R}^0_a + \delta\bm{R}_a) \approx \bm{E}_{\mathrm{e}}(\bm{R}^{0}_a)  + \bm{E}_{\mathrm{N}}(\bm{R}^{0}_a) + \left[-\varphi_{\mathrm{el}} (\bm{R}_a^0) + \frac{ e}{4\pi\varepsilon_0}
    \sum_{b\neq a}^{N_{\mathrm{N}}} Z_b \varphi_{a,b}(\bm{R}^0)\right]\delta \bm{R}_a\,,
    \label{eq:field-expansion}
\end{equation}
where we introduced the electric field gradient $\varphi_{\mathrm{el}}(\bm{R}^0_a)$ caused by all electrons at the position of nucleus $a$, as well as $\varphi_{a,b}(\bm{R}^0)$ as the charge-normalized contribution to the field gradient at position $a$ by nucleus $b$. These quantities themselves are given by
\begin{align}
    \varphi_{a,b}(\bm{R}^0) &=  \left[\frac{\mathbb{1}}{{R^0_{a,b}}^3} - \frac{3\bm{R}^0_{a,b} (\bm{R}^0_{a,b})^T}{{R^0_{a,b}}^5}\right],\label{eq:phi ab}\\
    \varphi_{\mathrm{el}}(\bm{R}^0_a) &= -\frac{e}{4\pi \varepsilon_0} \sum_i \left[  \frac{3\bm{r}^0_{i,a} (\bm{r}^0_{i,a})^T}{({r^0_{i,a}})^5} - \frac{\mathbb{1}}{({r^0_{i,a}})^3} \right]\,,
\end{align}
where $\mathbb{1}$ denotes a $3\times3$ unit matrix, and $\bm{x}_1\bm{x}_2^T$ is the outer product of two column vectors,  $\bm{x}_1$ and $\bm{x}_2$.
The expansion in Equation~\eqref{eq:field-expansion} is now used in the spin-orbit Hamiltonian.  Distinguishing between zeroth and first-order terms, as well as between the type of generating particle (electrons or nuclei), one obtains
\begin{equation} \label{eq:termlist}
    H_{\mathrm{NSOC}} = H_{\mathrm{NSOC}}^{e,(0)} + H_{\mathrm{NSOC}}^{N,(0)} + H_{\mathrm{NSOC}}^{e,(1)} + H_{\mathrm{NSOC}}^{N,(1)}\, ,
\end{equation}
with 
\begin{align}
    H_{\mathrm{NSOC}}^{e,(0)} &= -\sum_{a=1}^{N_{\mathrm{N}}} \frac{f_{a}} {m_{a} c^2} \gamma_{a} \bm{I}_{a} \cdot 
    (\bm{E}_{\mathrm{e}}(\bm{R}^0_a)  \times \bm{P}_{a})\,, \\
    H_{\mathrm{NSOC}}^{N,(0)} &= -\sum_{a=1}^{N_\mathrm{N}} \frac{f_{a}} {m_{a} c^2} \gamma_{a} \bm{I}_{a} \cdot 
   (\bm{E}_{\mathrm{N}}(\bm{R}^0_a) \times \bm{P}_{a})\,, 
   \label{eq:Nuclear-Spin-orbit-equi}\\
    H_{\mathrm{NSOC}}^{e,(1)} &= \sum_{a=1}^{N_\mathrm{N}} \frac{f_{a} }{m_{a}c^2} \gamma_{a} \bm{I}_{a} \cdot 
    (\varphi_{\mathrm{el}}(\bm{R}_a^0) \delta\bm{ R}_{a} ) \times \bm{P}_{a}\,, \\
    H_{\mathrm{NSOC}}^{N,(1)} &= -\sum_{a=1}^{N_\mathrm{N}} \sum_{b \neq a}^{N_\mathrm{N}}  \frac{f_{a}e\mu_0 Z_{b} }{ m_{a} 4\pi } \gamma_{a} \bm{I}_{a} \cdot
    \left( \varphi_{a,b}(\bm{R}^0) \delta\bm{ R}_{{a, b}} ) \right) \times \bm{P}_{a}\, ,
\end{align}
and the definition $\delta\bm{ R}_{a, b} =  \delta\bm{ R}_{a} - \delta\bm{ R}_{b}$.

Similarly, for spin-other-orbit coupling, we use Equation~\eqref{eq:Taylor-expansion} to expand $\frac{\bm{L_{b,a}}}{R_{a,b}^3} = \frac{\bm{R}_{b,a}}{R_{a,b}^3} \times \bm{P}_b$ and  $\frac{\bm{L_{i,a}}}{r_{i,a}^3} = \frac{\bm{r}_{i,a}}{{{r_{i,a}^3}}} \times \bm{p}_i$, arriving at
\begin{align}
        H_{\mathrm{NSOOC}}^{\mathrm{e},(0)} &=  \frac{\mu_0 e}{4 \pi m_e} \sum_{a=1}^{N_{\mathrm{N}}} \gamma_{a} \bm{I}_{a} \cdot  \sum_{i=1}^{N_{\mathrm{e}}} \frac{\bm{L}^0_{i,a}} {{(r^0_{i,a})}^3}  \,, \\
    H_{\mathrm{NSOOC}}^{\mathrm{N},(0)} &= -\frac{\mu_0e}{4 \pi} \sum_{a=1}^{N_{\mathrm{N}}} \gamma_{a} \bm{I}_{a} \cdot \sum_{b \neq a}^{N_{\mathrm{N}}}  \frac{Z_{b}}{m_{b}} \frac{\bm{L}^0_{b,a}} {{(R^0_{a,b})}^3}\,,
    \label{eq:Nuclear-Spin-Other-Orbit-equi}\\
    H_{\mathrm{NSOOC}}^{\mathrm {e},(1)} &= -\frac{\mu_0e}{4 \pi m_e} \sum_{a=1}^{N_{\mathrm{N}}} \sum_{i=1}^{N\mathrm{e}} \gamma_{a} \bm{I}_{a} \cdot (\varphi_{i,a}(\bm{R}^0) \delta\bm{ R}_{a}) \times \bm{p_i}\,, \\ 
    H_{\mathrm{NSOOC}}^{\mathrm{N},(1)} &=  \frac{\mu_0}{4 \pi} e\sum_{a=1}^{N_{\mathrm{N}}} \gamma_{a} \bm{I}_{a} \cdot 
    \sum_{b \neq a}^{N_{\mathrm{N}}}  \frac{Z_{b}}{m_{b}}  (\varphi_{a,b}(\bm{R}^0) \delta\bm{ R}_{a,b} ) \times 
    \bm{P}_{b}\,,
\end{align}
where $\varphi_{i,a}$ is defined as in Equation~\eqref{eq:phi ab}, with one coordinate vector corresponding to electron~$i$.
We may now take appropriate expectation values of $H_{\mathrm{NSOC}}+H_{\mathrm{NSOOC}}$ with respect to nuclear and electronic coordinates. This allows the introduction of an effective spin interaction Hamiltonian, 
\begin{equation}
   \bra\Psi  H_{\mathrm{NSOC}}+H_{\mathrm{NSOOC}}  \ket\Psi = -\sum_{a=1}^{N_{\mathrm{N}}} \gamma_{a} \bm{I}_{a} \cdot \bm{B}^{\mathrm{eff}}(\bm{R}_a)\,.
    \label{eq:effective-Hamiltonian}
\end{equation}
This interaction may be interpreted as the coupling of all nuclear spins to an effective motion-induced, total magnetic field $\bm{B}^{\mathrm{eff}}(\bm{R}_a)$.

We now have all the tools available to look at special cases of nuclear motion, in particular rotations and vibrations. We do this by an explicit evaluation of the expectation value in Equation~\eqref{eq:effective-Hamiltonian}, followed by the extraction of effective hyperfine couplings or site-dependent effective magnetic fields. Note that all relevant parameters are readily obtained from computational chemistry programs.

\subsubsection{Spin-rotation-coupling}

\label{sec:spin-rotation-coupling}
\hl{The coupling of nuclear spins to rotation is typically captured by the spin-rotation tensor $M^{\mathrm{rot}}_a$, defined as the negative second derivative of the energy with respect to $\langle \bm{J}\rangle$ and the nuclear spin $\bm{I}_a$ at site $\bm{R}_a$,
\begin{equation}
    M^{\mathrm{rot}}_a = -\frac{\partial^2E}{\partial\langle\bm{J}\rangle\partial\bm{I}_a} \Bigg|_{\langle \bm{J} \rangle = 0} \,.
    \label{eq:Spin-rotation-definition}
\end{equation}}
\hl{Note that differing sign conventions for the spin-rotation tensor are used in the literature. In order to determine $M^{\mathrm{rot}}_a$, we need to evaluate Equation~\eqref{eq:effective-Hamiltonian} with respect to the total wave-function of a rotating molecule. We write the latter as a product}
\begin{equation}
    \Psi(\bm{r},\bm{\phi}) = \chi^{\mathrm{rot}} (\bm{\phi}) \psi_0^{\mathrm{e,rot}}({\bm{r}}|{\bm{R}^0}) \,,
\end{equation}
with ${\psi}_0^{\mathrm{e,rot}}$ denoting the electronic ground state wavefunction, now corrected for rotational motion, and ${\chi^{\mathrm{rot}}}$ as the nuclear wavefunction of a purely rotating molecule, where $\bm{\phi} = (\phi_1,\phi_2,\phi_3)$ denotes the Euler angles. Here, we implicitly assume that the rotating molecule remains close to its equilibrium configuration, i.e., $\delta\bm{R}_a\approx 0$ in the rotating frame.  The perturbation of the electronic wave function is induced by $H_{\mathrm{I}} = -\bm{J} \Theta_0^{-1} \bm{L}_{\mathrm{e}}$ from the Watson Hamiltonian in Equation~\eqref{eq:Watson-Hamiltonian}. In its dependence on the electronic angular momentum, this term is similar to the well-known paramagnetic term $H^{\mathrm{para}} = \frac{e}{2 m_\mathrm{e}} \bm{B} \cdot \bm{L}_{\mathrm{e}}$. From this, one can link the electronic part of the spin-rotation tensor to calculations of magnetic shielding effects.
Since geometric distortion due to rotation is neglected, the expectation values of only the  four zeroth-order terms in $H_{\mathrm{NSOC}} + H_{\mathrm{NSOOC}}$ are relevant. Considering the two electronic parts,
\begin{align}
\begin{split}
   \bra{\Psi} H_{\mathrm{NSOOC}}^{\mathrm{e},(0)} & + H_{\mathrm{NSOC}}^{\mathrm{e},(0)} \ket{\Psi} 
         =  \\
        &\!\!\!\!\!=\sum_{a=1}^{N_\mathrm{N}} \gamma_a\bm{I}_a \cdot {\bra{\chi^{\mathrm{rot}}}}\bra{\psi^{\mathrm{e,rot}}_0}  \sum_{i=1}^{N_{\mathrm{e}}} \frac{\mu_0 e}{4 \pi} \left[\frac{1}{m_e} \frac{\bm{L}^0_{i,a}}{({r^0_{i,a})}^3}
      - \frac{f_{a}} {m_{a} } 
    \frac{\bm{r}^0_{i,a}}{{(r^0_{i,a})}^3}  \times \bm{P}_{a} \right]\ket{\psi^{\mathrm{e,rot}}_0}\ket{\chi^{\mathrm{rot}}} \,,
    \label{eq:electronic-spin-rotation}
\end{split}
\end{align}
the first is often referred to as `paramagnetic spin orbit coupling'. \hl{For this term, the dependence on the rotational angular momentum $\bm{J}$ stems only from the perturbed electronic state $\ket{\psi_0^{\mathrm{e,rot}}} $.} The second term in Equation~\eqref{eq:electronic-spin-rotation} describes the spin-orbit interaction of the nuclei with the surrounding electronic electric field. Formally, it is equivalent to Equation~(11) in Ref.~\cite{Gauss1996}, but includes the factor $(f_{a} \!-\! 1)$ for Thomas precession as in Ref.~\cite{Reid}. For its  evaluation we use $\bra{\chi^{\mathrm{rot}}} \bm{r}^0_{i,a}   \times \bm{P}_{a}\ket{\chi^{\mathrm{rot}}} = m_a\left[(\bm{r}^0_{i,a} \cdot {\bm{R}_{a}^0})\mathbb{1} - {\bm{R}_{a}^0} (\bm{r}^0_{i,a})^T\right]\Theta_0^{-1} \langle \bm{J} \rangle$, and that $\ket{\chi^{\mathrm{rot}}}$ only depends on the Euler angles. We can then evaluate the derivative of Equation~\eqref{eq:electronic-spin-rotation} with respect to $\langle \bm{J} \rangle$ and $\bm{I}_a$, and obtain the electronic part of the spin-rotation tensor
\begin{align}
    M_{a}^{\mathrm{rot,e}} = \gamma_a\Bigg(\frac{2m_{\mathrm{e}}}{e}\sigma^\mathrm{para}_a(\bm{R}_{a}^0) \Theta_0^{-1} {+}  \frac{f_{a} -1 }{c^2} \left[ \langle (\bm{E}_{\mathrm{e}} (\mathbf{R}^0_a) \rangle \cdot \bm{R}^0_{a} )\mathbb{1} - \bm{R}^0_{a} \langle \bm{E}_{\mathrm{e}} (\mathbf{R}^0_a) \rangle ^{T}\right] \Theta_0^{-1}\Bigg)\,.
    \label{eq:Spin-rotation-electronic-part}
\end{align}
which depends on the electronic part of the paramagnetic shielding tensor $\sigma^\mathrm{para}_a(\bm{R}_a)$, where the magnetic gauge origin is chosen at the position of the respective nucleus, as proven in Ref.~\cite{RAMSEY-1950}. The second term in Equation~\eqref{eq:Spin-rotation-electronic-part} represents the correction due to Thomas precession. Further details of this derivation are provided in Appendix~\ref{sec:appendix-spin-rotation}, \hl{where we also prove the relation to the paramagnetic shielding tensor}.
For the evaluation of the nuclear contributions 
\begin{equation}
    \bra{\Psi} H_{\mathrm{NSOOC}}^{\mathrm{N},(0)} + H_{\mathrm{NSOC}}^{\mathrm{N},(0)}\ket{\Psi} = \bra{\chi^{\mathrm{rot}}}   {-} \frac{\mu_0e}{4\pi}\sum_{a=1} ^{N_{\mathrm{N}}} \gamma_a \bm{I}_a\cdot\sum_{b\neq a}^{N_{\mathrm{N}}}  \frac{Z_b}{(R^0_{a,b})^3}\left[f_a\frac{\bm{L}^0_{a,b}}{m_a} + \frac{\bm{L}^0_{b,a}}{m_b}\right]\ket{\mathrm{\chi}^\mathrm{rot}}\,
    \label{eq:spin-rotation-H-exp-N}
\end{equation}
 we use $\bra{\chi^{\mathrm{rot}}} \bm{L}^0_{{a,b}}  \ket{\chi^{\mathrm{rot}}} = m_{{a}} [(\bm{R}^0_{a} - \bm{R}^0_{b}) \cdot \bm{R}^0_{{a}} \mathbb{1}- \bm{R}^0_{{a}} (\bm{R}^0_a - \bm{R}^0_b)^T]\Theta_0^{-1} \langle \bm{J} \rangle$.  \hl{After taking the derivatives with respect to the nuclear spin $I_a$ and the angular momentum $\langle \bm{J}\rangle$ of the energy expectation value in Equation~\eqref{eq:spin-rotation-H-exp-N} we get the nuclear contribution to the spin-rotation tensor}
\begin{equation}
    M_{a}^{\mathrm{rot,N}} = \gamma_a\Bigg(\frac{\mu_0 e}{4\pi}\sum_{b \neq a} \frac{Z_{b}}{m_{b}}\frac{\Theta_0^{({b,a})} \Theta_0^{-1} }{(R_{a,b}^0)^3} + \frac{\mu_0e}{4\pi} (f_{a} - 1) \sum_{b \neq a}^{N_\mathrm{N}}Z_{b}\frac{S_{a,b}^0}{(R_{a,b}^0)^3} \Theta_0^{-1}\Bigg)\,,
    \label{eq:spin-rot-nuclear}
\end{equation}
 with $\Theta_0^{({b,a})} = m_{{b}}\left[({R}^0_{a,b})^2 \mathbb{1}- \bm{R}^0_{a,b} (\bm{R}_{a,b}^0)^T\right]$, \hl{which can be interpreted as the moment of inertia of particle ${b}$ with respect to the position of particle ${a}$ at $\bm{R}_{{a}}$}, and $S_{a,b}^0=(\bm{R}^0_{a,b} \cdot \bm{R}^0_{a}) \mathbb{1} - \bm{R}^0_{a} (\bm{R}^0_{a,b})^{T}\,$.
\hl{Details on the steps from Equation~\eqref{eq:spin-rotation-H-exp-N} to Equation~\eqref{eq:spin-rot-nuclear} are provided in Appendix~B, Equation~\eqref{eq:Derivation-spin-rot-nuclear}.}
Combining the nuclear and electronic contribution eventually yields the well-known expression for the total, rotation-induced magnetic field
\begin{equation}
    \bm{B}_{\mathrm{rot}}^{\mathrm{eff}}(\bm{R}_a) =  \left [\frac{2m_{\mathrm{e}}}{e}\sigma^{\mathrm{para}}_a(\bm{R}_a^0) \Theta_0^{-1} + \frac{\mu_0 e}{4\pi}\sum_{b\neq a}^{N_\mathrm{N}} \frac{Z_{b}}{m_{b}}\frac{\Theta_0^{({b,a})} \Theta_0^{-1} }{(R_{a,b}^0)^3} + \frac{1}{\gamma_a}M_{a}^{\mathrm{rot,prec}}\right] \langle\bm{J} \rangle
    = \frac{1}{\gamma_a}M_{a}^{\mathrm{rot}} \langle \bm{J} \rangle   \,,
    \label{eq:B-Spin-rotation}
\end{equation}
with $M_{a}^{\mathrm{rot,prec}}$ denoting the correction of the spin-rotation constant with respect to Thomas precession,
\begin{equation}
    M_{a}^{\mathrm{rot,prec}} = \gamma_a(f_a - 1) \left[ \frac{\mu_0e}{4\pi}  \sum_{b \neq a}^{N_\mathrm{N}}Z_{b}\frac{S_{a,b}^0}{(R_{a,b}^0)^3} \Theta_0^{-1} + \frac{1}{c^2} \left[ ( \langle \bm{E}_{\mathrm{e}} (\mathbf{R}^0_a) \rangle \cdot \bm{R}^0_{a}) \mathbb{1} - \bm{R}^0_{a} \langle \bm{E}_{\mathrm{e}} (\mathbf{R}^0_a) \rangle ^{T}\right] \Theta_0^{-1}\right]\,.
\end{equation}
\hl{The relation of the electronic part of the spin-rotation tensor to the paramagnetic shielding tensor is particularly important, as it provides an experimentally accessible, absolute shielding scale.
\cite{Malkin2013} Furthermore, this relation allows the straightforward implementation of spin-rotation tensor calculations in modern quantum chemistry programs. \cite{Gauss1996} However, note the importance of being able to choose the magnetic gauge origin at the position of the respective nucleus, and that the paramagnetic contribution to the chemical shielding $\sigma_a$ is calculated via $\sigma^{\mathrm{para}}_a(\bm{R}_a) = \sigma_a - \sigma^{\mathrm{dia}}_a(\bm{R}_a)$. Also, for the calculation of the diamagnetic shielding tensor $\sigma^{\mathrm{dia}}_a(\bm{R}_a)$, conventional atomic orbitals must be used. A few additional comments on chemical shielding are summarized in Appendix \ref{sec:Chemical-shielding}.}

The contribution to the total spin-rotation tensor due to Thomas precession is small if the nuclear-induced and electronic-induced electric field at a specific nuclear site nearly cancel out each other. In such a case, the whole contribution due to SOC becomes very small and may be neglected. It is, however, common practice in many implementations, to take SOC into account by calculating the magnetic field in a co-rotating frame without considering Thomas precession.\cite{Gauss1996} We note that even if the total SOC contribution is negligibly small, a meaningful separation of electronic and nuclear contributions always requires the inclusion of Thomas precession.

\subsubsection{Spin-vibration-coupling}
\label{sec:Spin-vibration-coupling}
In the following, we now assume the excitation of a single pseudo-rotational motion, corresponding to the excitation of a single degenerate mode. 
The molecular state $\Psi$ can then be written as 
\begin{equation}
    \Psi(\bm{r},Q) = \chi^{\mathrm{vib}} ( Q) \psi_{0}^{\mathrm{e,vib}}({\bm{r}}|{\bm{R}^0}) \,,
\end{equation}
with $Q$ representing the set of all normal coordinates.
We further assume only small geometric deviations, such that the harmonic approximation remains valid and $\psi_{0}^{\mathrm{e,vib}}$ can again be evaluated at the equilibrium position. However, $\psi_{0}^{\mathrm{e,vib}}$ is corrected by the interaction of vibrational angular momentum and electronic angular momentum, according to Equation~\eqref{eq:Watson-Hamiltonian}. The Hamiltonian perturbing the electronic wave function is now $H_{\mathrm I} = (\bm{G} \Theta^{-1})\cdot \bm{L}_{\mathrm{e}}$.
The nuclear state can be written in tensor product notation as
\begin{equation}
    \ket{\chi^{\mathrm{vib}}_t} =  \bigotimes_{s}\ket{\nu_s} \bigotimes_{r\neq t} \ket{\nu_{r_1}} \ket{\nu_{r_2}} \otimes \ket{\nu_t,l_t}\,,
\end{equation}
with $s$ running over all non-degenerate states and the $r$ over all two-fold degenerate states except for one specific degenerate pair $t=(t_1,t_2)$. Only the latter may carry vibrational angular momentum, or, in other words, no other doubly degenerate vibration is assumed to show a suitable phase relation.
To evaluate the expectation value of the zeroth-order spin-orbit coupling operators, the steps are equivalent to the evaluation of the spin-rotation tensor, where it is sufficient to replace $\bm{J}$ with $-\bm{G}$ in Equation~\eqref{eq:B-Spin-rotation}. \hl{This then yields
\begin{equation}
    \bra{\Psi} H^{(0)}_{\mathrm{NSOC}}+H^{(0)}_{\mathrm{NSOOC}} \ket{\Psi} = \sum_{a=1}^{N_\mathrm{N}}\bm{I}_aM^{\mathrm{rot}}_a \bra{\chi_t^{\mathrm{vib}}}\bm{G}\ket{\chi_t^{\mathrm{vib}}} = \sum_{a=1}^{N_\mathrm{N}} \bm{I}_aM^{\mathrm{rot}}_a \bm{\zeta}_t \langle G_t\rangle\,,
    \label{eq:SVC-0-order}
\end{equation}
}
\hl{showing that the zeroth order terms in the case of spin-vibration coupling can be directly obtained from the spin-rotation tensor and the Coriolis coupling.} 
Furthermore, since $\delta\bm{R}_{\alpha} $ is no longer zero, the four remaining terms in the Taylor expansion of the spin interaction Hamiltonian need to be taken into consideration.

We first evaluate the expectation value of $H^{\mathrm{e},(1)}_{\mathrm{NSOC}}$ in the state $\ket{\Psi} = \ket{\chi^{\mathrm{vib}}_t}\ket{\psi_0^{\mathrm{e,vib}}}$ where only one two-fold pseudo-rotation is excited. The calculation for this case, carried out in Appendix~\ref{sec:spin-vibration-appendix} in Equation~\eqref{eq:Derivation-SVC-electronic-SOC}, yields
\begin{equation}
           \bra{\Psi} H^{\mathrm{e},(1)}_{\mathrm{NSOC}} \ket{\Psi}    = \frac{\mu_0e}{4\pi } \sum_{a=1}^{N_{\mathrm{N}}} \frac{f_a }{m_{a} } \gamma_a \bm{I}_a \cdot
              \bm{\Gamma}_{a,t}\, \langle G_t \rangle\,,
              \label{eq:SVC-e-1-exp}
\end{equation}
where we introduced the coupling vector
\begin{equation}
\bm{\Gamma}_{a,t} =  \frac{2 \pi \varepsilon_0}{e} \left[(\langle\varphi_{\mathrm{el}}(\bm{R}_a^0)\rangle \bm{l}_{a,t_1}) \times \bm{l}_{a,t_2} - (\langle\varphi_{\mathrm{el}}(\bm{R}_a^0)\rangle \bm{l}_{a,t_2}) \times \bm{l}_{a,t_1}\right]\,,
\label{eq:Coupling-vector-1}
\end{equation}
\hl{which involves the expectation value of the electronic contribution to the electric field gradient at the equilibrium geometry and the two atomic displacement vectors of the excited degenerate mode.}
As shown in Appendix~\ref{sec:spin-vibration-appendix} as well, the expectation value of the third term of Equation~\eqref{eq:termlist}, $H_{\mathrm{NSOOC}}^{\mathrm{e},(1)}$, may be neglected. For the remaining nuclear terms, $H_{\mathrm{NSOOC}}^{\mathrm{N},(1)}$  and $H_{\mathrm{NSOC}}^{\mathrm{N},(1)}$, we find
\begin{equation}
    \bra{\Psi} H_{\mathrm{NSOOC}}^{\mathrm{N},(1)} + H_{\mathrm{NSOC}}^{\mathrm{N},(1)} \ket{\Psi} = {+} \frac{\mu_0e}{4\pi}\sum_{a=1} ^{N_\mathrm{N}} \gamma_a \bm{I} _a \cdot\sum_{b\neq a}^{N_{\mathrm{N}}} Z_{b} \left[f_{a}\frac{\bm{\Gamma}_{a,b,t}}{m_{a}} {+} \frac{\bm{\Gamma}_{b,a,t}}{m_{b}} \right] \langle G_t \rangle\,,
    \label{eq:SVC-N-1-exp}
\end{equation}
with another vector coupling
\begin{equation}
    \bm{\Gamma}_{a,b,t} = \frac{1}{2}
    \ \left[(\varphi_{a,b}(\bm{R}^0)(\bm{l}_{b,t_1} - \bm{l}_{a,t_1})) \times \bm{l}_{a,t_2} - (\varphi_{a,b}(\bm{R}^0)(\bm{l}_{b,t_2} - \bm{l}_{a,t_2})) \times \bm{l}_{a,t_1}\right]\,.
    \label{eq:Coupling-vector-2}
\end{equation}
The corresponding derivation is provided in Equation~\eqref{eq:Derivation-SVC-nuclear} of Appendix~C. Finally, from the \hl{energy expectation values in Equations~\eqref{eq:SVC-0-order},~\eqref{eq:SVC-e-1-exp} and~\eqref{eq:SVC-N-1-exp}}, again an `effective' magnetic field can be determined \hl{as defined in Equation~\eqref{eq:effective-Hamiltonian}}, this time generated by the excitation of two degenerate modes with vibrational angular momentum $\langle G_t \rangle$,
\begin{equation}
    \bm{B}_{\mathrm{vib}}^{\mathrm{eff}}(\bm{R}_a) =  \frac{1}{\gamma_a}\bm{M}^{\mathrm{vib}}_{a,t}\, \langle G_t \rangle\,,
    \label{eq:spin-vib-B-field}
\end{equation}
with the spin-vibration vector given as
\begin{equation}
    \bm{M}^{\mathrm{vib}}_{a,t} = - M^{\mathrm{rot}}_{a} \bm{\zeta}_t {-} \gamma_a \frac{\mu_0e }{4\pi} \left [\sum_{b\neq a}^{N_{\mathrm{N}}} Z_{b} \left[f_{a}\frac{\bm{\Gamma}_{a,b,t}}{m_{a}} {+} \frac{\bm{\Gamma}_{b,a,t}}{m_{b}}\right] {+} f_{a} \frac{\bm{\Gamma}_{a,t}}{m_{a}}\right]\,.
    \label{eq:spin-vib-vector}
\end{equation}
\hl{Thus, the induced magnetic field depends on the spin-rotation tensor $M^{\mathrm{rot}}_a$ the Coriolis coupling $\bm{\zeta}_t$ 
as well as on the newly introduced couplings $\bm{\Gamma}_{a,t}$ and $\bm{\Gamma}_{a,b,t}$. The latter depend on the mode vectors and on the electric field gradients of the surrounding particles. 
}

\section{Results}
\subsection{Choice of molecular benchmark systems}
The formalism conceived in the previous section is now applied to a historically motivated selection of small molecules. Our main criterion is that a vibrational Zeeman effect has been confirmed by magnetic vibrational circular dichroism experiments \cite{Wang1993,Wang1994}. This also implies a sufficiently high molecular symmetry. From a computational benchmarking perspective, a second criterion is molecular size, since smaller systems allow the application of advanced electronic structure methods. We choose the three trihalomethanes chloroform, bromoform, and fluoroform, as well as the aromatic, planar systems benzene and 1-3-5 triazine. In a first attempt to also study the influence of chemical moieties, we further investigate the derivatives 1-3-5 trichlorbenzene,  1-3-5 tribromobenzene, and 1-3-5 trifluorbenzene. Note that C$^{13}$-substituted forms are studied exclusively, since this isotope carries a nuclear spin of $I\!=\!1/2$ which serves as a `local probe' of the effective intramolecular magnetic field in NMR experiments.

\subsection{Computational Details}
Due to the availability of convenient subroutines for the evaluation of magnetic properties via London or gauge-including atomic orbitals (GIAOs),\cite{London1937,Ditchfield1972,Wolinski1990} the ORCA program package is used for all electronic structure calculations in this work.\cite{orca,Orca-GIAO} For the small trihalomethanes, a single-reference, closed-shell variant of the coupled-cluster method with single and double excitations plus perturbative triples is performed in combination with the cc-pVTZ basis set~\cite{Dunning1989,Dunning1992,Dunning1993,Wilson1999} for geometry optimization and frequency calculations. However, due to current limitations of the code, magnetic properties such as shielding tensors are evaluated within a density functional theory (DFT) framework \cite{Stochyev2018} employing the $\omega{}$-B97X-V functional, a 12-parameter, range-separated hybrid, meta-GGA density functional with nonlocal correlation,\cite{Mardirossian2016} in combination with the larger aug-cc-pVQZ basis set of the same family.
To investigate the sensitivity with respect to variations in the geometry, we compare our results to those obtained for equilibrium geometries and frequencies calculated exclusively via DFT. Deviations of around 3\% (see Table~S1 in the SI for details) indicate sufficient accuracy also at the DFT level, making it our choice for the larger aromatic molecules. However, we note further that a very tightly converged geometry is crucial for consistent results. Wherever possible, constrained optimizations that conserve molecular symmetry are highly recommended to exactly reproduce the expected vibrational degeneracies.
\hl{Currently, all quantities needed for the explicit evaluation of properties, such as the spin-rotation tensor in Equation~\eqref{eq:B-Spin-rotation} or the vibrational induced magnetic field in Equation~\eqref{eq:spin-vib-B-field}, are extracted from ORCA output files via Python scripts.} 

\subsection{Spin-rotation tensor}
The interaction between a specific nuclear spin and the overall rotational motion of a molecule is captured by the spin-rotation tensor. This quantity is going to be evaluated for all relevant nuclei and analyzed with respect to spin-orbit and spin-other-orbit contributions. To our knowledge, this separation has not been considered in the literature yet. Results for the nuclear spin-rotation tensors $M^{\mathrm{rot}}_{a}$ of methane and the three trihalomethanes are summarized in Table~\ref{tab:spin-rot}, allowing to study the influence of the halogen substitution.
\hl{For all molecules treated in this section, the spin-rotation tensor is symmetric and a listing of the eigenvalues of $M^{\mathrm{rot}}_{a}$ is sufficient. The latter can be related to the diagonal elements of the spin-rotation tensor in the principal axis frame of the moment of inertia, which are accessible in experimental measurements of molecular ensembles.}

In the case of highly symmetric CH\textsubscript{4}, the three eigenvalues of $M^{\mathrm{rot}}$, evaluated at the C atom, are identical, whereas two distinct eigenvalues appear for the H atom positions. Similarly, for the hydrogen and carbon atoms in the CH$X$\textsubscript{3} ($X$ = Fl,Cl,Br) molecules, the reduced symmetry yields two distinct eigenvalues for C and H and three for the halogen atom. Note that, for spin-rotation coupling, SOOC contributions are always much larger than those of SOC, rendering the latter negligible in all cases except for methane. \hl{In case of a molecular rotation, none of the nuclei follow a small, local circular path that would couple well to the nuclear spin at that position, but rather a path with a radius approximately on the order of the entire molecule.}

\hl{Deviations between calculated and measured values of the spin-rotation tensor in Table~\ref{tab:spin-rot} may be caused by several reasons. Besides methodological improvements regarding the computation of magnetic properties, also the inclusion of vibrational corrections in the result for the spin-rotation tensor\cite{Benchmarking2013} and relativistic calculations for heavier nuclei\cite{Ruud2014, Malkin2013} might be steps to take.}

\begin{table}[!h]
\caption{Distinct eigenvalues of the spin-rotation tensor $M^{\mathrm{rot}}$ at a specific nuclear site, calculated with formula \eqref{eq:B-Spin-rotation}, and compared to known experimental results where possible. The aug-cc-pVQZ basis set was used in a DFT calculation employing the $\omega{}$-B97X-V functional. $M^{\mathrm{rot}}_{\mathrm{SOOC}}$ and  $M^{\mathrm{rot}}_{\mathrm{SOC}}$ denote pure nuclear spin-other-orbit and spin-orbit coupling contributions, respectively. Depending on the symmetry of the molecule, there are a different number of eigenvalues of $M^{\mathrm{rot}}$for each atom. }
\label{tab:spin-rot}

\centering

\begin{tabular}{lcccccc}
\toprule
Molecule & Atom & $M^{\mathrm{rot}}_{\mathrm{SOOC}}\left[\mathrm{kHz}\right]$ & $M^{\mathrm{rot}}_{\mathrm{SOC}}\left[\mathrm{kHz}\right]$ & $M^{\mathrm{rot}}\left[\mathrm{kHz}\right]$ & $M^{\mathrm{rot}}_{\mathrm{exp}}\left[\mathrm{kHz}\right]$ & Ref.\\
\midrule
\multirow{1}{*}{$^{13}\mathrm{CH}_4$}
& \textsuperscript{13}C & 17.664 & 0.000 & 17.665 & 15.94(2.37) & \cite{JAMESON1987461}\\
\cline{2-7} \\
\multirow{2}{*}{$^{12}\mathrm{CH}_4$} & \multirow{2}{*}{H} & -17.288 & -0.277 & -17.565 & -16.495(91) & \cite{Itano-1980}\\
& & 2.028 & 0.000 & 2.028 & 1.875(98) & \cite{Itano-1980} \\
\midrule
\multirow{7}{*}{$^{13}\mathrm{CH}\mathrm{F}_3$} & \multirow{2}{*}{\textsuperscript{13}C} & 1.757 & 0.000 & 1.757 & & \\
& & 2.974 & 0.000 & 2.974 & & \\
\cline{2-7}
& \multirow{2}{*}{H} & 0.455 & 0.000 & 0.455 & &\\
& & -1.002 & -0.021 & -1.023 & &\\
\cline{2-7}
& \multirow{3}{*}{F} & 6.383 & 0.000 & 6.370 & &\\
& & 11.778 & -0.039 & 11.752 & &\\
& & 10.956 & -0.069 & 10.887 & &\\
\midrule
\multirow{5}{*}{
$^{12}\mathrm{CH}\,^{35}\mathrm{Cl}_3$} 
& \multirow{2}{*}{H} & 0.141 & 0.000 & 0.141 & &\\
& & -0.306 & -0.004 & -0.310 & &\\
\cline{2-7}
& \multirow{3}{*}{\textsuperscript{35}Cl} & 0.294 & 0.000 & 0.294 & &\\
& & 0.936 & 0.000 & 0.936 & 0.84(18) & \cite{KISIEL2009177}\\
& & 1.213 & -0.001 & 1.212 & 1.62(7) & \cite{KISIEL2009177}\\
\midrule
\multirow{5}{*}{$^{12}\mathrm{CH}\,^{79}\mathrm{Br}_3$
} 
& \multirow{2}{*}{H} & 0.049 & 0.000 & 0.049 & & \\
& & -0.115 & -0.001 & -0.116 & & \\
\cline{2-7}
& \multirow{3}{*}{{\textsuperscript{79}}Br} & 0.530 & 0.000 & 0.530 & &\\
& & 2.372 & -0.002 & 2.370 & 1.57(7) &  \cite{KISIEL2009177} \\
& & 3.210 & -0.003 & 3.207 & 4.18(26) & \cite{KISIEL2009177}\\
\bottomrule
\end{tabular}
\end{table}

\subsection{Vibrationally induced magnetic fields}
We apply Equations~\eqref{eq:B-Spin-rotation}~and~\eqref{eq:spin-vib-B-field} to determine the totally induced vibrational magnetic field ${B}^{\mathrm{eff}}_{\mathrm{vib}}(\bm{R}_a)$ and the analogous induced hyperfine splitting $H^{\mathrm{hyp}}_{\mathrm{vib}} = -\gamma_a{B}^{\mathrm{eff}}_{\mathrm{vib}}(\bm{R}_a)$. This is, of course, only identical to the value of the induced hyperfine splitting if the nuclear spin is perfectly aligned with the vibrationally induced magnetic field. 
\hl{We also present the separate SOC and SOOC contributions to the total induced magnetic field, as they help to understand and interpret the origin of the local hyperfine interaction. The  corresponding equations are presented in Appendix~\ref{sec:appendix-spin-vib-splitting}}.
For the sake of brevity and relevance, only infrared-active, degenerate modes with an induced hyperfine splitting of at least $1\,\mathrm{kHz}$ on some atom are listed.

\subsubsection{Trihalomethanes}
Fluoroform, chloroform and bromoform belong to the $C_{3\mathrm{v}}$ molecular point group. They feature three non-degenerate A$_1$ vibrational modes and three doubly degenerate E modes. The latter can be categorized as a CH bend of highest frequency, a CX\textsubscript{3} d-deform, and a CX\textsubscript{3} d-stretch of lowest frequency. 

\begin{figure}[!h]
\centering
\subfloat[CCl\textsubscript{3} d-deform vibration at 261~cm$^{-1}$]{%
\resizebox*{7.8cm}{!}{\includegraphics{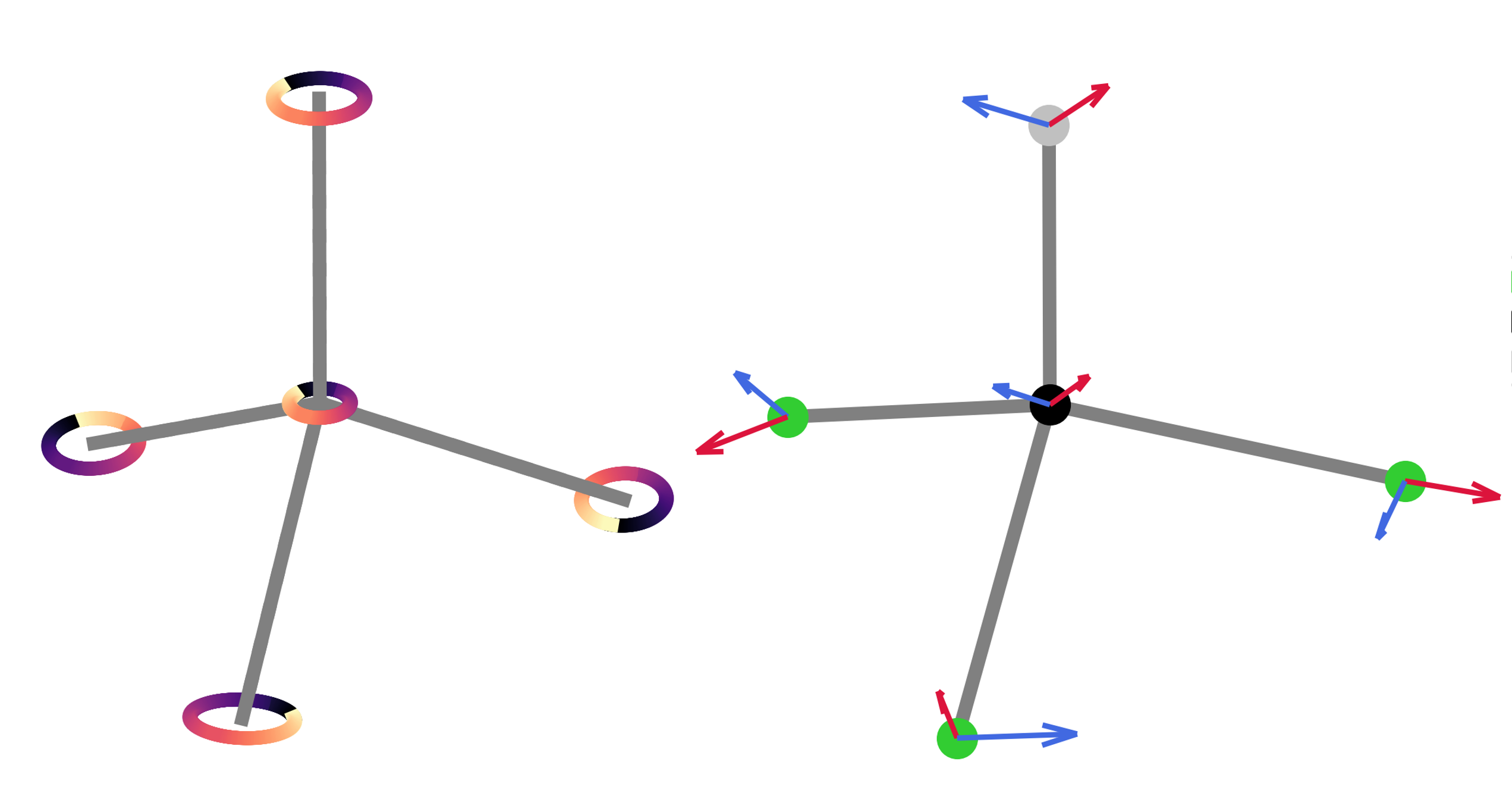}}}\hspace{5pt}
\subfloat[CCl\textsubscript{3} d-stretch vibration at 765~cm$^{-1}$]{%
\resizebox*{7.8cm}{!}{\includegraphics{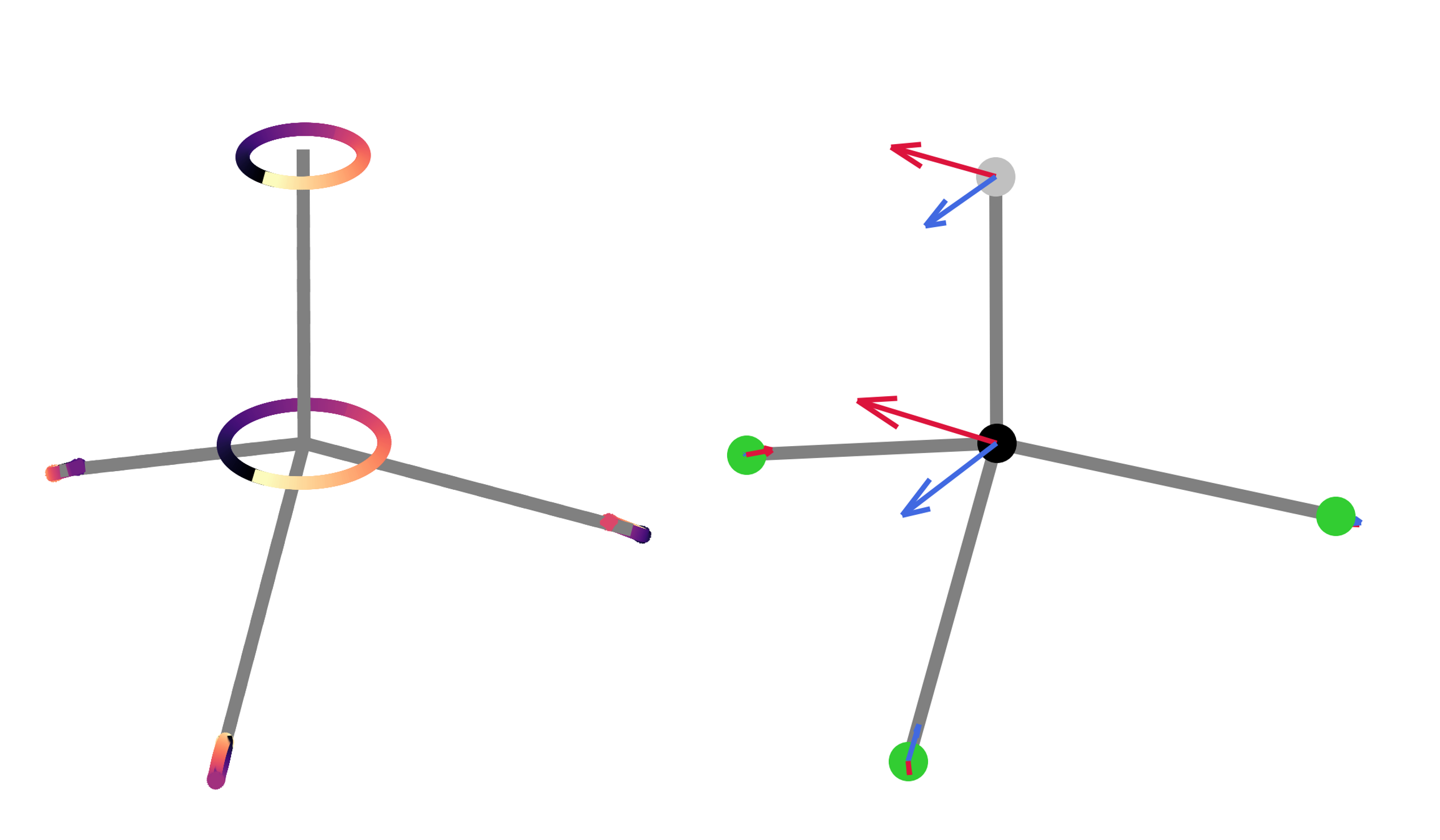}}}\par
\subfloat[CH-bend vibration at 1241~cm$^{-1}$]{%
\resizebox*{7.8cm}{!}{\includegraphics{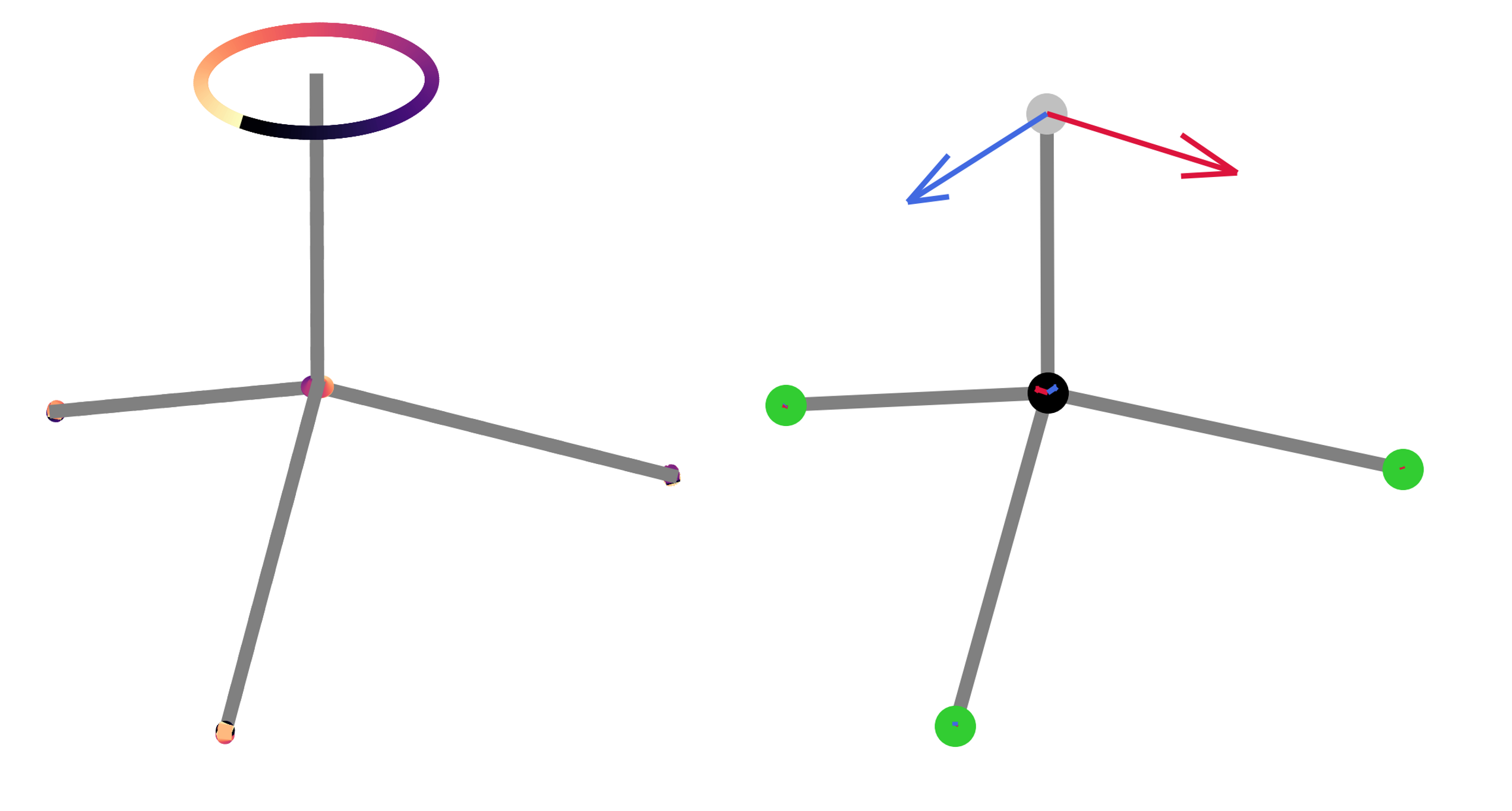}}}
\caption{Degenerate normal modes of chloroform as a representative of the trihalomethanes. Direction and relative size of the two eigenvectors are shown for each degenerate pair (red and blue), together with the corresponding pseudorotational motion. \hl{Nuclear trajectories are plotted with a dark-to-bright color gradient illustrating their relative evolution in time. } Cl atoms are green, H atoms gray, and C atoms black. } \label{fig:Chloroform-Modes}
\end{figure}

In Table~\ref{tab:spin-vib-trihalomethanes}, we display the induced splittings for all three degenerate modes of CHFl\textsubscript{3}, CHCl\textsubscript{3} and CHBr\textsubscript{3}. A graphical depiction of these modes is provided in Figure~\ref{fig:Chloroform-Modes} for chloroform as a representative. It shows the two underlying, degenerate vibrational eigenvectors for each mode, together with the more or less circular nuclear trajectories of the corresponding pseudorotational motion, obtained for a phase shift of $\pi/2$ between the two degenerate vibrations indicated by arrows in blue and red.

Regarding the CH-bend vibrational mode, which is almost exclusively affecting the position of the H atom, similar values for the effective magnetic field on the individual atoms are obtained for all trihalomethanes. The H atom features a comparably large pseudorotational angular momentum, which is obvious also from the graphical representation. The orbital plane of the H atom is perpendicular to the C-H bond. We can immediately understand that the large vibrational hyperfine splitting at the H atom, on average about {6} times larger than all other splittings caused by this most effective mode, must stem almost exclusively from SOC. On the other hand, the comparably moderate, but still significant splittings of approximately $10\,\text{kHz}$ at the \textsuperscript{13}C nuclei are mainly caused by SOOC between H and C. Within the classical interpretation, the large circular motion of the H atom creates a magnetic dipole field that is well aligned with the C-H bond, and the C atom is located closer to the magnetic pole and hence in an area of higher field line density than the halogens.

Contrary to the above, in the case of the low-frequency d-deform vibration, the angular momentum is concentrated on the halogen atoms, while an excitation of the d-stretch mode is best described as a joint circular motion of the entire C-H bond. Still, for the d-deform vibration, at least the hyperfine splitting on H is dominated by SOC: the light H atom gives rise to a relatively large effective current, as not angular momentum, but rather angular momentum divided by nuclear mass, determines the size of the individual splittings. For the d-stretch vibration, the hyperfine splittings on both H and C are dominated by SOC. Classically, one could argue that this is because rotational directions are of the same sign.

\begin{table}[!h]
\caption{Spin-vibration coupling in trihalomethanes. Coriolis coupling in all three degenerate pairs is parallel to the $C-H$ axis. For symmetry reasons it is sufficient to present absolute values for just one halogen atom per mode. Absolute values of vibrationally induced magnetic fields $B^{\mathrm{eff}}_{\mathrm{vib}}$ are given in $\mathrm{mT}$, vibration-induced hyperfine splittings $H_{\mathrm{vib}}^{\mathrm{hyp}}$ in kHz. $B^{\mathrm{SOOC}}_{\mathrm{vib}}$ and $B^{\mathrm{SOC}}_{\mathrm{vib}}$ refer to the contributions of nuclear spin-orbit and nuclear spin-other-orbit coupling. Absolute values of the Coriolis coupling $\bm{\zeta}$ are a measure for the total vibrational angular momentum, whereas absolute values of atomic contributions reveal the intramolecular distribution of angular momentum. \hl{For the geometry optimization and  frequency calculation CCSD(T) with the cc-pVTZ basis set was employed. The aug-cc-pVQZ basis set was used in a DFT calculation for magnetic properties employing the $\omega{}$-B97X-V functional. }}
\label{tab:spin-vib-trihalomethanes}
\begin{tabular}{lcccccccc}
  Mol.&$\bar{\nu}/\mathrm{cm}^{-1}$ & $|\bm{\zeta}|$ & Atom & $B^{\mathrm{eff}}_{\mathrm{vib}}/\mathrm{mT}$ & $B_{\mathrm{vib}}^{\mathrm{SOC}}/\mathrm{mT}$ & $B_{\mathrm{vib}}^{\mathrm{SOOC}}/\mathrm{mT}$ & $|\bm{\zeta}_{\alpha}|$ & $H_{\mathrm{vib}}^{\mathrm{hyp}}/\mathrm{kHz}$ \\
  \toprule
\multirow{9}{*}{CHF\textsubscript{3}}&\multirow{3}{*}{$514$} & \multirow{3}{*}{$0.790$} & F & 0.10 & 0.0{8} & 0.11 & 0.295 & {4.0} \\
  & &  & \textsuperscript{13}C & 0.{12} & 0.0{9} & 0.21 & 0.063 & {1.3} \\
  & & & H & 0.{04} & 0.{00} & 0.03 & 0.019 & {1.6} \\
\cline{2-9}
\noalign{\vskip 4pt}
& \multirow{3}{*}{$1177$} & \multirow{3}{*}{$0.713$} & F & 0.09 & 0.02 & 0.09 & 0.019 & 3.{6} \\
  & & & \textsuperscript{13}C & 0.{16} & 0.{25} & 0.08 & 0.559 & {1.7} \\
  & & & H & 0.{88} & {1.01} & 0.14 & 0.103 & {37.4} \\
\cline{2-9}
\noalign{\vskip 4pt}
& \multirow{3}{*}{$1414$} & \multirow{3}{*}{$0.986$} & F & 0.23 & 0.0{1} & 0.23 & 0.005 & 9.{3} \\
  & &  & \textsuperscript{13}C & 1.{10} & 0.01 & 1.09 & 0.161 & 11.{8} \\

  & & & H & {1.99} & {1.79} & 0.20 & 0.821 & {84.6} \\
\midrule
\multirow{9}{*}{CHCl\textsubscript3} & \multirow{3}{*}{$261$} & \multirow{3}{*}{$0.871$} & \textsuperscript{35}Cl & 0.15 & 0.01 & 0.16 & 0.310 & 0.6 \\
 & &  & \textsuperscript{13}C & 0.{00} & 0.07 & 0.07 & 0.048 & {0.0} \\
 & & & H & 0.{05} & 0.{05} & 0.01 & 0.008 & {1.9} \\
\cline{2-9}
\noalign{\vskip 4pt}
& \multirow{3}{*}{$765$} & \multirow{3}{*}{$0.846$} & \textsuperscript{35}Cl & 0.1{3} & 0.00 & 0.13 & 0.016 & 0.5 \\
  & &  & \textsuperscript{13}C & 0.2{3} & 0.2{0} & 0.03 & 0.771 & 2.{5} \\
  & & & H & 0.{46} & 0.{69} & 0.23 & 0.032 & {19.6} \\
\cline{2-9}
\noalign{\vskip 4pt}
& \multirow{3}{*}{$1241$} & \multirow{3}{*}{$0.988$} & \textsuperscript{35}Cl & 0.21 & 0.00 & 0.21 & 0.000 & 0.9 \\
 & & & \textsuperscript{13}C & 0.9{8} & 0.01 & 0.97 & 0.054 & 10.{5} \\

  & & & H & {1.48} & {1.37} & 0.11 & 0.936 & {63.1} \\
\midrule
\multirow{9}{*} {CHBr\textsubscript{3}} & \multirow{3}{*}{$153$} & \multirow{3}{*}{$0.939$} & \textsuperscript{81}Br & 0.1{4} & 0.04 & 0.17 & 0.323 & 1.{7} \\
  & &  & \textsuperscript{13}C & 0.{07} & 0.0{9} & 0.02 & 0.025 & {0.7} \\
  & & & H & 0.1{2} & 0.1{3} & 0.00 & 0.004 & {5.3} \\
\cline{2-9}
\noalign{\vskip 4pt}
& \multirow{3}{*}{$666$} & \multirow{3}{*}{$0.927$} & \textsuperscript{81}Br & 0.15 & 0.00 & 0.15 & 0.008 & 1.7 \\
  & &  & \textsuperscript{13}C & 0.{32} & 0.{23} & 0.09 & 0.864 & {3.4} \\
  & & & H & 0.{61} & 0.{87} & 0.25 & 0.042 & {26.1} \\
\cline{2-9}
\noalign{\vskip 4pt}
& \multirow{3}{*}{$1171$} & \multirow{3}{*}{$0.996$} & \textsuperscript{81}Br & 0.21 & 0.00 & 0.21 & 0.000 & 2.4 \\
  & &  & \textsuperscript{13}C & 0.9{6} & 0.0{1} & 0.95 & 0.052 & 10.{3} \\
  & & & H & {1.46} & {1.36} & 0.10 & 0.945 & {62.4} \\
\hline
\end{tabular}
\end{table}

For all modes in Table~\ref{tab:spin-vib-trihalomethanes}, one observes that the Coriolis coupling, which is proportional to the total vibrational angular momentum for the whole molecule, is approximately the same. However, the order of magnitude of the effective magnetic field values at individual nuclei differs significantly. This is due to the fact that not the total vibrational angular momentum, but its distribution over individual nuclei in the molecule is relevant for the magnitude of the vibrational hyperfine interaction. 
This clearly illustrates that the concept of an `effective' induced magnetic moment, which is assumed proportional to the total angular momentum built from all spins, orbital momenta, pseudorotation and rotation,~\cite{Moss1972,Wihelmer2024} is only applicable e.g. when discussing the energy of a molecule e.g. in an external field, or for the estimate of macroscopic properties such as total magnetization.

\subsubsection{Benzene and 1-3-5 substituted derivatives}
\label{sec:Benzene-1-3-5-derivatives}
\begin{figure}[!h]
\centering
\subfloat[degenerate mode at 1055 cm\textsuperscript{-1}]{%
\resizebox*{7.8cm}{!}{\includegraphics{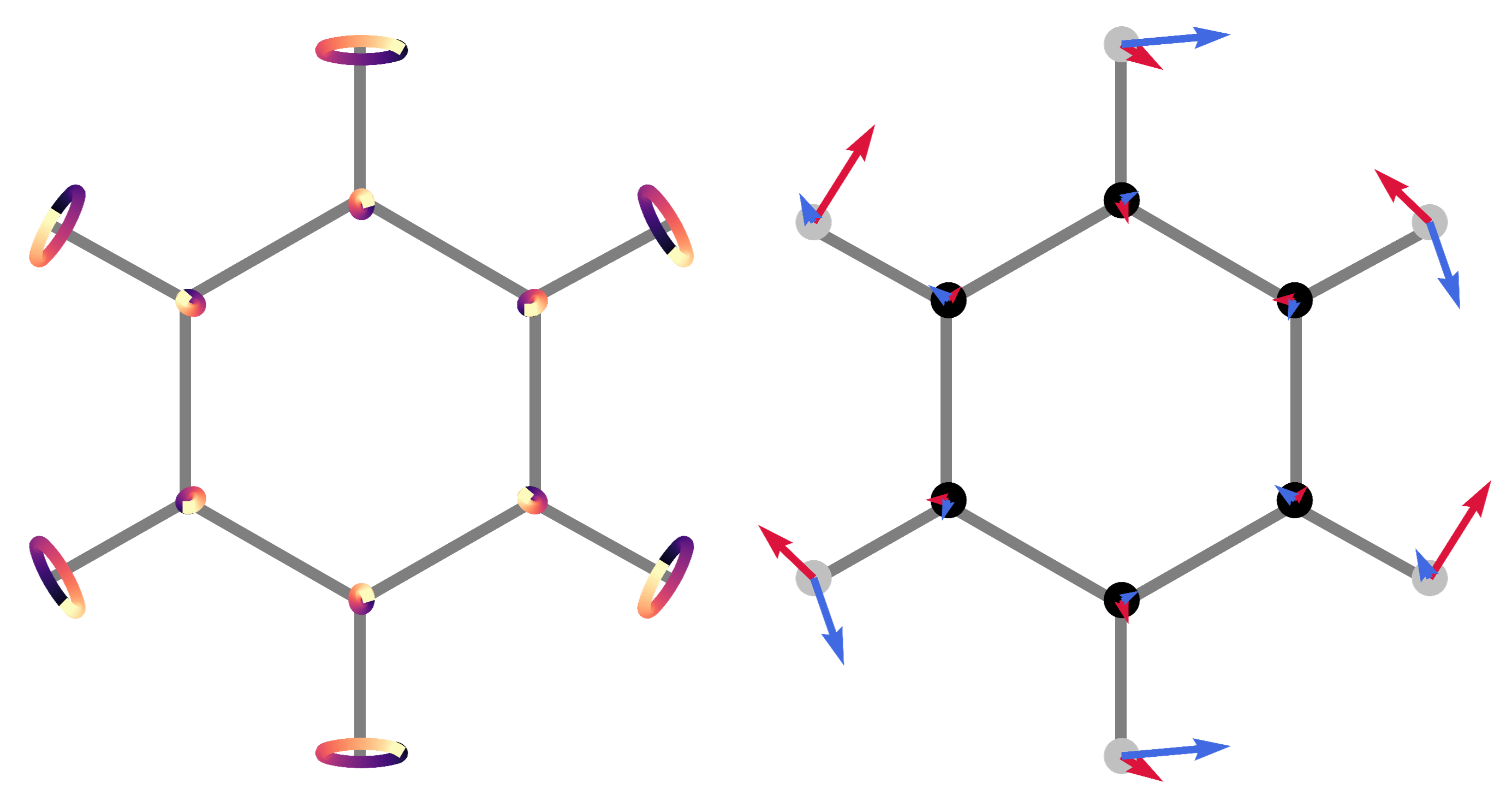}}}\hspace{5pt}
\subfloat[degenerate mode at 1498 cm\textsuperscript{-1}]{%
\resizebox*{7.8cm}{!}{\includegraphics{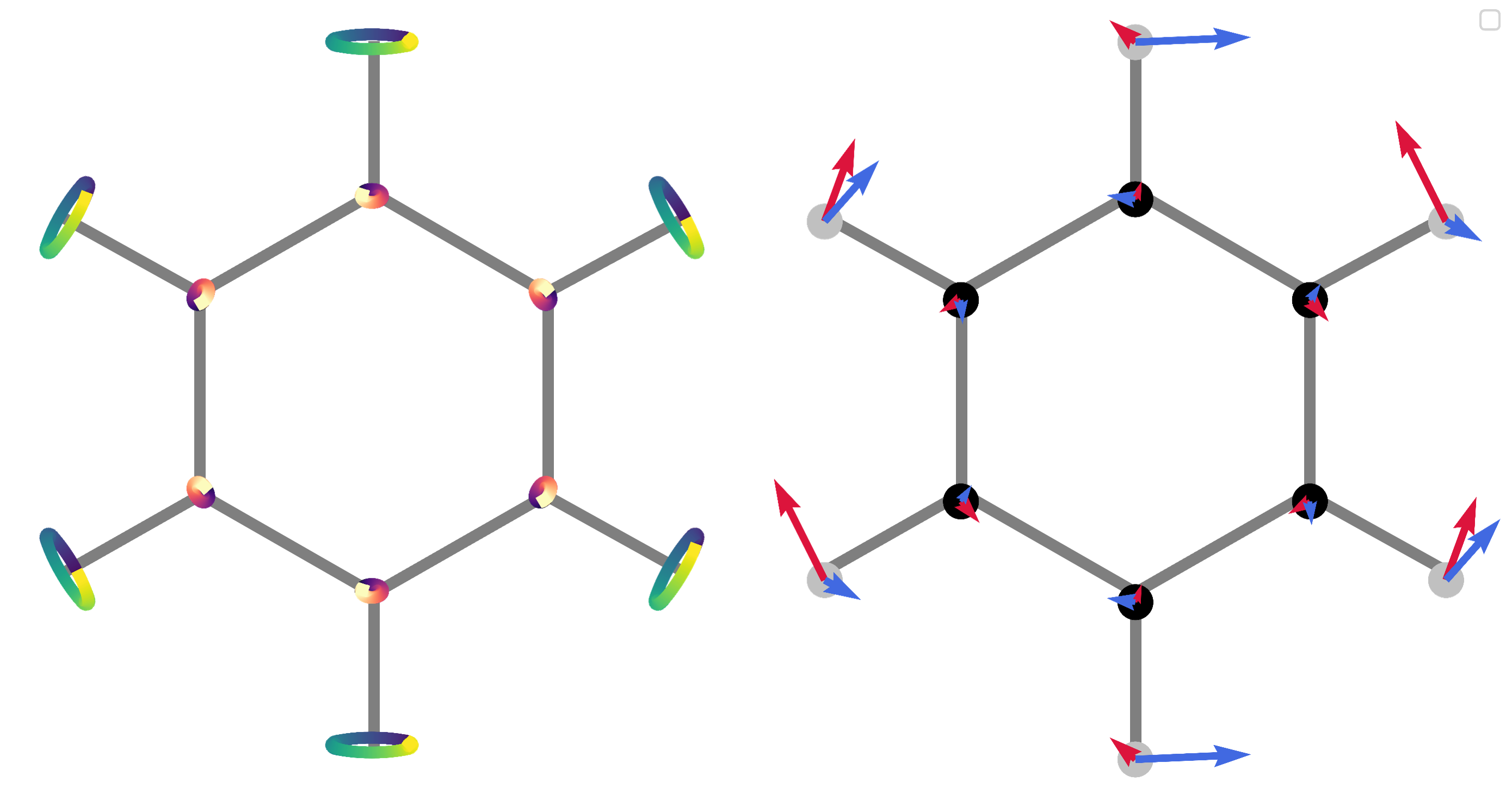}}}\par
\subfloat[degenerate mode at 3169 cm\textsuperscript{-1}]{%
\resizebox*{7.8cm}{!}{\includegraphics{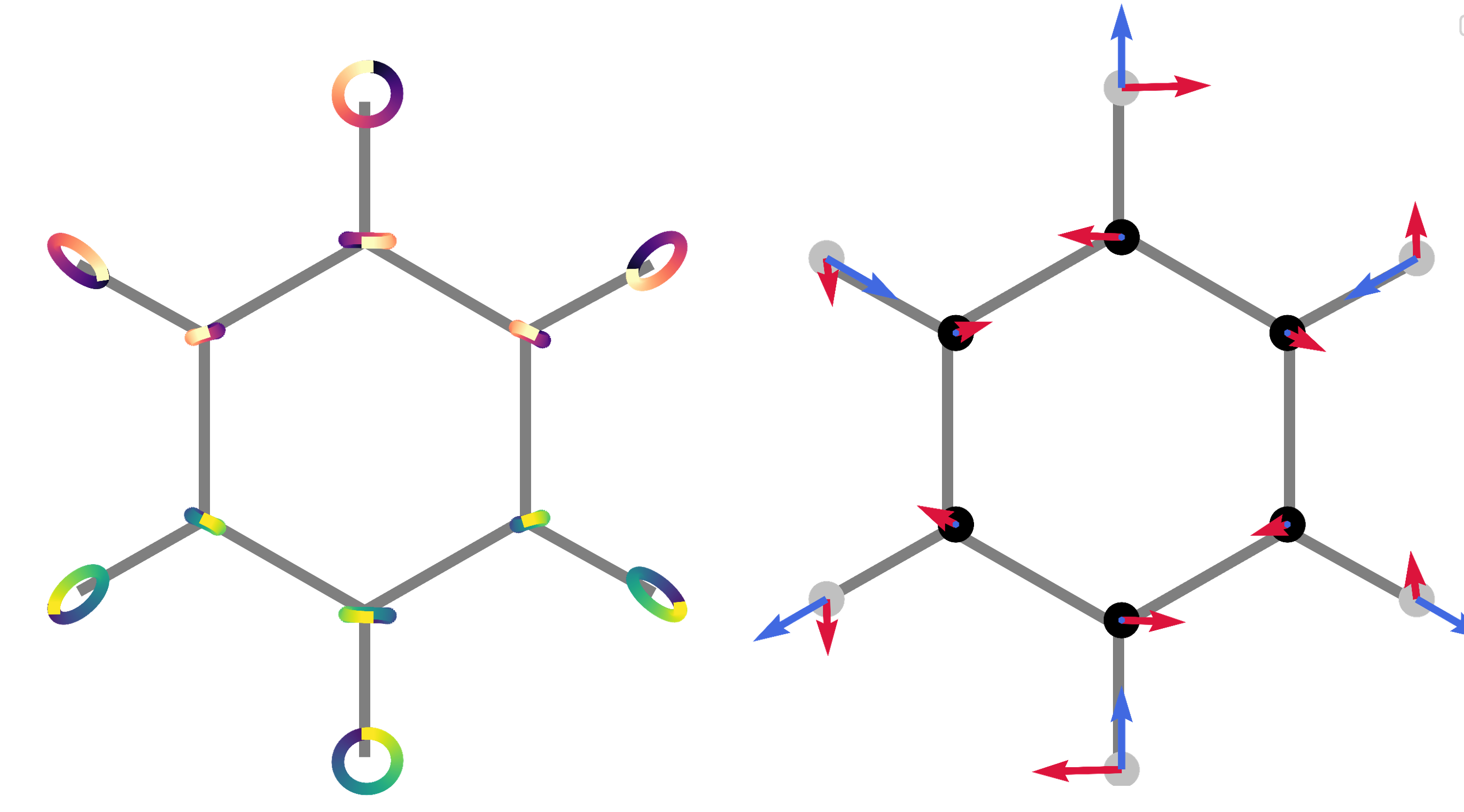}}}
\caption{Selected degenerate normal modes of benzene. Direction and relative size of the two eigenvectors are shown for each degenerate pair (red and blue), together with the corresponding pseudorotational motion. \hl{Nuclear trajectories are plotted with a dark-to-bright color gradient illustrating their relative evolution in time. 'Opposing directions' within a molecule are additionally emphasized by a different choice of colors.} H atoms are gray, and C atoms black. } \label{fig:Benzene-Modes}
\end{figure}

The infrared-active modes of benzene with considerable vibrational hyperfine splittings are listed in Table~\ref{tab:spin-vib-Benzene}, the corresponding data for the derivatives in Tables~S2 to~S4 in the SI. Largest magnetic field strengths are obtained at the hydrogen nucleus, and SOC is the dominant contributor. Nonetheless, these maxima are considerably smaller than in the case of the trihalomethanes. As can be seen in Figure~\ref{fig:Benzene-Modes}, which depicts the corresponding modes, this is mainly due to the fact that the vibrational angular momentum is equally distributed among all nuclei of the same type, such that local angular momenta, and hence the locally induced currents, are much smaller. For symmetry reasons (the Coriolis coupling is perpendicular to the molecular plane), it is sufficient to provide only values for C and H. The benzene-substituted derivatives provide more infrared-active vibrations, but otherwise, the resulting fields are of the same order of magnitude as obtained for benzene. Note that, due to the changing appearance of hydrogen and halogen atoms, two different results occur for the effective splittings at the carbon nuclei.

\begin{table}[!h]
\caption{Spin-vibration coupling in benzene. For symmetry reasons, field values provided for one C and one H atom are sufficient. Only infrared-active degenerate modes are displayed. Absolute values of vibrationally induced magnetic fields $B^{\mathrm{eff}}_{\mathrm{vib}}$ are given in $\mathrm{mT}$, the vibration-induced hyperfine splitting $H_{\mathrm{vib}}^{\mathrm{hyp}}$ in kHz. $B^{\mathrm{SOOC}}_{\mathrm{vib}}$ and $B^{\mathrm{SOC}}_{\mathrm{vib}}$ refer to contributions of nuclear spin-orbit and nuclear spin-other-orbit coupling. Absolute values of the Coriolis coupling $\bm{\zeta}$ are a measure for the total vibrational angular momentum, whereas absolute values of atomic contributions reveal the intramolecular distribution of angular momentum. \hl{ The aug-cc-pVQZ basis set was used in a DFT calculation  employing the $\omega{}$-B97X-V functional. }}

\label{tab:spin-vib-Benzene}
\begin{tabular}{lccccccc}
  $\bar{\nu}/\mathrm{cm}^{-1}$ & $|\bm{\zeta}|$ & Atom & $B^{\mathrm{eff}}_{\mathrm{vib}}/\mathrm{mT}$ & $B_{\mathrm{vib}}^{\mathrm{SOC}}/\mathrm{mT}$ & $B_{\mathrm{vib}}^{\mathrm{SOOC}}/\mathrm{mT}$ & $|\bm{\zeta}_{\alpha}|$ & $H_{\mathrm{vib}}^{\mathrm{hyp}}/\mathrm{kHz}$ \\
  \hline
  \noalign{\vskip 4pt}
\noalign{\vskip 4pt}  
\multirow{2}{*}{$1054$} & \multirow{2}{*}{$ 0.687$} & \textsuperscript{13}C & 0.0{5} & 0.0{0} & 0.06 & 0.075 & {0.6} \\

  & & H & 0.{34} & 0.{39} & 0.04 & 0.039 & 1{4.7} \\

\hline
\noalign{\vskip 4pt}
\multirow{2}{*}{$1498$} & \multirow{2}{*}{$ 0.288$} & \textsuperscript{13}C & 0.0{6} & 0.0{0} & 0.06 & 0.077 & 0.{6} \\

  & & H & 0.{27} & 0.{26} & 0.01 & 0.030 & {11.4} \\

\hline
\noalign{\vskip 4pt}
\multirow{2}{*}{$3193$} & \multirow{2}{*}{$ 0.025$} & \textsuperscript{13}C & 0.01 & 0.00 & 0.01 & 0.002 & 0.1 \\

  & & H & 0.04 & 0.0{3} & 0.01 & 0.002 & 1.{5} \\

\hline
\noalign{\vskip 4pt}
\end{tabular}
\end{table}

\subsubsection{Triazine}
For triazine, the results for the effective hyperfine splitting are of a similar order of magnitude as for benzene. Instead of three, we find four infrared-active degenerate modes. Results are summarized in Table~\ref{tab:spin-vib-triazine}. An interesting peculiarity is observed for the doubly degenerate vibration at $684~\mathrm{cm}^{-1}$. Despite an even four times smaller contribution of the H atoms to the total angular momentum in comparison to the doubly degenerate mode at $1456~\mathrm{cm}^{-1}$, their SOC is {nearly the same}. However, the plane of the traced-out circular motion of the nuclei differs significantly, and might be better aligned with the local electric field or its gradient at~$684~\mathrm{cm}^{-1}$. 
\begin{table}[!h]
\caption{Spin-vibration coupling in 1-3-5-triazine. For symmetry reasons, a field value provided for each atom type is sufficient. Only infrared-active degenerate modes are displayed. Absolute values of vibrationally induced magnetic fields $B^{\mathrm{eff}}_{\mathrm{vib}}$ are given in $\mathrm{mT}$, the vibration-induced hyperfine splitting $H_{\mathrm{vib}}^{\mathrm{hyp}}$ in kHz. $B^{\mathrm{SOOC}}_{\mathrm{vib}}$ and $B^{\mathrm{SOC}}_{\mathrm{vib}}$ refer to contributions of nuclear spin-orbit and nuclear spin-other-orbit coupling, the Coriolis coupling is denoted as $\bm{\zeta}$. \hl{ The aug-cc-pVQZ basis set was used in a DFT calculation  employing the $\omega{}$-B97X-V functional. }}
\label{tab:spin-vib-triazine}
\begin{tabular}{lccccccc}
$\bar{\nu}/\mathrm{cm}^{-1}$ & $|\bm{\zeta}|$ & Atom & $B^{\mathrm{eff}}_{\mathrm{vib}}/\mathrm{mT}$ & $B_{\mathrm{vib}}^{\mathrm{SOC}}/\mathrm{mT}$ & $B_{\mathrm{vib}}^{\mathrm{SOOC}}/\mathrm{mT}$ & $|\bm{\zeta}_{\alpha}|$ & $H_{\mathrm{vib}}^{\mathrm{hyp}}/\mathrm{kHz}$ \\
  \hline
  \noalign{\vskip 4pt}

\multirow{3}{*}{$684$} & \multirow{3}{*}{$ 0.652$} & N & 0.0{1} & 0.0{1} & 0.00 & 0.135 & 0.{0} \\
  &  & \textsuperscript{13}C & 0.00 & 0.0{2} & 0.03 & 0.093 & 0.0 \\
  & & H & 0.2{7} & 0.2{7} & 0.01 & 0.011 & {11.5} \\

\hline
\noalign{\vskip 4pt}

\multirow{3}{*}{$1200$} & \multirow{3}{*}{$ 0.751$} & N & 0.06 & 0.0{0} & 0.05 & 0.062 & 0.2 \\
  &  & \textsuperscript{13}C & 0.{12} & 0.0{0} & 0.12 & 0.132 & {1.3} \\
  & & H & 0.{48} & 0.{53} & 0.05 & 0.056 & {20.6} \\

\hline
\noalign{\vskip 4pt}
\multirow{3}{*}{$1456 $} & \multirow{3}{*}{$ 0.175$} & N & 0.0{6} & 0.0{0} & 0.06 & 0.075 & 0.{2} \\
  &  & \textsuperscript{13}C & 0.1{2} & 0.0{1} & 0.13 & 0.029 & 1.{3} \\
  & & H & 0.{31} & 0.{32} & 0.01 & 0.045 & {13.3} \\
\hline
\noalign{\vskip 4pt}
\multirow{3}{*}{$1616$} & \multirow{3}{*}{$ 0.583$} & N & 0.0{0} & 0.0{0} & 0.01 & 0.106 & 0.{0} \\
  &  & \textsuperscript{13}C & 0.0{8} & 0.0{0} & 0.08 & 0.101 & 0.{8} \\
  & & H & 0.{17} & 0.1{6} & 0.01 & 0.012 & {7.3} \\

\hline
\noalign{\vskip 4pt}
\end{tabular}
\end{table}

\section{Conclusion}
In this work, we provide a unified and generic theoretical description of nuclear spin-rotation and nuclear spin-vibration coupling, rigorously derived entirely from first principles and linked to modern computational chemistry codes through the calculation of selected expectation values of the electronic wavefunction of a given molecule, in combination with a standard evaluation of vibrational eigenmodes through Hessian diagonalization after geometry optimization.

In a straightforward analogy to the Breit-Pauli Hamiltonian for electrons, we distinguish between nuclear spin-orbit coupling (NSOC) and nuclear spin-other-orbit coupling (NSOOC). This way, all contributions to the effective intramolecular magnetic  field,  due to any type of nuclear motion are nicely captured. Our theoretical approach also allows to distinguish between electronic and nuclear contributions; although they are not accessible by experiment, the ability to evaluate them separately allows for a better understanding of more or less complicated intramolecular field geometries, and reveals problematic points of simplified, semi-classical models involving e.g. the motion of fractional charges.\cite{Juraschek17,Juraschek19,Juraschek20}  

Our theoretical framework is then applied to methane, benzene, and selected derivatives of these molecules, allowing for a direct comparison to experimental measurements of spin-rotation coupling, but primarily for a tentative analysis of effective intramolecular magnetic field geometries generated by suitable, degenerate vibrational modes. As a key result, the latter concept, an optical excitation of pseudorotational motions as suggested by us in a recent publication,\cite{Wihelmer2024} should give rise to vibrationally induced chemical shifts, representing a novel type of hyperfine interaction. These shifts, although small, should be accessible in modern NMR experiments and may offer new pathways for nuclear spin control via vibrational excitation.

We find that maximizing effective magnetic field strength at an arbitrary position within a given molecule is best achieved by singling out degenerate vibrations, which lead, in a classical sense, to large-amplitude circular motions of nuclei in direct proximity to that location. The highest vibrationally induced hyperfine coupling may thus be reached when the total vibrational angular momentum is concentrated on a single, preferably light nucleus with non-zero spin, which basically translates to a large circular motion of a single, magnetically sensitive atom. In this case, the nuclear spin-orbit interaction fully dominates the effect.
 
{We showed that the degenerate C-H bend vibration in trihalomethanes, when excited with vibrational angular momentum of $\hbar$, leads to particularly high effective magnetic fields of about $2~\mathrm{mT}$ or vibrationally induced hyperfine splittings up to $80~\text{kHz}$ at the hydrogen nucleus. The effect is largest for fluoroform, but shows a very weak dependence on halogen mass. Hence, for future experimental investigations, we recommend fluoroform as a potential substance of interest.} Another experimental outlook is the obvious possibility of bulk studies on adsorbed molecules, molecular monolayers or functionalized surfaces, especially in the light of the finding mentioned above, suggesting e.g. the adsorption of a diatomic molecule such as OH for a supposedly large hyperfine splitting at the H atom. 

Finally, we note that predicted values for magnetic field strength and hyperfine splittings are linearly dependent on the vibrational angular quantum number $l$. Thus, the question of maximally achievable vibrational angular momenta rises naturally. Furthermore, the details of suitable optical excitation mechanisms remain an open question and need to be addressed in future work. 

\appendix

\hl{
\section{Chemical shielding}
\label{sec:Chemical-shielding}
When molecules are placed in strong magnetic fields, such as in NMR experiments, local electronic currents are induced. These currents are themselves sources of a magnetic field, reducing the magnetic field on the individual nuclear sites. This shielding effect leads to a reduced nuclear spin Zeeman splitting and can be described by
\begin{equation}
    E^{\mathrm{Zee}} = -\sum_a\gamma_a\bm{I}_a \cdot (\mathbb{1}-\sigma_a) \bm{B}\,,
\end{equation}
with the chemical shielding tensor $\sigma_a$ for nucleus $a$. \cite{levitt2008spin}
The relation of this second order response property to the expectation value of the electronic Hamiltonian in an external magnetic field is derived in the next subsection of the appendix. The shielding tensor is important as its paramagnetic part is related to the electronic part of the spin-rotation tensor.\cite{Flygare1974,Gauss1996}
}
\hl{
\subsection{Electronic Hamiltonian in an external magnetic field}
The Hamiltonian of electrons in an external constant magnetic field $\bm{B}$, associated with a vector potential $\bm{A} = -\frac{1}{2} (\bm{r} - \bm{R}_O) \times \bm{B}$, reads \cite{Helgaker1991,Helgaker1998}
\begin{equation}
    H = \sum_{i=1}^{N_{\mathrm{e}}} \left[\frac{\bm{p}^2_i}{2m_i}  + \frac{q_i^2}{8m_i}[\bm{B}^2 \bm{r}_{i,O}^2 - (\bm{B} \cdot \bm{r}_{i,O})^2] \right]+\frac{e}{2m_{\mathrm{e}}}\bm{B} \cdot \bm{L}_{e,O}\,,
\end{equation}
with $\bm{R}_O$ denoting the gauge origin and $\bm{L}_{i,O}$ the total electronic angular momentum with respect to the position $\bm{R}_O$.
In this purely electronic Hamiltonian, the magnetic field stemming from the interaction with the nuclear spins is included by adding the dipole vector potential of $\bm{A}_{\mathrm{nuc}} = \sum_{a} \frac{\mu_0}{4\pi}\frac{\bm{\mu}_{a} \times \bm{r}_{a}}{r_{a}^3}$. Here, $\bm{\mu}_a = \gamma_a\bm{I}_a$ is the magnetic moment associated with nuclear spin.
This leads to additional terms in the electronic Hamiltonian,
\begin{multline}
    H_{\mathrm{B-NS}} = \sum_{i=1}^{N_{\mathrm{e}}} \frac{\mu_0 e}{4\pi m_{\mathrm{e}}} \sum_{a=1}^{N_\mathrm{N}} \bm{\mu}_{a} \cdot \frac{\bm{L}_{i,a}}{r_{i,a}^3} - \frac{e^2\mu_0} {8\pi m_{\mathrm{e}}} \sum_{i=1}^{N_{\mathrm{e}}} \sum_{a=1}^{N_\mathrm{N}} \bm{\mu}_{a}^T \frac{(\bm{r}_i - \bm{R}_O) \bm{r}_{i,a}^T - (\bm{r}_{i,a} \cdot (\bm{r}_i-\bm{R}_O)) \mathbb{1}}{r_{i,a}^3} \bm{B}\\
    +\frac{\mu_0^2 e^2}{32 m_e\pi^2} \sum_{i=1}^{N_{\mathrm{e}}}\sum_{a,b=1}^{N_\mathrm{N}} \bm{\mu}^T_{a} \frac{(\bm{r}_{i,a} \cdot \bm{r}_{i,b}) \mathbb{1} - \bm{r}_{i,b} \bm{r}_{i,a}^T
    }{r_{i,a}^3 r_{i,b}^3} \bm{\mu}_{b}\,,
\end{multline}
where the first term is called paramagnetic spin-orbit coupling or orbital hyperfine interaction, and the last term is referred to as diamagnetic spin-orbit coupling.
Chemical shielding is now defined as
\begin{equation}
    (\sigma_{a})_{\alpha\beta} = \frac{\partial}{\partial B_\alpha} \frac{\partial}{\partial \mu_{a,\beta}} \bra{\psi^{\mathrm{e}}_{B}(\bm{R}_O)} H \ket{\psi^{\mathrm{e}}_{B}(\bm{R}_O)}\,\bigg|_{\bm{B} = 0, \bm{\mu} = 0}\,,
\end{equation}
with $\ket{\psi^{\mathrm{e}}_{B}(\bm{R}_O)}$ denoting the electronic state in the presence of the magnetic field. The electronic state depends on the chosen gauge, in our case on the gauge origin $\bm{R}_O$. The total shielding $\sigma_{a}$ may be split in a diamagnetic contribution $\sigma_{a}^{\mathrm{dia}}(\bm{R}_O)$ and a paramagnetic contribution $\sigma_{a}^{\mathrm{para}} (\bm{R}_O)$. However, note that these terms are not separately observable due to their gauge dependence. Only the total shielding
\begin{equation}
    \sigma_{a} =
    \underbrace{
        - \frac{e^2\mu_0} {8\pi m_{\mathrm{e}}} 
        \bra{\psi^{\mathrm{e}}_B} 
        \sum_{i=1}^{N_{\mathrm{e}}}  
        \frac{\bm{r}_{i,O} \bm{r}_{i,a}^T - \bm{r}_{i,a} \cdot \bm{r}_{i,O} \mathbb{1}}{r_{i,a}^3} 
        \ket{\psi^{\mathrm{e}}_{B}}\bigg|_{\bm{B}=0}
    }_{\displaystyle \sigma^{\mathrm{dia}}_a(\bm{R}_O)}
    +
    \underbrace{
        \bm{\nabla}_{\!\bm{B}} 
        \bra{\psi^{\mathrm{e}}_{B}} 
        \sum_{i=1}^{N_{\mathrm{e}}} 
        \frac{\mu_0 e}{4\pi m_{\mathrm{e}}}   
        \frac{\bm{L}_{i,a}}{r_{i,a}^3} 
        \ket{\psi_{B}^{\mathrm{e}}} 
        \bigg|_{\bm{B}=0}
    }_{\displaystyle \sigma_a^{\mathrm{para}}(\bm{R}_O)}\,
    \label{eq:Shielding}
\end{equation}
is a gauge-invariant physical quantity.
}

\hl{\subsection{Paramagnetic and diamagnetic shielding calculations at arbitrary gauge origins}
Most computational chemistry programs choose the nuclear position as gauge origin for the calculation of the diamagnetic and paramagnetic shielding at a specific nuclear site. However, we can relate the diamagnetic shielding for different gauge origins $O_1$ and $O_{{2}}$ via the expectation value of the electronic electric field,
\begin{multline}
    (\sigma^{\mathrm{dia}}_{a}(\bm{R}_{O_1}) - \sigma^{\mathrm{dia}}_{a}(\bm{R}_{O_2}))  = -\frac{e^2 \mu_0}{8 \pi m_{\mathrm{e}}} \bra{\psi^{\mathrm{e}}_{B}} \sum_{i=1}^{N_{\mathrm{e}}} \frac{\bm{R}_{21} \bm{r}_{i,a}^T - (\bm{r}_{i,a}\cdot \bm{R}_{21}) \mathbb{1} }{r_{i,a}^3} \ket{\psi^{\mathrm{e}}_{B}} \Bigg|_{\bm{B}=0} \\
    = \frac{e}{2m_\mathrm{e} c^2} \left[\bm{R}_{12} \langle \bm{E}^{\mathrm{e}}(\bm{R}_{a}) \rangle^T - (\bm{R}_{12} \cdot \langle\bm{E}^{\mathrm{e}}(\bm{R}_{a}) \rangle )\mathbb{1 } \right]\,.
    \label{eq:shielding-efield}
\end{multline}
 From this, one can then reconstruct the diamagnetic and the paramagnetic part of the shielding with an arbitrary gauge origin. Alternatively, this may be written as
\begin{equation}
    (\sigma^{\mathrm{dia}}_{a}(\bm{R}_{O_1}) - \sigma^{\mathrm{dia}}_{a}(\bm{R}_{O_2}))_{\alpha\beta} = -\frac{e}{2m_{\mathrm{e}}c^2} \varepsilon_{\alpha \rho \kappa} \varepsilon_{\kappa \mu\beta} (\bm{R}_{21})_{{\mu}} \langle E_{{\rho}}^{\mathrm{e}}(\bm{R}_{a})\rangle\,,
\end{equation}
where the Einstein sum convention is used for Cartesian indices. Note that this separation only works if conventional atomic orbitals are used. In the case of London or gauge-including atomic orbitals, this is not applicable since the separation into the paramagnetic and diamagnetic parts is typically performed differently.\cite{Helgaker1998, Stochyev2018}}
\hl{
\section{Spin-rotation coupling}
\label{sec:appendix-spin-rotation}
The spin-rotation tensor $M^{\mathrm{rot}}_a$ is defined as the negative second derivative of the energy with respect to $\langle \bm{J}\rangle$ and the nuclear spin $\bm{I}_a$ at site $\bm{R}_a$,
\begin{equation}
    M^{\mathrm{rot}}_a = -\nabla _{\!\langle\bm{J}\rangle}\!\nabla_{\!\bm{I}_a}E \,.
\end{equation}
Note that there are differing sign conventions for the spin-rotation tensor. We first look at $H_{\mathrm{NSOC}}^{\mathrm{e},(0)}$, appearing in Equation~\eqref{eq:electronic-spin-rotation} 
\begin{equation}
   \bra{\Psi}  H_{\mathrm{NSOC}}^{\mathrm{e},(0)} \ket{\Psi} 
        = \sum_{a=1}^{N_\mathrm{N}} \gamma_a\bm{I}_a \cdot {\bra{\chi^{\mathrm{rot}}}}\bra{\psi^{\mathrm{e,rot}}_0}  \sum_{i=1}^{N_{\mathrm{e}}} \frac{\mu_0 e}{4 \pi}  \frac{f_{a}} {m_{a} } 
    \frac{\bm{R}^0_{a} - \bm{r}_i}{(r^0_{i,a})^3}  \times \bm{P}_{a} \ket{\psi^{\mathrm{e,rot}}_0}\ket{\chi^{\mathrm{rot}}} \,.
    \label{eq:electronic-spin-rotation-NSOC}
\end{equation}
Assuming that $\psi^{\mathrm{e,rot}}_0$ is determined for fixed $\bm{R}^0$, $\bm{P}_a$ does not act on $\psi^{\mathrm{e,rot}}_0$. Thus, we can use  $\bra{\chi^{\mathrm{rot}}} (\bm{R}^0_{a} - \bm{r}_i)  \times \bm{P}_{a}\ket{\chi^{\mathrm{rot}}} = m_a[(\bm{R}^0_{a} - \bm{r}_{i}) \cdot {\bm{R}_{a}^0} - {\bm{R}_{a}^0} (\bm{R}^0_{a} - \bm{r}_{i})^T]\Theta_0^{-1} \langle \bm{J} \rangle$, which yields
\begin{equation}
   \bra{\Psi}  H_{\mathrm{NSOC}}^{\mathrm{e},(0)} \ket{\Psi} 
        = \frac{\mu_0e}{4\pi}\sum_{a=1}^{N_\mathrm{N}} \gamma_a\bm{I}_a \cdot  f_{a}  \sum_{i=1}^{N_\mathrm{N}} \bra{\psi_{0}^\mathrm{e,rot}}   \frac{\bm{R}^0_{a} (\bm{r}^0_{i,a})^T - (\bm{R}^0_{a} \cdot \bm{r}^0_{i,a}) \mathbb{1}}{(r_{i,a}^0)^3} \ket{\psi^{\mathrm{e,rot}}_{0}} \Theta_0^{-1} \langle \bm{J} \rangle \,.
    \label{eq:electronic-spin-rotation-NSOC-2}
\end{equation}
For the electronic part of the spin-rotation constant we find
\begin{equation}
\begin{aligned}
    M^{\mathrm{rot,e}}_a &= -\nabla_{\!\langle\bm{J}\rangle} \nabla_{\!\bm{I}_a} \bra{\chi^{\mathrm{rot}}}\bra{\psi_0^{\mathrm{e,rot}}} H_{\mathrm{NSOC}}^{\mathrm{e},(0)} + H_{\mathrm{NSOOC}}^{\mathrm{e},(0)}\ket{\psi_0^{\mathrm{e,rot}}}\ket{\chi^{\mathrm{rot}}}\Bigg|_{\langle\bm{J}\rangle=0} \\
    &= -\frac{\mu_0e}{4\pi} \gamma_a \Bigg( f_{a}  \sum_{i=1}^{N_\mathrm{e}} \bra{\psi_{0}^\mathrm{e,rot}}   \frac{\bm{R}_{a}^0 (\bm{r}_{i,a}^0)^T - (\bm{R}_{a}^0 \cdot \bm{r}_{i,a}^0) \mathbb{1}}{(r_{i,a}^0)^3} \ket{\psi^{\mathrm{e,rot}}_{0}} \Theta_0^{-1} \\
    &\qquad\qquad\qquad\qquad\qquad\qquad\qquad+ \bm{\nabla}_{\langle\bm{J}\rangle} \bra{\psi^{\mathrm{e,rot}}_{0}} \sum_{i=1}^{N_\mathrm{e}} \frac{1}{ m_{\mathrm{e}}}   \frac{\bm{L}_{i,a}^0}{(r_{i,a}^0)^3} \ket{\psi^{\mathrm{e,rot}}_{0}} \Bigg)\Bigg|_{\langle\bm{J}\rangle=0} \\
    &= \frac{2m_{\mathrm{e}}}{e}  \gamma_a \left[f_{a}(\sigma_{a}^{\mathrm{dia}}(\bm{0}) - \sigma_a^{\mathrm{dia}}( \bm{R}_{a}^0)) +\sigma^{\mathrm{para}}_{a} (\bm{0})\right] \Theta_0^{-1} \\
    &= \frac{2m_{\mathrm{e}}}{e} \gamma_a \left[ (\sigma_{a} - \sigma_{a}^{\mathrm{dia}}(\bm{R}_a^0))\Theta_0^{-1}+ (f_{a} -1) (\sigma_{a}^{\mathrm{dia}}(\bm{0}) - \sigma_{a}^{\mathrm{dia}}(\bm{R}_a^0))) \Theta_0^{-1} \right]\\
    &= \frac{2m_{\mathrm{e}}}{e} \gamma_a \left[  \sigma_{a}^{\mathrm{para}}(\bm{R}_{a}^0)\Theta_0^{-1}+ (f_{a} -1) (\sigma_{a}^{\mathrm{dia}}(\bm{0}) - \sigma_{a}^{\mathrm{dia}}(\bm{R}_a^0))) \Theta_0^{-1}\right]\,.
    \label{eq:relation-spinrot-electronic}
\end{aligned}
\end{equation}
Herein, we used the similarity of the perturbation terms $H^{\mathrm{para}} = \frac{e}{2m_{\mathrm{e}}}\bm{B} \cdot \bm{L}_{\mathrm{e}}$ and $H_{\mathrm I} = - \bm{J}\Theta_0^{-1}\bm{L}_{\mathrm{e}}$ to relate the second term in Equation~\eqref{eq:Shielding} to the second term in the second line of Equation~\eqref{eq:relation-spinrot-electronic} as we can prove in the following way. }
\hl{The wavefunction $\ket{\psi^{\mathrm{e,rot}}}$ is the solution of the electronic Schr\"odinger equation
\begin{equation}
    \left[H_\mathrm{e}(\bm{R}^0) - \bm{J} \Theta_0^{-1} \bm{L}_{\mathrm{e}} \right] \ket{\psi^{\mathrm{e,rot}}} = E^{\mathrm{e,rot}} \ket{\psi^{\mathrm{e,rot}}}\,,
\end{equation}
where the angular momentum $\bm{J}$ is seen as a parameter. Similarly, the electronic Schr\"odinger equation, to first order in a magnetic field $\bm B$, and neglecting the nuclear magnetic moments, reads
\begin{equation}
    \left[H_\mathrm{e}(\bm{R}^0) + \frac{e}{2m_\mathrm{e}} \bm{B} \cdot \bm{L}_{\mathrm{e}} \right] \ket{\psi^{\mathrm{e}}_B(\bm{0})} = E^{\mathrm{e}}_{B} \ket{\psi^{\mathrm{e}}_B(\bm{0})}\,,
\end{equation}
with $\ket{\psi^{\mathrm{e}}_B(\bm{0})}$ the electronic state in the presence of the magnetic field $\bm B$ at gauge origin $\bm R_O = \bm 0$, as discussed in Appendix~\ref{sec:Chemical-shielding}.
Thus, it is obvious that we have
\begin{align}
    \begin{split}
    -\bm{\nabla}_{\!\langle\bm{J}\rangle} \bra{\psi^{\mathrm{e,rot}}_{0}} \sum_{i=1}^{N_\mathrm{e}} \frac{1}{ m_{\mathrm{e}}}   \frac{\bm{L}_{i,a}^0}{(r_{i,a}^0)^3} \ket{\psi^{\mathrm{e,rot}}_{0}} \Bigg|_{\langle\bm{J}\rangle=0} \!\!\!\!\!\!&= \frac{2m_{\mathrm{e}}}{e} \nabla_{\!\bm{B}} \bra{\psi^{\mathrm{e}}_{B}(\bm{0})} \sum_{i=1}^{N_\mathrm{e}} \frac{1}{ m_{\mathrm{e}}}   \frac{\bm{L}_{i,a}^0}{(r_{i,a}^0)^3} \ket{\psi^{\mathrm{e}}_{B}(\bm{0})} \Theta_0^{-1}\Bigg|_{\bm{B} = 0}  \\ &=\frac{8\pi m_e}{\mu_0 e^2} \sigma^{\mathrm{para}}_a(\bm{0}){\Theta_0^{-1}}\,.
    \end{split}
\end{align}}

\hl{Note also the equivalence
\begin{equation}
   \frac{2m_{\mathrm{e}}}{e} (\sigma_{a}^{\mathrm{dia}}(\bm{0}) - \sigma_{a}^{\mathrm{dia}}(\bm{R}_a^0)))  =\frac{1}{c^2} \left[ (\langle \bm{E}_{\mathrm{e}} (\mathbf{R}^0_a) \rangle \cdot \bm{R}^0_{a}) \mathbb{1} - \bm{R}^0_{a} \langle \bm{E}_{\mathrm{e}} (\mathbf{R}^0_a) \rangle ^{T}\right]\,,
\end{equation}
which is obtained from Equation~\eqref{eq:shielding-efield}. Note that in the above the center of mass is always chosen as the coordinate origin, thus $\sigma_a^{\mathrm{dia}}(\bm{0})$ is the diamagnetic shielding at the center of mass. For the nuclear contribution to the spin-rotation tensor we find
\begin{equation}
    \begin{aligned}
        M^{\mathrm{rot,N}}_a &= -\nabla_{\!\langle\bm{J}\rangle} \nabla_{\!\bm{I}_a} \bra{\chi^{\mathrm{rot}}} H_{\mathrm{NSOC}}^{\mathrm{N},(0)} + H_{\mathrm{NSOOC}}^{\mathrm{N},(0)}\ket{\mathrm{\chi}^\mathrm{rot}}\\
        &= \nabla_{\!\langle\bm{J}\rangle} \nabla_{\!\bm{I}_a} \bra{\chi^{\mathrm{rot}}} \frac{\mu_0e}{4\pi}\sum_{a=1} ^{N_{\mathrm{N}}} \gamma_a \bm{I}_a\cdot\sum_{b\neq a}^{N_{\mathrm{N}}}  \frac{Z_b}{(R^0_{a,b})^3}\left[f_a\frac{\bm{L}^0_{a,b}}{m_a} + \frac{\bm{L}^0_{b,a}}{m_b}\right]\ket{\mathrm{\chi}^\mathrm{rot}} \\
   & = {+} \frac{\mu_0e}{4\pi} \nabla_{\!\langle\bm{J}\rangle} \nabla_{\!\bm{I}_a} \sum_{a=1} ^{N_{\mathrm{N}}} \gamma_a \bm{I}_a\cdot\sum_{b\neq a}^{N_{\mathrm{N}}}  \frac{Z_b}{(R^0_{a,b})^3}\left[f_a S_{a,b}^0 + S_{b,a}^0\right] \Theta_0^{-1}\langle \bm{J} \rangle\\
 & =      \gamma_a \Bigg( \frac{\mu_0 e}{4\pi}\sum_{b \neq a} \frac{Z_{b}}{m_{b}}\frac{\Theta_0^{(b,a)} \Theta_0^{-1} }{(R_{a,b}^0)^3} {+} \frac{\mu_0e}{4\pi} (f_{a} - 1) \sum_{b \neq a}^{N_\mathrm{N}}Z_{b}\frac{S_{a,b}^0}{(R_{a,b}^0)^3} \Theta_0^{-1}\Bigg)\,,
 \end{aligned}
 \label{eq:Derivation-spin-rot-nuclear}
\end{equation}
where we used $\bra{\chi^{\mathrm{rot}}} \bm{L}^0_{{a,b}}  \ket{\chi^{\mathrm{rot}}} = m_{{a}} [(\bm{R}^0_{{a}} - \bm{R}^0_{b}) \cdot \bm{R}^0_{{a}} \mathbb{1}- \bm{R}^0_{{a}} (\bm{R}^0_{a} - \bm{R}^0_{b})^T]\Theta_0^{-1} \langle \bm{J} \rangle\,$.}

\hl{
\section{Spin-vibration vector}
\label{sec:spin-vibration-appendix}
This section fills in details on the derivation of the effective magnetic field and the spin-vibration vector in Section~\ref{sec:Spin-vibration-coupling}. 
First, we discuss the expectation value $ \bra{\chi_t^{\mathrm{vib}}}  (\Lambda \delta\bm{ R}_{a}) \times \bm{P}_{b} \ket{\chi_t^{\mathrm{vib}}}$ with a constant symmetric matrix $\Lambda$, as it appears in the expectation values of the spin interaction Hamiltonians with respect to $\ket{\Psi} = \ket{\chi^{\mathrm{vib}}_t}\ket{\psi_0^{\mathrm{e,vib}}}$. For $\ket{\chi^{\mathrm{vib}}_t}$ we have, according to Equation~(S20) in the SI, 
\begin{equation}
\bra{\chi^{\mathrm{vib}}_t} Q_{r} P_{r} \ket{\chi^{\mathrm{vib}}_t} = \frac{i}{2}\,,
\end{equation}
and $\bra{\chi^{\mathrm{vib}}_t} Q_r P_s \ket{\chi_t^{\mathrm{vib}}}$ with $r\!\neq\! s$ can only be non-zero if both $r$ and $s$ correspond to the same degenerate vibration $t$, and this degenerate mode is excited with angular momentum, as can be seen from Equations~(S22) and (S23).
Combining all these results and using normal coordinates, we deduce
\begin{equation}
    \begin{aligned}
\bra{\chi_t^{\mathrm{vib}}}&  (\Lambda \delta\bm{ R}_{a}) \times \bm{P}_{b} \ket{\chi_t^{\mathrm{vib}}} =  \sum_{r=1}^{3N_{\mathrm{N}}-6} \sum_{s=1}^{3N_{\mathrm{N}}-6} ((\Lambda \bm{l}_{a,r}) \times \bm{l}_{b,s})   \bra{\chi_t^{\mathrm{vib}}} Q_{r} P_{s} \ket{\chi_t^{\mathrm{vib}}}\\
    &= \sum_{r\neq s} ((\Lambda \bm{l}_{a,r}) \times \bm{l}_{b,s})   \bra{\chi_t^{\mathrm{vib}}} Q_{r} P_{s}\ket{\chi_t^{\mathrm{vib}}}  + \sum_r ((\Lambda \bm{l}_{a,r}) \times \bm{l}_{b,r})   \bra{\chi_t^{\mathrm{vib}}} Q_{r} P_{r}\ket{\chi_t^{\mathrm{vib}}}\\
    & =  \sum_{t_r \neq t_s} (\Lambda \bm{l}_{a,t_r})  \times \bm{l}_{b,t_s} \bra{\nu_t,l_t} Q_{t_r} P_{t_s} \ket{\nu_t,l_t} \\  &= \frac{1}{2} ((\Lambda \bm{l}_{a,t_1}) \times \bm{l}_{b,t_2} - (\Lambda \bm{l}_{a,t_2}) \times \bm{l}_{b,t_1})
    \langle G_t \rangle\,,
    \end{aligned}
    \label{eq:Symmetric-cross-product-relation}
\end{equation}
where we used 
\begin{equation}
    \begin{aligned}
     \left[\sum_{r=1} ^{3N_{\mathrm{N}}-6} \Lambda(\bm{l}_{a,r} ) \times \bm{l}_{b,r} \right]_\alpha & = \sum_{\beta,\gamma=1}^3 \sum_{r=1}^{3N_{\mathrm{N}}-6} \sum_{\delta=1}^3 \varepsilon_{\alpha \beta \gamma} \Lambda_{\beta \delta} [l_{a,r}]_\delta [l_{b,r}]_\gamma\\
    &=  \sum_{\beta ,\gamma=1}^3  \sum_{\delta=1}^3 \varepsilon_{\alpha \beta \gamma} \Lambda_{\beta \delta} \!\!\!\!\underbrace{\sum_{r= 1}^{3N_{\mathrm{N}}-6}v_{3(a-1)+\delta,r} \,v_{3(b-1)+\gamma,r}}_{\displaystyle=\delta_{3(a-1)+\delta,3(b-1)+\gamma} =\delta_{a,b}\delta_{\delta,\gamma}} = 0\,,
    \end{aligned}
\end{equation}
and applied $[l_{a,r}]_\delta = v_{3(a-1)+\delta,r}$ from Equation~(S4), which can be found in the SI.
We are now equipped to derive in detail the expectation values of the spin interaction Hamiltonians in Equations~\eqref{eq:SVC-e-1-exp} 
\begin{equation}
           \begin{aligned}
            \bra{\Psi} H^{\mathrm{e},(1)} _{\mathrm{NSOC}} \ket{\Psi}& =\bra{\chi^{\mathrm{vib}}_t} \bra{\psi^{\mathrm{e,vib}}_0}H^{\mathrm{e},(1)}_{\mathrm{NSOC}} \ket{\psi^{\mathrm{e,vib}}_0}\ket{\chi^{\mathrm{vib}}_t}  \\
                &= \sum_{a=1}^{N_\mathrm{N}} \frac{f_{a} }{m_{a}c^2} \gamma_{a} \bm{I}_{a} \cdot 
    \bra{\chi^{\mathrm{vib}}_t}(\bra{\psi^{\mathrm{e,vib}}_0}\varphi_{\mathrm{el}}(\bm{R}_a^0)\ket{\psi^{\mathrm{e,vib}}_0} \delta\bm{ R}_{a} ) \times \bm{P}_{a}\ket{\chi^{\mathrm{vib}}_t}\,, \\
     & = \frac{1}{2}\sum_{a=1}^{N_\mathrm{N}}  \frac{f_a}{m_ac^2}\gamma_a \bm{I}_a \cdot \left[(\langle\varphi_{\mathrm{el}}(\bm{R}_a^0)\rangle \bm{l}_{a,t_1}) \times \bm{l}_{a,t_2} - (\langle\varphi_{\mathrm{el}}(\bm{R}_a^0)\rangle \bm{l}_{a,t_2}) \times \bm{l}_{a,t_1}\right]\,,\\
               & = \frac{e}{4\pi \varepsilon_0} \sum_{a=1}^{N_{\mathrm{N}}} \frac{f_a }{m_{a} c^2} \gamma_a \bm{I}_a \cdot
                  \bm{\Gamma}_{a,t} \langle G_t \rangle\,.
            \end{aligned}
            \label{eq:Derivation-SVC-electronic-SOC}
\end{equation}
where we used \eqref{eq:Symmetric-cross-product-relation} and the definition of $\bm{\Gamma}_{a,t}$ from Equation~\eqref{eq:Coupling-vector-1}.
Next, we look at the other first order electronic term
\begin{equation}
    \begin{aligned}
        \bra\Psi H_{\mathrm{NSOOC}}^{\mathrm {e},(1)} \ket\Psi &= \bra{\chi^{\mathrm{vib}}_t} \bra{\psi^{\mathrm{e,vib}}_0}H_{\mathrm{NSOOC}}^{\mathrm {e},(1)} \ket{\psi^{\mathrm{e,vib}}_0} \ket{\chi^{\mathrm{vib}}_t} \\&= -\frac{\mu_0e}{4 \pi m_e} \sum_{a=1}^{N_{\mathrm{N}}} \sum_{i=1}^{N\mathrm{e}} \gamma_{a} \bm{I}_{a} \cdot \bra{\chi^{\mathrm{vib}}_t} \bra{\psi^{\mathrm{e,vib}}_0}(\varphi_{i,a}(\bm{R}^0) \delta\bm{ R}_{a}) \times \bm{p_i}\ket{\psi^{\mathrm{e,vib}}_0} \ket{\chi^{\mathrm{vib}}_t}\,.
        \end{aligned}    
\end{equation}
This expression vanishes as we consider $\ket{\psi^{\mathrm{e,vib}}_0}$ for fixed nuclear position only. Thus, this term is proportional to $\bra{\chi^{\mathrm{vib}}_t}\delta\bm{R}_a\ket{\chi^{\mathrm{vib}}_t}= \frac{1}{\sqrt{m_a}}\sum_{r=1}^{N_{\mathrm{N}}-6} \bm{l}_{a,r }\bra{\chi^{\mathrm{vib}}_t}Q_r\ket{\chi^{\mathrm{vib}}_t}$ = 0. Furthermore, we derive expectation values of the nuclear contributions in Equation~\eqref{eq:SVC-N-1-exp},
\begin{equation}
\begin{aligned}
    \bra\Psi H_{\mathrm{NSOOC}}^{\mathrm{N},(1)} &+ H_{\mathrm{NSOC}}^{\mathrm{N},(1)} \ket{\Psi}  = \bra{\chi^{\mathrm{vib}}_t} \bra{\psi^{\mathrm{e,vib}}_0}H_{\mathrm{NSOOC}}^{\mathrm{N},(1)} + H_{\mathrm{NSOC}}^{\mathrm{N},(1)}\ket{\psi^{\mathrm{e,vib}}_0}\ket{\chi^{\mathrm{vib}}_t}\\
    &= \frac{\mu_0e}{4\pi} \sum_{a=1}^{N_\mathrm{N}} \gamma_{a} \bm{I}_{a} \cdot \sum_{b \neq a}^{N_\mathrm{N}}\bra{\chi_t^{\mathrm{vib}}}    (\varphi_{a,b}(\bm{R}^0) \delta\bm{ R}_{a,b} ) \times  \left[ {-}\frac{f_{a} Z_{b} }{ m_{a} } 
      \bm{P}_{a} +     
     \frac{Z_{b}}{m_{b}}   
    \bm{P}_{b}\right]\ket{\mathrm{\chi_t^\mathrm{vib}}}\\
    & = {+}\frac{\mu_0e}{4\pi}\sum_{a=1} ^{N_\mathrm{N}} \gamma_a \bm{I} _a \cdot\sum_{b\neq a}^{N_{\mathrm{N}}} Z_{b} \left[f_{a}\frac{\bm{\Gamma}_{a,b,t}}{m_{a}} {+} \frac{\bm{\Gamma}_{b,a,t}}{m_{b}} \right] \langle G_t \rangle\,,
    \label{eq:Derivation-SVC-nuclear}
\end{aligned}
\end{equation}
where we used $\bra{\chi_t^{\mathrm{vib}}}  (\Lambda \delta\bm{ R}_{a}) \times \bm{P}_{b} \ket{\chi_t^{\mathrm{vib}}}= \frac{1}{2} ((\Lambda \bm{l}_{a,t_1}) \times \bm{l}_{b,t_2} - (\Lambda \bm{l}_{a,t_2}) \times \bm{l}_{b,t_1})
\langle G_t \rangle\,$, which is valid for any symmetric matrix $\Lambda$ and arbitrary indices $a$ and $b$, as proven in Equation~\eqref{eq:Symmetric-cross-product-relation}.
The coupling constant  $\bm{\Gamma}_{a,b,t}$ is defined in Equation~\eqref{eq:Coupling-vector-2}.
}
\hl{
\section{Splitting of induced B-field in spin-orbit and spin-other-orbit coupling}
\label{sec:appendix-spin-vib-splitting}
We can distinguish between the part of the effective magnetic field induced by the motion of all other charged particles (NSOOC), and the part due to the particle's own motion (NSOC). For a vibrationally induced magnetic field, the former part reads 
\begin{equation}
     \bm{B}^{\mathrm{vib}}_{\mathrm{NSOOC}}(\bm{R}_a) =  \left( \left[-\frac{\mu_0 e}{4\pi} \sum_{b \neq a} Z_{b} \frac{S^0_{b a}}{(R^0_{a b})^3}  - \frac{2m_{\mathrm{e}}}{e} \sigma^{\mathrm{para}}_{a}({\bm{0}}) \right] \Theta_0^{-1} \bm{\zeta}_t  - \frac{\mu_0e }{4\pi} \sum_{b\neq a}^{N_\mathrm{N}} Z_{b} \frac{\bm{\Gamma}_{b,a,t}}{m_{b}} \right) \langle G_t\rangle\,,
\end{equation}
while the latter is given by
\begin{multline}
    \bm{B}^{\mathrm{vib}}_{\mathrm{NSOC}}(\bm{R}_a) = {-}f_{a} \Bigg(\frac{e\mu_0 }{4\pi }  \left [\frac{1}{m_{a}} \left(\sum_{b\neq a}^{N_\mathrm{N}}  Z_{b} \bm{\Gamma}_{a,b,t}  {+}  \bm{\Gamma}_{a,t} \right) {+} \sum_{b \neq a}^{N_{\mathrm{N}}} Z_{b} \frac{S^0_{ab}}{(R^0_{a,b})^3} \Theta_0^{-1} \bm{\zeta}_t\right]  \\ {+} \frac{1}{c^2}  \langle \bm{E}^{\mathrm{e}}_{a} \rangle \times  ((\Theta_0^{-1} \bm{\zeta}_t) \times \bm{R}_{a}^0) \Bigg) \langle G_t\rangle\,
\end{multline}
with $S^0_{a b} = (\bm{R}^0_{ab}\cdot\bm{R}_{a}^0) \mathbb{1} - \bm{R}_{a}^0 \left(\bm{R}^0_{a b}\right)^{\!T}$.
A similar classification can be done in the case of rotation, where we obtain
\begin{align}
    \bm{B}^{\mathrm{rot}}_{\mathrm{NSOOC}}(\bm{R}_a) &=  \left[{\frac{2m_{\mathrm{e}}}{e}} \sigma_a^{\mathrm{para}}(\bm{0}) +\frac{\mu_0e}{4\pi} \sum_{b\neq a}^{N_{\mathrm{N}}} \frac{Z_b}{(R^0_{a,b})^3} S^0_{b,a} \right ]\Theta_0^{-1}\langle \bm{J} \rangle\,,
      \\
    \bm{B}^{\mathrm{rot}}_{\mathrm{NSOC}}(\bm{R}_a) &= f_a\left[{\frac{2m_{\mathrm{e}}}{e}}(\sigma^{\mathrm{dia}}(\bm{0})-\sigma^{\mathrm{dia}}(\bm{R}_a)) + \frac{\mu_0e}{4\pi} \sum_{b\neq a}^{N_{\mathrm{N}}} \frac{Z_b}{(R^0_{a,b})^3} S^0_ {a,b} \right ]\Theta_0^{-1}\langle \bm{J} \rangle\,.
\end{align}
}

\section*{Acknowledgements}
Support by NAWI Graz and the use of HPC resources provided by the IT services of Graz University of Technology (ZID) as well as the Vienna Scientific Cluster (VSC) is gratefully acknowledged.

\section*{Disclosure statement}
There are no relevant financial or non-financial competing interests to report. 

\section*{Data availability statement}
All data that support the findings of this study are included within the article (and the supplementary file). 

\section*{Funding}
This research was funded in whole by the Austrian Science Fund (FWF) [10.55776/P36903]. For the purpose of open access, the author has applied a CC BY public copyright license to any Author Accepted Manuscript version arising from this submission.

\bibliographystyle{tfo}  
\bibliography{main}

@article{Mattioni2024,
	author = {Mattioni, Andrea and Staab, Jakob K. and Blackmore, William J. A. and Reta, Daniel and Iles-Smith, Jake and Nazir, Ahsan and Chilton, Nicholas F.},
	doi = {10.1038/s41467-023-44486-3},
	id = {Mattioni2024},
	isbn = {2041-1723},
	journal = {Nature Communications},
	number = {1},
	pages = {485},
	title = {Vibronic effects on the quantum tunnelling of magnetisation in Kramers single-molecule magnets},
	url = {https://doi.org/10.1038/s41467-023-44486-3},
	volume = {15},
	year = {2024}
}

@article{Zhou2024nat,
	author = {Zhou, Aimei and Li, Denan and Tan, Mingshu and Lv, Yanpei and Pang, Simin and Zhao, Xinxing and Shi, Zhifu and Zhang, Jun and Jin, Feng and Liu, Shi and Sun, Lei},
	doi = {10.1038/s41467-024-54989-2},
	isbn = {2041-1723},
	journal = {Nature Communications},
	number = {1},
	pages = {10763},
	title = {Phononic modulation of spin-lattice relaxation in molecular qubit frameworks},
	url = {https://doi.org/10.1038/s41467-024-54989-2},
	volume = {15},
	year = {2024}
}

@article{Wihelmer2024,
author = {Wilhelmer, Raphael and Diez, Matthias and Krondorfer, Johannes K. and Hauser, Andreas W.},
title = {Molecular Pseudorotation in Phthalocyanines as a Tool for Magnetic Field Control at the Nanoscale},
journal = {Journal of the American Chemical Society},
volume = {146},
number = {21},
pages = {14620-14632},
year = {2024},
doi = {10.1021/jacs.4c01915},
note ={PMID: 38743819},
URL = {https://doi.org/10.1021/jacs.4c01915},
eprint = {https://doi.org/10.1021/jacs.4c01915}
}

@article{Mardirossian2016,
    author = {Mardirossian, Narbe and Head-Gordon, Martin},
    title = {ωB97M-V: A combinatorially optimized, range-separated hybrid, meta-GGA density functional with VV10 nonlocal correlation},
    journal = {The Journal of Chemical Physics},
    volume = {144},
    number = {21},
    pages = {214110},
    year = {2016},
    month = {06},
    issn = {0021-9606},
    doi = {10.1063/1.4952647},
    url = {https://doi.org/10.1063/1.4952647}
}

@article{Dunning1989,
    author = {Dunning, Thom H., Jr.},
    title = {Gaussian basis sets for use in correlated molecular calculations. I. The atoms boron through neon and hydrogen},
    journal = {The Journal of Chemical Physics},
    volume = {90},
    number = {2},
    pages = {1007-1023},
    year = {1989},
    month = {01},
    issn = {0021-9606},
    doi = {10.1063/1.456153},
    url = {https://doi.org/10.1063/1.456153}
}

@article{Dunning1992,
    author = {Kendall, Rick A. and Dunning, Thom H., Jr. and Harrison, Robert J.},
    title = {Electron affinities of the first‐row atoms revisited. Systematic basis sets and wave functions},
    journal = {The Journal of Chemical Physics},
    volume = {96},
    number = {9},
    pages = {6796-6806},
    year = {1992},
    month = {05},
    issn = {0021-9606},
    doi = {10.1063/1.462569},
    url = {https://doi.org/10.1063/1.462569}
}

@article{Dunning1993,
    author = {Woon, David E. and Dunning, Thom H., Jr.},
    title = {Gaussian basis sets for use in correlated molecular calculations. III. The atoms aluminum through argon},
    journal = {The Journal of Chemical Physics},
    volume = {98},
    number = {2},
    pages = {1358-1371},
    year = {1993},
    month = {01},
    issn = {0021-9606},
    doi = {10.1063/1.464303},
    url = {https://doi.org/10.1063/1.464303}
}

@article{Wilson1999,
    author = {Wilson, Angela K. and Woon, David E. and Peterson, Kirk A. and Dunning, Thom H., Jr.},
    title = {Gaussian basis sets for use in correlated molecular calculations. IX. The atoms gallium through krypton},
    journal = {The Journal of Chemical Physics},
    volume = {110},
    number = {16},
    pages = {7667-7676},
    year = {1999},
    month = {04},
    issn = {0021-9606},
    doi = {10.1063/1.478678}
}

@article{London1937,
    author = {London, F.},
    title = {{Théorie quantique des courants interatomiques dans les combinaisons aromatiques}},
    journal = {J. Phys. Radium},
    volume = {8},
    number = {10},
    pages = {397--404},
    year = {1937},
    month = {10},
    doi = {10.1051/jphysrad:01937008010039700}
}

@article{Wolinski1990,
	author = {Wolinski, Krzysztof and Hinton, James F. and Pulay, Peter},
	doi = {10.1021/ja00179a005},
	isbn = {0002-7863},
	journal = {Journal of the American Chemical Society},
	month = {11},
	number = {23},
	pages = {8251--8260},
	publisher = {American Chemical Society},
	title = {Efficient implementation of the gauge-independent atomic orbital method for NMR chemical shift calculations},
	url = {https://doi.org/10.1021/ja00179a005},
	volume = {112},
	year = {1990}
}

@article{Ditchfield1972,
    author = {Ditchfield, R.},
    title = {Molecular Orbital Theory of Magnetic Shielding and Magnetic Susceptibility},
    journal = {The Journal of Chemical Physics},
    volume = {56},
    number = {11},
    pages = {5688-5691},
    year = {1972},
    month = {06},
    issn = {0021-9606},
    doi = {10.1063/1.1677088}
    }

@article{Wick1948,
  title = {On the Magnetic Field of a Rotating Molecule},
  author = {Wick, G. C.},
  journal = {Phys. Rev.},
  volume = {73},
  issue = {1},
  pages = {51--57},
  numpages = {0},
  year = {1948},
  month = {Jan},
  publisher = {American Physical Society},
  doi = {10.1103/PhysRev.73.51},
  url = {https://link.aps.org/doi/10.1103/PhysRev.73.51}
}

@article{Juraschek17,
  title = {Dynamical multiferroicity},
  author = {Juraschek, Dominik M. and Fechner, Michael and Balatsky, Alexander V. and Spaldin, Nicola A.},
  journal = {Phys. Rev. Materials},
  volume = {1},
  issue = {1},
  pages = {014401},
  numpages = {9},
  year = {2017},
  month = {Jun},
  publisher = {American Physical Society},
  doi = {10.1103/PhysRevMaterials.1.014401},
  url = {https://link.aps.org/doi/10.1103/PhysRevMaterials.1.014401}
}

@article{Juraschek19,
  title = {Orbital magnetic moments of phonons},
  author = {Juraschek, Dominik M. and Spaldin, Nicola A.},
  journal = {Phys. Rev. Materials},
  volume = {3},
  issue = {6},
  pages = {064405},
  numpages = {8},
  year = {2019},
  month = {Jun},
  publisher = {American Physical Society},
  doi = {10.1103/PhysRevMaterials.3.064405},
  url = {https://link.aps.org/doi/10.1103/PhysRevMaterials.3.064405}
}

@article{Juraschek20,
  title = {Phono-magnetic analogs to opto-magnetic effects},
  author = {Juraschek, Dominik M. and Narang, Prineha and Spaldin, Nicola A.},
  journal = {Phys. Rev. Research},
  volume = {2},
  issue = {4},
  pages = {043035},
  numpages = {11},
  year = {2020},
  month = {Oct},
  publisher = {American Physical Society},
  doi = {10.1103/PhysRevResearch.2.043035},
  url = {https://link.aps.org/doi/10.1103/PhysRevResearch.2.043035}
}

@article{Moss1972,
author = { R.E.   Moss  and  A.J.   Perry},
title = {The vibrational Zeeman effect},
journal = {Molecular Physics},
volume = {25},
number = {5},
pages = {1121-1134},
year  = {1973},
publisher = {Taylor & Francis},
doi = {10.1080/00268977300100971},
URL = {https://doi.org/10.1080/00268977300100971},
eprint = {https://doi.org/10.1080/00268977300100971   
}}

@article{Braun1981,
author = {Braun, Petr and Kiselev, A and Rebane, T},
year = {1981},
month = {06},
pages = {2163},
title = {Contribution of vibrations to the magnetic moment of molecules of the symmetric top type},
volume = {80},
journal = {Journal of Experimental and Theoretical Physics}
}

@article{Watson1968,
author = {James K.G. Watson},
title = {Simplification of the molecular vibration-rotation hamiltonian},
journal = {Molecular Physics},
volume = {15},
number = {5},
pages = {479-490},
year = {1968},
publisher = {Taylor & Francis},
doi = {10.1080/00268976800101381},
URL = {https://doi.org/10.1080/00268976800101381},
eprint = {   https://doi.org/10.1080/00268976800101381}}

@book{wilson1980molecular,
  title={Molecular Vibrations: The Theory of Infrared and Raman Vibrational Spectra},
  author={Wilson, E.B. and Decius, J.C. and Cross, P.C.},
  isbn={9780486639413},
  lccn={79055912},
  series={Dover Books on Chemistry Series},
  url={https://books.google.at/books?id=CPkvsDrPiv0C},
  year={1980},
  publisher={Dover Publications}
}

@article{Gauss1996,
    author = {Gauss, Jürgen and Ruud, Kenneth and Helgaker, Trygve},
    title = "{Perturbation‐dependent atomic orbitals for the calculation of spin‐rotation constants and rotational g tensors}",
    journal = {The Journal of Chemical Physics},
    volume = {105},
    number = {7},
    pages = {2804-2812},
    year = {1996},
    month = {08},
    issn = {0021-9606},
    doi = {10.1063/1.472143},
    url = {https://doi.org/10.1063/1.472143},
    eprint = {https://pubs.aip.org/aip/jcp/article-pdf/105/7/2804/19254245/2804\_1\_online.pdf},
}

@article{Flygare1974,
author = {Flygare, W. H.},
title = {Magnetic interactions in molecules and an analysis of molecular electronic charge distribution from magnetic parameters},
journal = {Chemical Reviews},
volume = {74},
number = {6},
pages = {653-687},
year = {1974},
doi = {10.1021/cr60292a003},
URL = {https://doi.org/10.1021/cr60292a003},
eprint = {https://doi.org/10.1021/cr60292a003}}

@book{Bersuker2006,
  TITLE = {The Jahn-Teller effect},
  AUTHOR = {Isaac Bersuker},
  YEAR = {2006},
  ISBN = {9780511524769},
  DOI = {10.1017/CBO9780511524769},
  PUBLISHER = {Cambridge University Press},
  place={Cambridge}
}

@book{Domcke15,
author = {Domcke, Wolfgang and Yarkony, David R and Köppel, Horst},
title = {Advanced Series in Physical Chemistry: Conical Intersections},
publisher = {World Scientific},
VOLUME = {15},
year = {2004},
doi = {10.1142/5406},
address = {},
edition   = {},
URL = {https://www.worldscientific.com/doi/abs/10.1142/5406},
eprint = {https://www.worldscientific.com/doi/pdf/10.1142/5406}
}

@book{Domcke17,
author = {Domcke, Wolfgang and Yarkony, David R and Köppel, Horst},
title = {Advanced Series in Physical Chemistry: Conical Intersections},
publisher = {World Scientific},
VOLUME = 17,
year = {2011},
doi = {10.1142/7803},
address = {},
edition   = {},
URL = {https://www.worldscientific.com/doi/abs/10.1142/7803},
eprint = {https://www.worldscientific.com/doi/pdf/10.1142/7803}
}

@article{Berry1960,
author = {Berry, R. Stephen},
title = {Correlation of Rates of Intramolecular Tunneling Processes, with Application to Some Group V Compounds},
journal = {The Journal of Chemical Physics},
volume = {32},
number = {3},
pages = {933-938},
year = {1960},
doi = {10.1063/1.1730820},
URL = {https://doi.org/10.1063/1.1730820},
eprint = {https://doi.org/10.1063/1.1730820}
}

@article{Wang1993,
  title = {Vibrational Zeeman effect for the ${\ensuremath{\nu}}_{4}$ mode of haloforms (HC${\mathit{X}}_{3}$) determined by magnetic vibrational circular dichroism},
  author = {Wang, Baoliang and Tam, Cheok N. and Keiderling, Timothy A.},
  journal = {Phys. Rev. Lett.},
  volume = {71},
  issue = {7},
  pages = {979--982},
  numpages = {0},
  year = {1993},
  month = {Aug},
  publisher = {American Physical Society},
  doi = {10.1103/PhysRevLett.71.979},
  url = {https://link.aps.org/doi/10.1103/PhysRevLett.71.979}
}

@article{Wang1994,
    author = {Wang, Baoliang and Keiderling, Timothy A.},
    title = "{Measurement of the vibrational Zeeman effect for HCF3 using magnetic vibrational circular dichroism}",
    journal = {The Journal of Chemical Physics},
    volume = {101},
    number = {2},
    pages = {905-911},
    year = {1994},
    month = {07},
    issn = {0021-9606},
    doi = {10.1063/1.467744},
    url = {https://doi.org/10.1063/1.467744},
    eprint = {https://pubs.aip.org/aip/jcp/article-pdf/101/2/905/10774569/905\_1\_online.pdf},
}

@article{Devine1984,
author = {Devine, T. R. and Keiderling, T. A.},
title = {Magnetic vibrational circular dichroism of benzene and 1,3,5-substituted derivatives},
journal = {The Journal of Physical Chemistry},
volume = {88},
number = {3},
pages = {390-394},
year = {1984},
doi = {10.1021/j150647a013},
URL = { https://doi.org/10.1021/j150647a013},
eprint = {  https://doi.org/10.1021/j150647a013}
}

@article{RAMSEY-1950,
  title = {Magnetic Shielding of Nuclei in Molecules},
  author = {Ramsey, Norman F.},
  journal = {Phys. Rev.},
  volume = {78},
  issue = {6},
  pages = {699--703},
  numpages = {0},
  year = {1950},
  month = {Jun},
  publisher = {American Physical Society},
  doi = {10.1103/PhysRev.78.699},
  url = {https://link.aps.org/doi/10.1103/PhysRev.78.699}
}

@article{Reid,
  title = {Spin-rotation interaction and magnetic shielding in $^{1}\ensuremath{\Sigma}$ molecules},
  author = {Reid, Roderick V. and Chu, Amy Hwia-May},
  journal = {Phys. Rev. A},
  volume = {9},
  issue = {2},
  pages = {609--613},
  numpages = {0},
  year = {1974},
  month = {Feb},
  publisher = {American Physical Society},
  doi = {10.1103/PhysRevA.9.609},
  url = {https://link.aps.org/doi/10.1103/PhysRevA.9.609}
}

@article{orca,
author = {Neese, Frank},
title = {Software Update: The ORCA Program System—Version 6.0},
journal = {WIREs Computational Molecular Science},
volume = {15},
number = {2},
pages = {e70019},
doi = {https://doi.org/10.1002/wcms.70019},
url = {https://wires.onlinelibrary.wiley.com/doi/abs/10.1002/wcms.70019},
eprint = {https://wires.onlinelibrary.wiley.com/doi/pdf/10.1002/wcms.70019},
note = {e70019 CMS-1186.R1},
year = {2025}
}

@article{Orca-GIAO,
author = {Stoychev, Georgi L. and Auer, Alexander A. and Izsák, Róbert and Neese, Frank},
title = {Self-Consistent Field Calculation of Nuclear Magnetic Resonance Chemical Shielding Constants Using Gauge-Including Atomic Orbitals and Approximate Two-Electron Integrals},
journal = {Journal of Chemical Theory and Computation},
volume = {14},
number = {2},
pages = {619-637},
year = {2018},
doi = {10.1021/acs.jctc.7b01006},
    note ={PMID: 29301077},

URL = { 
    
        https://doi.org/10.1021/acs.jctc.7b01006
    
    

},
eprint = { 
    
        https://doi.org/10.1021/acs.jctc.7b01006
    
    

}

}

@article{PhysRev.85.24,
  title = {Rotational Magnetic Moments of $^{1}\ensuremath{\Sigma}$ Molecules},
  author = {Eshbach, J. R. and Strandberg, M. W. P.},
  journal = {Phys. Rev.},
  volume = {85},
  issue = {1},
  pages = {24--34},
  numpages = {0},
  year = {1952},
  month = {Jan},
  publisher = {American Physical Society},
  doi = {10.1103/PhysRev.85.24},
  url = {https://link.aps.org/doi/10.1103/PhysRev.85.24}
}

@article{PhysRev.112.1929,
  title = {Radio-Frequency Spectra of Hydrogen Deuteride in Strong Magnetic Fields},
  author = {Quinn, W. E. and Baker, J. M. and LaTourrette, J. T. and Ramsey, N. F.},
  journal = {Phys. Rev.},
  volume = {112},
  issue = {6},
  pages = {1929--1940},
  numpages = {0},
  year = {1958},
  month = {Dec},
  publisher = {American Physical Society},
  doi = {10.1103/PhysRev.112.1929},
  url = {https://link.aps.org/doi/10.1103/PhysRev.112.1929}
}

@article{Keiderling-Bour,
  title = {Theory of Molecular Vibrational Zeeman Effects as Measured with Circular Dichroism},
  author = {Keiderling, Timothy A. and Bou\ifmmode \check{r}\else \v{r}\fi{}, Petr},
  journal = {Phys. Rev. Lett.},
  volume = {121},
  issue = {7},
  pages = {073201},
  numpages = {6},
  year = {2018},
  month = {Aug},
  publisher = {American Physical Society},
  doi = {10.1103/PhysRevLett.121.073201},
  url = {https://link.aps.org/doi/10.1103/PhysRevLett.121.073201}
}

@article{Uehara-Kiyoji,
author = {Uehara ,Kiyoji and Shimoda ,Koichi},
title = {Hyperfine Interactions in the v3=1 Exited State of Methane},
journal = {Journal of the Physical Society of Japan},
volume = {36},
number = {2},
pages = {542-551},
year = {1974},
doi = {10.1143/JPSJ.36.542},

URL = { 
    
        https://doi.org/10.1143/JPSJ.36.542
    
    

},
eprint = { 
    
        https://doi.org/10.1143/JPSJ.36.542
    
    

}

}

@article{Benchmarking2013,
    author = {Teale, Andrew M. and Lutnæs, Ola B. and Helgaker, Trygve and Tozer, David J. and Gauss, Jürgen},
    title = {Benchmarking density-functional theory calculations of NMR shielding constants and spin–rotation constants using accurate coupled-cluster calculations},
    journal = {The Journal of Chemical Physics},
    volume = {138},
    number = {2},
    pages = {024111},
    year = {2013},
    month = {01},
   
    issn = {0021-9606},
    doi = {10.1063/1.4773016},
    url = {https://doi.org/10.1063/1.4773016},
    eprint = {https://pubs.aip.org/aip/jcp/article-pdf/doi/10.1063/1.4773016/14790759/024111\_1\_online.pdf},
}

@article{Breit-1932,
  title = {Dirac's Equation and the Spin-Spin Interactions of Two Electrons},
  author = {Breit, G.},
  journal = {Phys. Rev.},
  volume = {39},
  issue = {4},
  pages = {616--624},
  numpages = {0},
  year = {1932},
  month = {Feb},
  publisher = {American Physical Society},
  doi = {10.1103/PhysRev.39.616},
  url = {https://link.aps.org/doi/10.1103/PhysRev.39.616}
}

@article{Howard1970,
author = {B.J. Howard and R.E. Moss},
title = {The molecular hamiltonian},
journal = {Molecular Physics},
volume = {19},
number = {4},
pages = {433--450},
year = {1970},
publisher = {Taylor \& Francis},
doi = {10.1080/00268977000101471},


URL = { 
    
        https://doi.org/10.1080/00268977000101471
    
    

},
eprint = { 
    
        https://doi.org/10.1080/00268977000101471
    
    

}

}

@article{KISIEL2009177,
title = {Assignment and analysis of the rotational spectrum of bromoform enabled by broadband FTMW spectroscopy},
journal = {Journal of Molecular Spectroscopy},
volume = {257},
number = {2},
pages = {177-186},
year = {2009},
issn = {0022-2852},
doi = {https://doi.org/10.1016/j.jms.2009.08.006},
url = {https://www.sciencedirect.com/science/article/pii/S0022285209001933},
author = {Zbigniew Kisiel and Adam Kraśnicki and Lech Pszczółkowski and Steven T. Shipman and Leonardo Alvarez-Valtierra and Brooks H. Pate},
}

@article{Itano-1980,
    author = {Itano, Wayne M. and Ozier, Irving},
    title = {Avoided‐crossing molecular‐beam spectroscopy of methane},
    journal = {The Journal of Chemical Physics},
    volume = {72},
    number = {6},
    pages = {3700-3711},
    year = {1980},
    month = {03},
    
    issn = {0021-9606},
    doi = {10.1063/1.439581},
    url = {https://doi.org/10.1063/1.439581},
    eprint = {https://pubs.aip.org/aip/jcp/article-pdf/72/6/3700/18921505/3700\_1\_online.pdf},
}

@article{JAMESON1987461,
title = {Gas-phase 13C chemical shifts in the zero-pressure limit: refinements to the absolute shielding scale for 13C},
journal = {Chemical Physics Letters},
volume = {134},
number = {5},
pages = {461-466},
year = {1987},
issn = {0009-2614},
doi = {https://doi.org/10.1016/0009-2614(87)87173-7},
url = {https://www.sciencedirect.com/science/article/pii/0009261487871737},
author = {A.Keith Jameson and Cynthia J. Jameson},
}

@article{Gregory-2016,
  title = {Controlling the rotational and hyperfine state of ultracold $^{87}\mathrm{Rb}^{133}\mathrm{Cs}$ molecules},
  author = {Gregory, Philip D. and Aldegunde, Jesus and Hutson, Jeremy M. and Cornish, Simon L.},
  journal = {Phys. Rev. A},
  volume = {94},
  issue = {4},
  pages = {041403},
  numpages = {5},
  year = {2016},
  month = {Oct},
  publisher = {American Physical Society},
  doi = {10.1103/PhysRevA.94.041403},
  url = {https://link.aps.org/doi/10.1103/PhysRevA.94.041403}
}

@article{Ramsey1951,
  title = {Dependence of Magnetic Shielding of Nuclei upon Molecular Orientation},
  author = {Ramsey, Norman F.},
  journal = {Phys. Rev.},
  volume = {83},
  issue = {3},
  pages = {540--541},
  numpages = {0},
  year = {1951},
  month = {Aug},
  publisher = {American Physical Society},
  doi = {10.1103/PhysRev.83.540},
  url = {https://link.aps.org/doi/10.1103/PhysRev.83.540}
}

@article{Ruud2014,
    author = {Ruud, Kenneth and Demissie, Taye B. and Jaszuński, Michał},
    title = {Ab initio and relativistic DFT study of spin–rotation and NMR shielding constants in XF6 molecules, X = S, Se, Te, Mo, and W},
    journal = {The Journal of Chemical Physics},
    volume = {140},
    number = {19},
    pages = {194308},
    year = {2014},
    month = {05},
    issn = {0021-9606},
    doi = {10.1063/1.4875696},
    url = {https://doi.org/10.1063/1.4875696},
    eprint = {https://pubs.aip.org/aip/jcp/article-pdf/doi/10.1063/1.4875696/9542198/194308\_1\_online.pdf},
}

@article{Helgaker1991,
    author = {Helgaker, Trygve and J{\o}rgensen, Poul},
    title = {An electronic Hamiltonian for origin independent calculations of magnetic properties},
    journal = {The Journal of Chemical Physics},
    volume = {95},
    number = {4},
    pages = {2595-2601},
    year = {1991},
    month = {08},

    issn = {0021-9606},
    doi = {10.1063/1.460912},
    url = {https://doi.org/10.1063/1.460912},
    eprint = {https://pubs.aip.org/aip/jcp/article-pdf/95/4/2595/18994709/2595\_1\_online.pdf},
}

@article{Helgaker1998,
author = {Helgaker, Trygve and Jaszuński, Michał and Ruud, Kenneth},
title = {Ab Initio Methods for the Calculation of NMR Shielding and Indirect Spin−Spin Coupling Constants},
journal = {Chemical Reviews},
volume = {99},
number = {1},
pages = {293-352},
year = {1999},
doi = {10.1021/cr960017t},
    note ={PMID: 11848983},

URL = { 
    
        https://doi.org/10.1021/cr960017t
    
    

},
eprint = { 
    
        https://doi.org/10.1021/cr960017t
    
    

}

}

@article{Wong2023,
    author = {Wong, Jonathan and Ganoe, Brad and Liu, Xiao and Neudecker, Tim and Lee, Joonho and Liang, Jiashu and Wang, Zhe and Li, Jie and Rettig, Adam and Head-Gordon, Teresa and Head-Gordon, Martin},
    title = {An in-silico NMR laboratory for nuclear magnetic shieldings computed via finite fields: Exploring nucleus-specific renormalizations of MP2 and MP3},
    journal = {The Journal of Chemical Physics},
    volume = {158},
    number = {16},
    pages = {164116},
    year = {2023},
    month = {04},
    issn = {0021-9606},
    doi = {10.1063/5.0145130},
    url = {https://doi.org/10.1063/5.0145130},
    eprint = {https://pubs.aip.org/aip/jcp/article-pdf/doi/10.1063/5.0145130/17105222/164116\_1\_5.0145130.pdf},
}

@article{Born1927,
author = {Born, M. and Oppenheimer, R.},
title = {Zur Quantentheorie der Molekeln},
journal = {Annalen der Physik},
volume = {389},
number = {20},
pages = {457-484},
doi = {https://doi.org/10.1002/andp.19273892002},
url = {https://onlinelibrary.wiley.com/doi/abs/10.1002/andp.19273892002},
eprint = {https://onlinelibrary.wiley.com/doi/pdf/10.1002/andp.19273892002},
year = {1927}
}

@article{Eckart1935,
  title = {Some Studies Concerning Rotating Axes and Polyatomic Molecules},
  author = {Eckart, Carl},
  journal = {Phys. Rev.},
  volume = {47},
  issue = {7},
  pages = {552--558},
  numpages = {0},
  year = {1935},
  month = {Apr},
  publisher = {American Physical Society},
  doi = {10.1103/PhysRev.47.552},
  url = {https://link.aps.org/doi/10.1103/PhysRev.47.552}
}

@article{Meyer2002,
   author = "Meyer, Henning",
   title = "THE
MOLECULAR
HAMILTONIAN", 
   journal= "Annual Review of Physical Chemistry",
   year = "2002",
   volume = "53",
   number = "Volume 53, 2002",
   pages = "141-172",
   doi = "https://doi.org/10.1146/annurev.physchem.53.082201.124330",
   url = "https://www.annualreviews.org/content/journals/10.1146/annurev.physchem.53.082201.124330",
   publisher = "Annual Reviews",
   issn = "1545-1593",
   type = "Journal Article",

  }

@article{Fisher1971,
    author = {Fisher, George P.},
    title = {The Electric Dipole Moment of a Moving Magnetic Dipole},
    journal = {American Journal of Physics},
    volume = {39},
    number = {12},
    pages = {1528-1533},
    year = {1971},
    month = {12},
    issn = {0002-9505},
    doi = {10.1119/1.1976708},
    url = {https://doi.org/10.1119/1.1976708},
    eprint = {https://pubs.aip.org/aapt/ajp/article-pdf/39/12/1528/11842022/1528_1_online.pdf},
}

@article{THOMAS1926,
  author    = {L. H. Thomas},
  title     = {The Motion of the Spinning Electron},
  journal   = {Nature},
  year      = {1926},
  volume    = {117},
  number    = {2945},
  pages     = {514--514},
  doi       = {10.1038/117514a0},
  url       = {https://doi.org/10.1038/117514a0},
  issn      = {1476-4687}
}

@article{Stochyev2018,
author = {Stoychev, Georgi L. and Auer, Alexander A. and Izsák, Róbert and Neese, Frank},
title = {Self-Consistent Field Calculation of Nuclear Magnetic Resonance Chemical Shielding Constants Using Gauge-Including Atomic Orbitals and Approximate Two-Electron Integrals},
journal = {Journal of Chemical Theory and Computation},
volume = {14},
number = {2},
pages = {619-637},
year = {2018},
doi = {10.1021/acs.jctc.7b01006},
    note ={PMID: 29301077},

URL = { 
    
        https://doi.org/10.1021/acs.jctc.7b01006
    
    

},
eprint = { 
    
        https://doi.org/10.1021/acs.jctc.7b01006
    
    

}

}

@article{DEVINE1986,
title = {Magnetic vibrational circular dichroism of the haloforms},
journal = {Chemical Physics Letters},
volume = {124},
number = {4},
pages = {341-344},
year = {1986},
issn = {0009-2614},
doi = {https://doi.org/10.1016/0009-2614(86)85030-8},
url = {https://www.sciencedirect.com/science/article/pii/0009261486850308},
author = {T.R. Devine and T.A. Keiderling}
}

@book{levitt2008spin,
  title={Spin dynamics: basics of nuclear magnetic resonance},
  author={Levitt, Malcolm H},
  year={2008},
  publisher={John Wiley \& Sons}
}

@article{Malkin2013,
author = {Malkin, Elena and Komorovsky, Stanislav and Repisky, Michal and Demissie, Taye B. and Ruud, Kenneth},
title = {The Absolute Shielding Constants of Heavy Nuclei: Resolving the Enigma of the 119Sn Absolute Shielding},
journal = {The Journal of Physical Chemistry Letters},
volume = {4},
number = {3},
pages = {459-463},
year = {2013},
doi = {10.1021/jz302146m},
    note ={PMID: 26281741},

URL = { 
    
        https://doi.org/10.1021/jz302146m
    
    

},
eprint = { 
    
        https://doi.org/10.1021/jz302146m
    
    

}

}

@article{
ChiralPhonons2023,
author = {Jiaming Luo  and Tong Lin  and Junjie Zhang  and Xiaotong Chen  and Elizabeth R. Blackert  and Rui Xu  and Boris I. Yakobson  and Hanyu Zhu },
title = {Large effective magnetic fields from chiral phonons in rare-earth halides},
journal = {Science},
volume = {382},
number = {6671},
pages = {698-702},
year = {2023},
doi = {10.1126/science.adi9601},
URL = {https://www.science.org/doi/abs/10.1126/science.adi9601},
eprint = {https://www.science.org/doi/pdf/10.1126/science.adi9601}}

\clearpage
\section*{Supplementary Information}
\addcontentsline{toc}{section}{Supplementary Information}
\setcounter{section}{0}
\setcounter{equation}{0}
\setcounter{figure}{0}
\setcounter{table}{0}
\makeatletter
\let\@seccntformat\@sseccntformat
\@removefromreset{equation}{section}
\@removefromreset{figure}{section}
\@removefromreset{table}{section}
\makeatother
\renewcommand{\thesection}{S\arabic{section}}
\renewcommand{\thesubsection}{S\arabic{section}.\arabic{subsection}}
\renewcommand{\theequation}{S\arabic{equation}}
\renewcommand{\thefigure}{S\arabic{figure}}
\renewcommand{\thetable}{S\arabic{table}}

In this Supplementary Information, we present a short introduction on normal coordinates and the isotropic harmonic oscillator. Furthermore we describe the nuclear wavefunction describing pseudorotation in the harmonic approximation and present additional results for the vibrationally induced hyperfine splitting and induced magnetic fields for a selection of molecules.

\section{Normal coordinates}
\label{sec:Normal-coordinates}
This section provides a short introduction to normal coordinates. For further details we recommend reading the textbook of Wilson, Decius and Cross.\cite{wilson1980molecular}
The total nuclear Hamiltonian in the Born-Oppenheimer approximation, written in the harmonic expansion, is
\begin{equation}
    H_{\mathrm{N}} = \sum_{a = 1}^{N_\mathrm{N}}\frac{\bm{P}_a^2}{m_a} + \sum_{a,b = 1}^{N_\mathrm{N}}\sum_{\alpha,\beta = 1}^{3} V_{a\alpha,b\beta}\,\delta R_{a,\alpha}\delta R_{b,\beta}\,,
\end{equation}
where $\delta R_{a,\alpha} = (\bm{R}_a-\bm{R}^0_a)_{\alpha}$. Using mass weighted coordinates $\delta R_{a,\alpha}^{\mathrm{mw}} = \sqrt{m_a}\delta R_{a,\alpha}$ and $\bm{P}^{\mathrm{mw}}_a=\frac{1}{\sqrt{m_a}}\bm{P}_a$ we get
\begin{equation}
    H_{\mathrm{N}} = \sum_{a=1}^{N_\mathrm{N}} (\bm{P}^{\mathrm{mw}}_a)^2 + \sum_{a,b=1}^{N_\mathrm{N}} \sum_{\alpha, \beta=1}^3\tilde{V}_{a\alpha,b\beta}\,\delta R^{\mathrm{mw}}_{a,\alpha}\delta R^{\mathrm{mw}}_{b,\beta}\,,
\end{equation}
where $\tilde{V}_{a\alpha,b\beta} = \frac{1}{\sqrt{m_a}\sqrt{m_b}} {V}_{a\alpha,b\beta}$ is the mass weighted Hessian matrix. Via the solution of the eigenvalue problem
\begin{equation}
    \tilde{V}\bm{v}_k = \omega_k^2\bm{{v}}_k\,,
\end{equation}
we define the normal mode vectors $\bm{v}_k$ of the Hessian, which obey $\bm{v}_k\cdot \bm{v}_j = \delta_{kj}$. Note that the $3N_{\mathrm{N}}$ dimensional vector  $\bm{v}_k$ can be written as
\begin{equation}
    \bm{v}_k= \begin{pmatrix}
        \bm{l}_{1,k}\\
        \bm{l}_{2,k}\\
        \vdots\\
        \bm{l}_{N_{\mathrm{N}},k}
    \end{pmatrix}\,.
    \label{eq:mode-vector-displacement-relation}
\end{equation}
Here, $\frac{1}{\sqrt{m_a}} \bm{l}_{a,k}$ is the displacement vector of atom $a$ for the $k$-th mode. We can then introduce normal coordinates $Q_r$ and their conjugate momenta $P_r$, satisfying the commutation relation $[Q_r,P_s] = i \hbar\ \delta_{rs}$, and the equations
\begin{align}
    \delta\bm{R}_{a} &= \frac{1}{\sqrt{m_a}} \sum_{r=1}^{3N_{\mathrm{N}}-6}\bm{l}_{a,r} Q_r\,,\\
    \bm{P}_a & = {\sqrt{m_a}} \sum_{r=1}^{3N_{\mathrm{N}}-6}\bm{l}_{a,r} P_r\,.
\end{align}
The sum runs only over $3N_{\mathrm{N}}-6$, as there are six trivial zero mode vectors which correspond to three translations and rotations. In normal coordinates, neglecting translation and rotational degrees of freedom, the nuclear Hamiltonian becomes
\begin{equation}
    H_{\mathrm{N}} = \sum_{r=1}^{3N_\mathrm{N}-6} \frac{1}{2}\left[P_r^2+ \omega_r^2Q_r^2\right]\,.
\end{equation}

\section{Isotropic harmonic oscillator in arbitrary dimensions}
\label{sec:appendix-harmonic-oscillator}
In the harmonic approximation, the different vibrations do not couple and the nuclear state is a product of the individual vibrational eigenfunctions. However, in the case of degenerate vibrations, linear combinations of the degenerate eigenstates are eigenstates as well. We are interested in the subspace of the degenerate mode $t$, in which the Hamiltonian is an $N$-dimensional isotropic harmonic oscillator and reads
\begin{equation}
    H_t = \frac{1}{2}\sum_{r\in\{t\}} P_r^2 + \frac{1}{2}  \omega_t^2 \sum_{r \in \{t\}} Q_r^2 = \sum_{r\in\{t\}}  \hbar \omega_t  \left[\frac{1}{2} + a_r^{\dagger} a_r\right]  \,.
\end{equation}
The eigenstates of this Hamiltonian are tensor products of 1-dimensional harmonic oscillator states $\Phi_{n_r}(Q_r)$
\begin{equation}
    \chi_{n_1,...,n_N}(Q_1,...,Q_N) = \prod_{r\in\{t\}} \Phi_{n_r}(Q_r)\,.
\end{equation}
In the two- or three-dimensional subspaces of degenerate vibrational modes, it is possible to choose simultaneous eigenstates of the Hamiltonian and one component of the angular momentum operator.

\subsection{Two-dimensional harmonic oscillator}
\label{sec:appendix-2-d-oscillator}
We look now explicitly at simultaneous eigenstates of $H_t$ and $L_z$ for twofold degeneracy. The angular momentum operator $L_z$, expressed in terms of the creation and annihilation operators, reads
\begin{equation}
    L_z = Q_{t_1}P_{t_2} - Q_{t_2}P_{t_1} = i \hbar [a_1 a_2^{\dagger} - a_2 a_1^{\dagger}]\,.
\end{equation}
This expression is not diagonal in the number operator $a^\dagger a$. One can, however, perform a unitary transformation and define two new annihilation operators
\begin{align}
    b_1 &= \frac{a_1 + i a_2}{\sqrt{2}}, \\
    b_2 & = \frac{a_1 - i a_2}{\sqrt{2}},
\end{align}
which also fulfill $[b_1,b_1^{\dagger}] = [b_2,b_2^{\dagger}] = 1$, with all other commutators being zero. In terms of these new operators, the angular momentum operator and the Hamiltonian take the form
\begin{align}
    L_z &= \hbar \left[b_2^{\dagger}b_2 - b_1^{\dagger}b_1\right]\,,\\
    H_t &= \hbar \omega_t \left[1 + b_1^{\dagger}b_1 + b_2^{\dagger}b_2 \right] \,.
\end{align}
The simultaneous eigenstates of $H$ and $L_z$ are then obtained by application of the raising operators $b_1^\dagger$ and $b_2^{\dagger}$, $n_1^\dagger$ and $n_2$ times, respectively
\begin{equation}
    \ket{n_1,n_2} = (b_1^{\dagger})^{n_1} (b_2^{\dagger})^{n_2} \ket{0}\,,
\end{equation}
with the corresponding energy
\begin{equation}
    E_{\nu} = \hbar\omega_t(1+\nu)\,.
\end{equation}
The principal quantum number $\nu$ is then a sum $\nu = n_1 + n_2$ of $n_1$ and $n_2$. For the eigenvalues of $L_z$ we find $l=n_2 - n_1$. Therefore, we can label these states $\ket{\nu,l}$. It is straightforward that the possible values for $l$ are evenly spaced integer values with spacing $\Delta l=2$. Thus, the possible $l$ values are $-\nu,\nu+2,...\nu$.

\subsection{Three-dimensional harmonic oscillator}
The three dimensional isotropic harmonic oscillator can be described in spherical coordinates. The eigenfunctions are 
\begin{equation}
    \chi_{\nu,l,m} = Y_{l,m}(\theta ,\phi) u_{\nu,l}(r)\,,
\end{equation}
where $Y_{l,m}(\theta,\phi)$ are the spherical harmonics. $Y_{l,m}(\theta,\phi)$ are eigenfunctions of $L_z$ and $\bm{L}^2$, with integer eigenvalues $m$ and $l(l+1)$, where $m$ runs from $-l$ to $l$, and $l$ from $0$ to $\nu$. The eigenenergy is $E_\nu= \left[\frac{3}{2} + \nu\right] \hbar\omega_t$. 

\subsection{Expectation values}
We are interested in expectation values of $Q_rP_r$ (sum convention does not apply here) with regard to arbitrary energy eigenstates $\ket{\chi}$ of the $n$-dimensional isotropic oscillator. First, we look at the commutator 
\begin{equation}
    \left[H_t,Q_r^2\right] = \frac{1}{2}[P_r^2,Q_r^2]  = -i\hbar \underbrace{\{P_r,Q_r\}}_{\displaystyle P_rQ_r+Q_rP_r}\,,
\end{equation}
where we used $\left[AB,C\right] = A\left[B,C\right] + \left[B,C\right]A$ several times. Thus, we can deduce
\begin{equation}
    \bra{\chi}\left[H_t,Q_r^2\right] \ket{\chi} = - i \hbar \bra{\chi}\{P_r,Q_r\} \ket{\chi} = 0\,,
\end{equation}
which leads us to
\begin{equation}
        \bra{\chi}Q_r P_r\ket{\chi} = \frac{1}{2} \bra{\chi}[Q_r,P_r]+\{Q_r,P_r\}\ket{\chi} = \frac{1}{2} i\hbar \,
        \label{eq:appendix-q-p-same-index-expec}
\end{equation}
for arbitrary energy eigenfunctions $\ket{\chi}$ of $H_t$. Additionally, we calculate the expectation value of $Q_{r}$ and $P_{s}$ explicitly for the three- and twofold degeneracy, where $r\neq s$ both correspond to the same degenerate mode.

\subsubsection{Expectation values for twofold degeneracy}
We express $Q_{t_1} P_{t_2}$ via cylindrical ladder operators and find
\begin{equation}
    Q_{t_1} P_{t_2} = \frac{\hbar}{4} \left[(b_2^{\dagger})^2 + b_2^2 - [(b_1^{\dagger})^2 + b_1^2] +\frac{2L_z}{\hbar} \right ]\,,
\end{equation}
and a similar expression for $Q_{t_2} P_{t_1} = Q_{t_1}P_{t_2} - L_z$. Thus, we obtain
\begin{equation}
    \bra{\nu,l} Q_{t_1} P_{t_2} \ket{\nu,l} = -\bra{\nu,l} Q_{t_2} P_{t_1} \ket{\nu,l} =\frac{1}{2} \hbar l = \frac{1}{2}
    \langle G_t \rangle\,. 
    \label{eq:appendix-ang-momentum-2d-harmonic}
\end{equation}

\subsubsection{Expectation values for threefold degeneracy}
For threefold degeneracy we find 
\begin{equation}
    \bra{\nu,l,m}Q_{t_1}P_{t_2}\ket{\nu,l,m} = -\bra{\nu,l,m}Q_{t_2}P_{t_1}\ket{\nu,l,m} = \frac{1}{2}m \hbar
        \label{eq:appendix-ang-momentum-3d-harmonic}
\end{equation}
and 
\begin{equation}
    \bra{\nu,l,m}Q_{t_2}P_{t_3}\ket{\nu,l,m} = \bra{\nu,l,m}Q_{t_1}P_{t_3}\ket{\nu,l,m} = 0.
\end{equation}

\section{DFT results for trihalomethanes}
For the sake of completeness we also present results entirely based on density functional theory, while data presented in Table~2 of the main manuscript relies on CCSD(T) geometries and frequencies for the vibrationally induced hyperfine splittings and magnetic fields of fluoroform, bromoform and chloroform. The data presented in Tables \ref{tab:spin-vib-trihalomethanes-DFT}, \ref{tab:spin-vib-trichlorobenzene}, \ref{tab:spin-vib-trifluorobenzene}, \ref{tab:spin-vib-tribromobenzene} is exclusively based on DFT results obtained with the $\omega{}$-B97X-V functional in combination with the cc-pVTZ basis for geometry optimization and frequency calculations. For the calculation of magnetic shielding tensors, the same functional with the aug-cc-pVQZ was employed. The calculated magnetic fields display only minimal deviations from the data in Table~2 of the main text.  
\newpage
\begin{table}[!h]
\caption{Spin-vibration coupling in trihalomethanes. Absolute values of vibrationally induced magnetic fields $B^{\mathrm{eff}}_{\mathrm{vib}}$ are given in $\mathrm{mT}$, vibration-induced hyperfine splittings $H_{\mathrm{vib}}^{\mathrm{hyp}}$ in kHz. $B^{\mathrm{SOOC}}_{\mathrm{vib}}$ and $B^{\mathrm{SOC}}_{\mathrm{vib}}$ refer to the contributions of nuclear spin-orbit and nuclear spin-other-orbit coupling. Absolute values of the Coriolis coupling $\bm{\zeta}$ are a measure for the total vibrational angular momentum, whereas absolute values of atomic contributions reveal the intramolecular distribution of angular momentum. The relative error describes the difference to the values in Tab.~2 of the main text, where the geometry optimization and frequency calculation was done with CCSD(T).}

\begin{tabular}{lcccccccc}
\label{tab:spin-vib-trihalomethanes-DFT}
   Molecule & $\bar{\nu}\left[\mathrm{cm}^{-1}\right]$ & $|\bm{\zeta}|$ & Atom & $B^{\mathrm{eff}}_{\mathrm{vib}}\left[\mathrm{mT}\right]$  & $|\bm{\zeta}_{\alpha}|$ & $H_{\mathrm{vib}}^{\mathrm{hyp}}\left[\mathrm{kHz}\right]$  & rel. Err. in \%\\
  \toprule
\multirow{9}{*}{\textsuperscript{13}CHF\textsubscript{3}}& \multirow{3}{*}{$510$} & \multirow{3}{*}{$0.792$} & F & 0.{102}  & 0.296 & {4.093} & {1.56}\\
  & &  & \textsuperscript{13}C & 0.{122}  & 0.062 & {1.308} & 0.{27}\\
  & & & H & 0.{037} & 0.0187  & {1.556} & 1.{35}\\
\cline{2-8}
\noalign{\vskip 4pt}

& \multirow{3}{*}{$1159$} & \multirow{3}{*}{$0.712$} & F & 0.09{1} & 0.020 & 3.6{56} & 2.{96}\\
  & &  & \textsuperscript{13}C & 0.{161}  & 0.561 & {1.720} & {1.30} \\
  & & & H & 0.{856}  & 0.102 & {36.434} & 2.{49}\\
  \cline{2-8}
\noalign{\vskip 4pt}
& \multirow{3}{*}{$1396$} & \multirow{3}{*}{$0.986$} & F &  0.23{4}  & 0.005 & 9.3{76} & 0.4{6}\\
  & &  & \textsuperscript{13}C & 1.08{8}  & 0.160 & 11.{645} & 1.{10}\\

  & & & H & {1.940}  & 0.822 & {82.620} & {2.40}\\
\midrule
\multirow{9}{*}{\textsuperscript{13}CH\textsuperscript{35}Cl\textsubscript{3}} & \multirow{3}{*}{$267$} & \multirow{3}{*}{$0.868$} & \textsuperscript{35}Cl & 0.1{49}  & 0.310 & 0.62{1} & 1.6{7}\\
  & &  & \textsuperscript{13}C & 0.{005}  & 0.049 & {0.053} & {61.13}\\
  & & & H & 0.{045} & 0.008  & {1.925}  & 0.{41}\\
\cline{2-8}
\noalign{\vskip 4pt}
& \multirow{3}{*}{$791$} & \multirow{3}{*}{$0.84$} & \textsuperscript{35}Cl & 0.121  & 0.015& 0.5{11} & 2.{35} \\

  & &  & \textsuperscript{13}C & 0.2{38}  & 0.77 & 2.{551} & 2.{08} \\

  & & & H & 0.{500} & 0.036 & 2{1.302} & {8.80}\\
\cline{2-8}
\noalign{\vskip 4pt}
& \multirow{3}{*}{$1261$} & \multirow{3}{*}{$0.989$} & \textsuperscript{35}Cl & 0.213  &  0.001 & 0.891 & 0.39\\
 & &  & \textsuperscript{13}C & 0.9{81}  & 0.058 & 10.{508} & 0.1{9}\\
  & & & H & {1.507}  & 0.932 & {64.157} & {1.67}\\
\midrule
\multirow{9}{*}{\textsuperscript{13}CH\textsuperscript{79}Br\textsubscript{3}} & \multirow{3}{*}{$156$} & \multirow{3}{*}{$0.935$} & \textsuperscript{81}Br & 0.1{46}  & 0.323 & 1.{567} & 5.{13}\\
  & &  & \textsuperscript{13}C & 0.{071} & 0.026 & {0.763} & {3.74}\\

  & & & H & 0.1{24} & 0.004  & {5.282}  & 0.{33}\\
\cline{2-8}
\noalign{\vskip 4pt}
& \multirow{3}{*}{$699$} & \multirow{3}{*}{$0.923$} & \textsuperscript{81}Br & 0.15{2}  & 0.007 & 1.6{27} & 6.1{4}\\
  & &  & \textsuperscript{13}C & 0.{327}  & 0.856 & {3.497} & {2.14} \\
  & & & H & 0.{670} & 0.047 & {28.513} & {9.39}\\
\cline{2-8}
\noalign{\vskip 4pt}
& \multirow{3}{*}{$1200$} & \multirow{3}{*}{$0.996$} & \textsuperscript{81}Br & 0.206  & 0.000 & 2.386 & 0.29\\
  & &  & \textsuperscript{13}C & 0.9{61}  & 0.058 & 10.2{92} & 0.{17}\\
  & & & H & {1.511}  & 0.939 & {64.316} & {3.13}\\
\bottomrule
\end{tabular}
\end{table}

\newpage
\section{DFT-results for benzene-substituted derivatives}
Here we present the calculated vibrational induced fields and splittings for 1-3-5 trichlorobenzene, tribromobenzene and trifluorobenzene, as discussed in Section~3.4.2 of the main text.
\begin{table}[!h]
\caption{Spin-vibration coupling in 1-3-5 trichlorobenzene. Coriolis coupling in all displayed degenerate pairs is perpendicular to the plane of the atoms. For symmetry reasons, it is sufficient to present the absolute values on one H atom, one Cl atom and two C atoms. For more details on the displayed quantities see Tab.~S1.  Only infrared active degenerate modes are displayed. \hl{The second carbon atom is bound to a hydrogen atom and the first to a chlorine atom.}}

\begin{tabular}{lccccccc}
\label{tab:spin-vib-trichlorobenzene}

  $\bar{\nu}\left[\mathrm{cm}^{-1}\right]$ & $|\bm{\zeta}|$ & Atom & $B^{\mathrm{eff}}_{\mathrm{vib}}\left[\mathrm{mT}\right]$ & $B_{\mathrm{vib}}^{\mathrm{SOC}}\left[\mathrm{mT}\right]$ & $B_{\mathrm{vib}}^{\mathrm{SOOC}}\left[\mathrm{mT}\right]$ & $|\bm{\zeta}_{\alpha}|$ & $H_{\mathrm{vib}}^{\mathrm{hyp}}\left[\mathrm{kHz}\right]$  \\
  \hline
\multirow{4}{*}{$429$} &  \multirow{4}{*}{$0.115$} &\textsuperscript{35}Cl&  0.02{3} & 0.001 & 0.024 & 0.005 & 0.{098}  \\
  &  & \textsuperscript{13}C & 0.0{10} & 0.0{42} & 0.034 & 0.032 & 0.{103} \\
   & & \textsuperscript{13}C & 0.0{29} & 0.0{01} & 0.030 & 0.050 & 0.{315} \\
  & & H & 0.1{85} & 0.{200} & 0.017 & 0.016 & {7.869} \\
\hline
\multirow{4}{*}{$811$} & \multirow{4}{*}{$ 0.372$} & \textsuperscript{35}Cl & 0.00{3} & 0.000 & 0.003 & 0.012 & 0.01{2} \\
  &  & \textsuperscript{13}C & 0.0{11} & 0.0{14} & 0.003 & 0.010 & 0.{117} \\
    & & \textsuperscript{13}C & 0.0{10} & 0.0{17} & 0.027 & 0.124 & 0.{108} \\

  & & H & 0.07{9} & 0.0{68} & 0.011 & 0.003 & 3.{363} \\

\hline
\multirow{4}{*}{$1122$} & \multirow{4}{*}{$0.409$} & \textsuperscript{35}Cl & 0.010 & 0.000 & 0.010 & 0.001 & 0.041 \\
    & & \textsuperscript{13}C & 0.0{33} & 0.0{02} & 0.036 & 0.024 & 0.{355} \\
  & & \textsuperscript{13}C & 0.{089} & 0.0{07} & 0.096 & 0.061 & {0.952} \\
  & & H & 0.{502} & 0.{538} & 0.036 & 0.049 & {21.386} \\
\hline
\multirow{4}{*}{$1428$} & \multirow{4}{*}{$0.516$} &\textsuperscript{35}Cl & 0.008 & 0.000 & 0.008 & 0.001 & 0.035 \\
  &  & \textsuperscript{13}C & 0.{013} & 0.0{13} & 0.027 & 0.146 & {0.142} \\
  & & \textsuperscript{13}C & 0.1{31} & 0.0{08} & 0.139 & 0.067 & 1.{404} \\
  & & H & 0.{359} & 0.{359} & 0.001 & 0.042 & {15.303} \\
\hline
\multirow{4}{*}{$1597$} & \multirow{4}{*}{$0.652$} & \textsuperscript{35}Cl & 0.013 & 0.000 & 0.013 & 0.000 & 0.053 \\
  & & \textsuperscript{13}C & 0.0{38} & 0.0{15} & 0.022 & 0.122 & 0.{406} \\
  & & \textsuperscript{13}C & 0.0{99} & 0.0{04} & 0.095 & 0.108 & {1.055} \\
  & & H & 0.{194} & 0.1{85} & 0.009 & 0.013 & {8.261} \\
\hline
\end{tabular}
\end{table}

\begin{table}[!h]
\caption{Spin-vibration coupling in 1-3-5trifluorobenzene. Coriolis coupling in all displayed degenerate pairs is perpendicular to the plane of the atoms. For symmetry reasons, it is sufficient to present the absolute values on one H atom, one F atom and two C atoms. For more details on the displayed quantities see Tab.~S1. Only infrared active degenerate modes are displayed. \hl{The first carbon atom is bound to a hydrogen atom and the second to a fluorine atom.}}

\begin{tabular}{lccccccc}
\label{tab:spin-vib-trifluorobenzene}

  $\bar{\nu}\left[\mathrm{cm}^{-1}\right]$ & $|\bm{\zeta}|$ & Atom & $B^{\mathrm{eff}}_{\mathrm{vib}}\left[\mathrm{mT}\right]$ & $B_{\mathrm{vib}}^{\mathrm{SOC}}\left[\mathrm{mT}\right]$ & $B_{\mathrm{vib}}^{\mathrm{SOOC}}\left[\mathrm{mT}\right]$ & $|\bm{\zeta}_{\alpha}|$ & $H_{\mathrm{vib}}^{\mathrm{hyp}}\left[\mathrm{kHz}\right]$  \\
  \hline

\multirow{4}{*}{$499$} & \multirow{4}{*}{$ 0.321$} & F & 0.0{17} & 0.01{5} & 0.032 & 0.018 & {0.671} \\
  & & \textsuperscript{13}C & 0.0{07} & 0.0{45} & 0.036 & 0.091 & 0.{080} \\
  & & \textsuperscript{13}C & 0.04{4} & 0.00{3} & 0.047 & 0.015 & 0.4{69} \\
  & & H & 0.{308} & 0.{321} & 0.012 & 0.018 & 1{3.132} \\
\hline
\multirow{4}{*}{$1005$} & \multirow{4}{*}{$0.501$} & F & 0.0{04} & 0.00{8} & 0.011 & 0.018 & 0.{141} \\
  &  & \textsuperscript{13}C & 0.01{8} & 0.00{3} & 0.015 & 0.008 & 0.1{91} \\
  & & \textsuperscript{13}C & 0.{063} & 0.0{01} & 0.081 & 0.161 & {0.677} \\
  & & H & 0.{313} & 0.{347} & 0.042 & 0.045 & {13.308} \\
\hline
\multirow{4}{*}{$1140$} & \multirow{4}{*}{$0.093$} & F & 0.016 & 0.001 & 0.013 & 0.009 & 0.6{50} \\
  & & \textsuperscript{13}C & 0.0{01} & 0.0{03} & 0.002 & 0.027 & 0.{013} \\
  & & \textsuperscript{13}C & 0.00{9} & 0.00{2} & 0.005 & 0.003 & 0.0{94} \\
  & & H & 0.{106} & 0.{108} & 0.003 & 0.006 & {4.492} \\
\hline
\multirow{4}{*}{$1479$} & \multirow{4}{*}{$0.722$} &  F & 0.02{4} & 0.001 & 0.024 & 0.007 & {0.981} \\

  &  & \textsuperscript{13}C & 0.{001} & 0.{019} & 0.020 & 0.195 & {0.009} \\
  & & \textsuperscript{13}C & 0.{115} & 0.0{03} & 0.118 & 0.070 & {1.229} \\
  & & H & 0.{276} & 0.{267} & 0.009 & 0.031 & {11.746} \\
\hline
\multirow{4}{*}{$1639$} & \multirow{4}{*}{$0.731$} & F & 0.01{3} & 0.00{0} & 0.014 & 0.001 & 0.5{38} \\

  & & \textsuperscript{13}C & 0.0{35} & 0.{017} & 0.018 & 0.144 & 0.{379} \\
  & & \textsuperscript{13}C & 0.0{98} & 0.0{07} & 0.091 & 0.111 & {1.045} \\
  & & H & 0.{180} & 0.1{67} & 0.013 & 0.012 & {7.645} \\
\hline
\end{tabular}
\end{table}

\begin{table}[!h]
\caption{Spin-vibration coupling in 1-3-5 tribromobenzene. Coriolis coupling in all displayed degenerate pairs is perpendicular to the plane of the atoms. For symmetry reasons, it is sufficient to present the absolute values on one H atom, one Br atom and two C atoms. For more details on the displayed quantities see Tab.~S1. Only infrared active degenerate modes are displayed. \hl{The first carbon atom is bound to a hydrogen atom and the second to a bromine atom.}}
\begin{tabular}{lccccccc}
\label{tab:spin-vib-tribromobenzene}

  $\bar{\nu}\left[\mathrm{cm}^{-1}\right]$ & $|\bm{\zeta}|$ & Atom & $B^{\mathrm{eff}}_{\mathrm{vib}}\left[\mathrm{mT}\right]$ & $B_{\mathrm{vib}}^{\mathrm{SOC}}\left[\mathrm{mT}\right]$ & $B_{\mathrm{vib}}^{\mathrm{SOOC}}\left[\mathrm{mT}\right]$ & $|\bm{\zeta}_{\alpha}|$ & $H_{\mathrm{vib}}^{\mathrm{hyp}}\left[\mathrm{kHz}\right]$  \\
  \hline
\multirow{4}{*}{$355$} & \multirow{4}{*}{$0.486$} & \textsuperscript{79}Br & 0.00{7} & 0.00{0} & 0.006 & 0.016 & 0.0{75} \\
 & & \textsuperscript{13}C & 0.0{24} & 0.0{02} & 0.022 & 0.103 & 0.{256} \\
  &  & \textsuperscript{13}C & 0.{026} & 0.0{54} & 0.028 & 0.028 & {0.280} \\

  & & H & 0.{149} & 0.{173} & 0.023 & 0.016 & {6.361} \\
\hline
\multirow{4}{*}{$746$} & \multirow{4}{*}{$ 0.500$} & \textsuperscript{79}Br & 0.012 & 0.00{1} & 0.012 & 0.004 & 0.1{33} \\
 & & \textsuperscript{13}C & 0.02{2} & 0.0{18} & 0.040 & 0.132 & 0.2{33} \\
  &  & \textsuperscript{13}C & 0.0{14} & 0.0{12} & 0.002 & 0.040 & 0.{152} \\
  & & H & 0.1{37} & 0.12{6} & 0.010 & 0.001 & {5.815} \\
\hline
\multirow{4}{*}{$1127$} & \multirow{4}{*}{$ 0.473$} &\textsuperscript{79}Br & 0.014 & 0.000 & 0.014 & 0.000 & 0.15{2} \\

    & & \textsuperscript{13}C & 0.{093} & 0.0{07} & 0.094 & 0.076 & {0.998} \\
  & & \textsuperscript{13}C & 0.0{39} & 0.0{02} & 0.042 & 0.030 & 0.{419} \\
  & & H & 0.{525} & 0.{564} & 0.038 & 0.051 & {22.363} \\
\hline
\multirow{4}{*}{$1421$} & \multirow{4}{*}{$0.463$} & \textsuperscript{79}Br & 0.013 & 0.000 & 0.013 & 0.000 & 0.13{8} \\

 & & \textsuperscript{13}C
 & 0.1{32} & 0.0{09} & 0.142 & 0.063 & 1.{414} \\
  & & \textsuperscript{13}C & 0.{015} & 0.{013} & 0.029 & 0.135 & {0.163} \\

  & & H & 0.{364} & 0.{365} & 0.001 & 0.044 & {15.511} \\
\hline
\multirow{4}{*}{$1586$} & \multirow{4}{*}{$ 0.640$} & \textsuperscript{79}Br & 0.019 & 0.000 & 0.019 & 0.000 & 0.200 \\
 & & \textsuperscript{13}C & 0.{100} & 0.0{03} & 0.097 & 0.108 & {1.075} \\
  &  & \textsuperscript{13}C & 0.0{39} & 0.0{15} & 0.024 & 0.119 & 0.{423} \\
  & & H & 0.{197} & 0.{189} & 0.008 & 0.014 & {8.408} \\
\hline
\end{tabular}
\end{table}


\end{document}